\documentclass[aps,physrev,preprint,superscriptaddress]{revtex4-2}

\usepackage{amsmath}
\usepackage{bm}
\usepackage[version=3]{mhchem}
\usepackage{siunitx}
\usepackage[caption=false]{subfig}
\usepackage{nicefrac}
\usepackage{booktabs}
\usepackage[usenames]{xcolor}
\usepackage{tikz}
\usepackage{graphicx}
\usepackage{float}
\usepackage{verbatim}

\usepackage[acronyms]{glossaries}
\newacronym{co2}{CO2}{carbon dioxide}
\newacronym{sco2}{sCO2}{Supercritical carbon dioxide}
\newacronym{les}{LES}{large eddy simulation}
\newacronym{fda}{FDA}{Food and Drug Administration}
\newacronym{dns}{DNS}{direct numerical simulation}
\newacronym{egs}{EGS}{Enhanced Geothermal Systems}
\newacronym{cfd}{CFD}{computational fluid dynamics}
\newacronym{rans}{RANS}{Reynolds-averaged Navier-Stokes}
\newacronym{sgs}{SGS}{subgrid-scale}
\newacronym{srk}{SRK EoS}{Soave-Redlich-Kwong Equation of State}
\newacronym{pr}{PR EoS}{Peng-Robinson Equation of State}
\newacronym{eos}{EoS}{Equation of State}
\newacronym{nist}{NIST}{National Institute of Standards and Technology}
\newacronym{hpc}{HPC}{high-performance computing}
\newacronym{smd}{SMD}{dynamic Smagorinsky}
\newacronym{ppm}{PPM}{piecewise parabolic method}
\newacronym{nrel}{NREL}{National Renewable Energy Laboratory}
\newacronym{rms}{rms}{root mean square}
\newacronym{amr}{AMR}{adaptive mesh refinement}
\newacronym{ecp}{ECP}{Exascale Computing Project}
\newacronym{doe}{DOE}{Department of Energy}
\newacronym{hmhw}{HMHW}{half-mean half-width}
\newacronym{tke}{TKE}{turbulent kinetic energy}
\newacronym{sdc}{SDC}{Spectral Deferred Correction}
\newacronym{mol}{MOL}{Method of Lines}
\newacronym{ode}{ODE}{ordinary differential equation}
\newacronym{pde}{PDE}{partial differential equation}
\newacronym{misdc}{MISDC}{multi-implicit spectral deferred correction}
\newacronym{cfl}{CFL}{Courant-Friedrichs-Lewy}
\newacronym{nscbc}{NSCBC}{Navier-Stokes Characteristic Boundary Conditions}
\newacronym{qssa}{QSSA}{quasi-steady state assumptions}
\newacronym{mpi}{MPI}{message passing interface}

\makeglossaries
\glsdisablehyper

\definecolor{c1med}{HTML}{F15A60} 
\definecolor{c2med}{HTML}{7AC36A} 
\definecolor{c3med}{HTML}{5A9BD4} 
\definecolor{c4med}{HTML}{FAA75B} 
\definecolor{c5med}{HTML}{9E67AB} 
\definecolor{c6med}{HTML}{CE7058} 
\definecolor{c7med}{HTML}{D77FB4} 
\definecolor{c8med}{HTML}{737373} 

\definecolor{c1brt}{HTML}{EE2E2F} 
\definecolor{c2brt}{HTML}{008C48} 
\definecolor{c3brt}{HTML}{185AA9} 
\definecolor{c4brt}{HTML}{F47D23} 
\definecolor{c5brt}{HTML}{662C91} 
\definecolor{c6brt}{HTML}{A21D21} 
\definecolor{c7brt}{HTML}{B43894} 
\definecolor{c8brt}{HTML}{010202} 

\begin{document}

\title{\textbf{Comparison of turbulence statistics in isothermal and non-isothermal large eddy simulations of supercritical carbon dioxide jets}}

\author{Julia Ream}
\email{Contact author: Julia.Ream@nrel.gov}
\affiliation{Department of Mathematics, Florida State University, Tallahassee, FL 32306, USA}
\affiliation{Computational Science Center, National Renewable Energy Laboratory, Golden, CO 80401, USA}

\author{Marc T. Henry de Frahan}
\email[]{Marc.HenrydeFrahan@nrel.gov}
\affiliation{Computational Science Center, National Renewable Energy Laboratory, Golden, CO 80401, USA}

\author{Shashank Yellapantula}
\email[]{Shashank.Yellapantula@nrel.gov}
\affiliation{Computational Science Center, National Renewable Energy Laboratory, Golden, CO 80401, USA}

\author{Michael J. Martin}
\email[]{Michael.Martin@nrel.gov}
\affiliation{Computational Science Center, National Renewable Energy Laboratory, Golden, CO 80401, USA}

\author{Mark Sussman}
\email[]{sussman@math.fsu.edu}
\affiliation{Department of Mathematics, Florida State University, Tallahassee, FL 32306, USA}

\author{Ray Grout}
\email[]{Ray.Grout@nrel.gov}
\affiliation{Computational Science Center, National Renewable Energy Laboratory, Golden, CO 80401, USA}

\begin{abstract}
Supercritical carbon dioxide is of interest in a wide range of engineering problems, including carbon capture, utilization, and storage as well as advanced cycles for power generation. Non-ideal variations in physical properties of supercritical carbon dioxide impact the physics of these systems. It is important to understand how drastic changes in thermodynamic properties influence these flow physics to aid and optimize the design of future technologies related to carbon capture and sequestration. In this study, we simulate turbulent supercritical carbon dioxide jets to gain a better understanding of these physics. Of particular interest is the impact of pseudo-boiling on supercritical flow dynamics. We use a second-order finite volume discretization method with adaptive mesh refinement as implemented in the reacting flow solver, \textit{PeleC}, to perform a large eddy simulation of three turbulent jets of supercritical carbon dioxide. We use the Soave-Redlich-Kwong equation of state to close the system and more accurately incorporate the departure from ideal gas behavior into the turbulent flow physics. We find that the isothermal supercritical jet exhibits many similar flow characteristics compared to ideal gas round turbulent jets, with minor differences seen in the decay and spreading rate of the jet and in a noticeable anisotropy between resolved turbulent kinetic energy components. The non-isothermal jet excluding the pseudo-boiling point exhibits only small difference compared to the isothermal case. The non-isothermal case involving the pseudo-boiling point displays markedly different behavior, with evidence indicative of increased Kelvin-Helmholtz-like instabilities and much faster jet decay and disintegration. These factors impact the degree of mixing in the transition region of the jet, leading to finer-scale vortices and faster transition to ambient properties, indicating a potential for larger heat transfer and more rapid combustion dynamics.
\end{abstract}



\date{\today}

\maketitle

\section{Introduction}

\gls{sco2} is a highly coveted alternative working fluid in many different applications, and is one of the most widely used supercritical fluids along with water \cite{SCF2}. One application of interest is the use of \gls{sco2} as the working fluid in advanced cycles for power generation. In the traditional Brayton cycle, air is used as the working fluid through which the cycle of adiabatic compression and expansion and isobaric heat addition and rejection occurs \cite{alma991020089049703276}. The Allam cycle is a specialized, high-pressure Brayton cycle that utilizes transcritical carbon dioxide as the working fluid \cite{AllamPatent}. When compared to the conventional Brayton cycle, studies show that the Allam cycle has much higher efficiency \cite{ALLAMCOMP, ALLAM20175948}. Additionally, the carbon footprint for the Allam cycle is negligible, allowing for \gls{co2} produced from the system to be stored underground or used elsewhere, thereby aiding in carbon sequestration efforts \cite{cleantechnol1010022}. Another significant area of research includes using \gls{sco2} as the working fluid for high flux thermal management \cite{10.1115/1.4034053}. The use of \gls{sco2} in jet impingement cooling for micro-structures can eliminate dry-out issues and enhance heat transfer rates \cite{CHEN201845}. To further develop these supercritical fluid technologies, a growing importance has been placed on understanding the underlying physics of these systems.

Physics relating to the round turbulent jet are particularly important for applications involving injection technologies. The high densities associated with the liquid-like behavior of supercritical fluids coupled with the relatively low gas-like viscosity associated with them typically results in a high Reynolds flow, resulting in a turbulent system. Understanding the turbulence physics of these jets is crucial in developing machinery for these systems. Current experimental research is mainly application-oriented. Research into the \gls{sco2} jet's rock breaking ability has been of primary importance to \gls{egs} applications \cite{EGScomp, EGS2, very_rock, experiment, rb}, with additional focus being given to pipeline leakage and flow dynamics upon wall impact \cite{WANG2015210, WANG201977}. These studies do not explore the underlying turbulence statistics of the flow. Chemical engineering design aspects of \gls{sco2} injection are more focused on solubility dynamics, in the context of molecular beam extraction \cite{freejet} and polymer collapse \cite{pulse_jet}, as opposed to turbulence. Other experiments focus on similar application specific quantities of interest, such as heat transfer and mixing, which is related in part to the turbulence dynamics \cite{heated_cyl, 10.1115/FEDSM2022-87029}, but they also note the difficulty in experimental design for investigating these aspects of the flow under the conditions needed to replicate those in real applications \cite{freejet}. Thus, numerical simulations are necessary to further explore the turbulence statistics of these flows.

The challenges of simulating \gls{sco2} flows, as well as other new combustion technologies \cite{10947586} has led to an increased interest in the development of appropriate numerical methods for these systems \cite{Ries_2021}. However, model validation is often complicated by the lack of experimental data at these extreme pressures and temperatures. Studies using \gls{dns} have been implemented to help establish benchmark test cases for other types of numerical schemes. Ruiz \emph{et al.} use 2D \gls{dns} to simulate a mixing layer created by two streams of supercritical oxygen and gaseous nitrogen, using two different \gls{cfd} solvers to add confidence to their results \cite{article}. A 3D \gls{dns} is used by Ries \emph{at al.} to simulate a round nitrogen jet for comparison with experimental data produced by Mayer \emph{et al.} \cite{DNS_N}. However, this study requires a reduction in Reynolds number from $1.62 \times 10^5$, based on the injection diameter, to $5300$ to feasibly execute the computations. Li also utilizes a low Reynolds number of $1750$ to study a round turbulent \gls{sco2} jet with a preconditioning scheme \cite{Li2012}. The \gls{rans} approach has also been implemented utilizing theory from the ideal gas case \cite{RANS}, but with the goal of ascertaining a more general understanding of why specifically \gls{sco2}'s rock-breaking ability is better than that of water. 

To maintain a high Reynolds flow and better capture the effects of the supercritical nature of the fluid on the turbulence dynamics, the use of \gls{les} has been explored. The impact of \gls{sgs} models in capturing transcritical and supercritical dynamics of cryogenic nitrogen have been analyzed through comparison with the Mayer \emph{et al.} experiment and highly accurate \gls{nist} data \cite{PETIT201361, doi:10.1080/00102200500287613, doi:10.1063/1.1795011, doi:10.1063/1.4937948, Same_LES}. Schmitt \emph{et al.} does a similar investigation using \gls{les}, then extending their investigation to include \gls{sco2} after validation with the Mayer \emph{et al.} data \cite{LES_N}. However, this investigation uses low-pressure jets and does note the \gls{sgs} models might need additional contributions to handle non-linearities and the pressure regime. 

Many of the numerical investigations cited thus far consider cryogenic nitrogen in order to compare with the Mayer \emph{et al.} experiment on supercritical jet turbulence \cite{mayer2003raman}. However, molecular structure can have a significant impact on flow characteristics; for example, heat-transfer correlations vary across species and experiment with no general trend having been identified for the supercritical fluid regime \cite{heat_transfer}.  A wide variety of numerical investigations into jet turbulence using \gls{sco2} exist but typically explore other parameter regimes of interest or application-specific quantities of interest. Examples of turbulent adjacent quantities of interest include fluctuation characteristics based on inlet conditions \cite{ZHANG2022124125}, effects of nozzle and aperture differences on pressure and velocity decay \cite{en13102627} and wave features \cite{LIU2021108422}, mixing between \gls{sco2} and other fluid phases \cite{RAMAN2018}, and energy dissipation \cite{LI2020103650}. These studies all involve high-pressure jets and are commonly found in applications involving rock fracturing. Related configurations are also studied, such as the swirling-round \gls{sco2} jet \cite{en14010106}, turbulent jet-in-crossflows \cite{ZHANG2022}, slot jet impingement \cite{ALKANDARI2022122949}, and channel flow \cite{ROGALEV2020}.

While much of the literature thus far has explored the impact of different numerical methods on modeling supercritical fluid flows and has aimed to strengthen the validity of these simulations in spite of the lack of experimental data available in the current landscape, a general consensus has still not been reached on how the supercritical nature of these fluids impacts the turbulence physics of these models. Thus, there remain open questions for understanding the fundamental flow behavior of turbulent jets in a supercritical environment, especially near the supercritical point, where both experimental and numerical investigations are still a challenge.

Our objective is to use \gls{les} to further explore the pseudo-boiling region of the pseudo-critical zone and analyze the influence of extreme thermodynamic fluctuations on turbulence statistics and flow dynamics within the flow field. Using the compressible Navier-Stokes equation solver, \textit{PeleC} \cite{PeleC1,PeleC2,PeleSoftware, 10.1145/3581784.3607065, PhysRevFluids.8.110511, PhysRevFluids.6.110504, PERRY2022112286, HassanalyQSSA}, closed with the \gls{srk} \cite{SOAVE1972}, we consider three cases to examine various quantities of interest associated with classical turbulence mechanics. These three cases are chosen to capture different areas around a peak in specific heat that is associated with the pseudo-critical region. The rest of the paper is as follows. We first present the physical model and numerical methods used for the investigation. We then present the simulation setup of the round turbulent jet as implemented in the code. Next we present the results of the turbulent \gls{sco2} jet and our analysis of the turbulence statistics. 

\section{Physical model and numerical methods}
Simulation of turbulent supercritical flows requires combining appropriate \textit{governing equations} and \textit{turbulent closures} withe an \textit{equation of state} that accurately represents the thermodynamics properties of supercritical carbon dioxide in the regime being simulated, and \textit{transport property correlations} that are accurate in the regime being simulated.
\subsection{Governing equations}
We consider the three-dimensional compressible Navier-Stokes equations as interpreted in their Favre-filtered form:
\begin{subequations} 
\label{filtered_NSE_FINAL}
\begin{eqnarray}
  \frac{\partial\overline{\rho}}{\partial t} + \frac{\partial }{\partial x_j} \left( \overline{\rho}\widetilde{ u_j} \right) = 0, \label{NSE_mass_FINAL}
  \end{eqnarray}
  \begin{eqnarray}
  \frac{\partial}{\partial t} \left( \overline{\rho}\widetilde{ u_i }\right) + \frac{\partial}{\partial x_j} \left(\overline{\rho}\widetilde{ u_i} \widetilde{u_j} + \overline{p }\delta_{ij} - \widetilde{\sigma_{ij}} \right) = - \frac{\partial \tau_{ij}}{\partial x_j}, \label{NSE_mom_FINAL}
  \end{eqnarray}
  \begin{eqnarray}
  \frac{\partial}{\partial t} \left( \overline{\rho}\widetilde{ E} \right) + \frac{\partial}{\partial x_j} \left(\left( \overline{\rho}\widetilde{ E}+\overline{p} \right)\widetilde{u_j} + \widetilde{q_j} - \widetilde{\sigma_{ij}}\widetilde{ u_i}\right)&& \nonumber \\ = - \frac{\partial}{\partial x_j } \left( \gamma c_v \mathcal{Q}_j + \dfrac{1}{2} \mathcal{J}_j \right)&&,  \label{NSE_E_FINAL}
\end{eqnarray}
\end{subequations}
where $\rho$ is the density (g/cm$^3$), $u_j$ is the velocity (cm/s) for the $x_j$ direction, $p$ is the pressure (Ba), $E = e + \frac{u_i u_i}{2}$ is the total energy, $e = c_v T$ is the internal energy, $T$ is the temperature (K), $c_v$ is the heat capacity at constant volume (Erg/(g $\cdot$ K)), and $\gamma = \nicefrac{c_p}{c_v}$, with $c_p$ being the specific heat capacity at constant pressure. $\overline{\cdot}$ denotes the filtered variables, with $\widetilde{\cdot}  = \nicefrac{\overline{\rho~\cdot}}{\overline{\rho}}$ defining the Favre-filter operation. 
\subsection{Turbulence Models}
Under appropriate assumptions \cite{batchelor_2000, PIOMELLI202183}, the filtered diffusive fluxes are given by:
\begin{equation} \label{filtered_trans}
  \widetilde{\sigma_{ij}} = 2\widetilde{\mu}\widetilde{ S_{ij}} - \frac{2}{3}\widetilde{\mu} \delta_{ij} \widetilde{ S_{kk}} \\, \quad \quad
  \widetilde{q_j} = -\widetilde{\lambda} \frac{\partial \widetilde{T}}{\partial x_j} \\,
\end{equation}
where
$S_{ij} = \frac{1}{2}\left(\frac{\partial u_i}{\partial x_j} + \frac{\partial u_j}{\partial x_i} \right)$ is the strain-rate tensor, $\mu$ is the dynamic viscosity (P), $\lambda$ is the thermal conductivity (Erg/(cm $\cdot$ s $\cdot$ K)), and $\delta_{ij}$ is the Kronecker delta. Terms arising from the Favre filtering procedure include the \gls{sgs} stress $\tau_{ij}$, \gls{sgs} heat flux $\mathcal{Q}_{j}$, and \gls{sgs} turbulent diffusion $\mathcal{J}_j$. These \gls{sgs} terms are modeled with the \gls{smd} as detailed in \cite{PIOMELLI202183}, with use of a three-point box filter \cite{filter} and filter-grid ratio of two for calculating model coefficients. Note that while the equations given in \eqref{filtered_NSE_FINAL} represent the system discretized within \textit{PeleC}, in practice, we evolve these filtered quantities without explicitly filtering the solution. Here, the grid acts as the filter for the \gls{les}, while the impact of an explicit filter choice is incorporated instead through these model coefficients for the \gls{sgs} terms. External forces such as gravity are not included in this study. 
\subsection{Equation of state}
The system is closed using the \gls{srk} \cite{SOAVE1972} to relate pressure, density, and temperature as follows: 
\begin{eqnarray} \label{SRK_EOS}
	p = \dfrac{RT}{V_m - b} - \dfrac{a \alpha}{V_m(V_m + b)}, \nonumber \\
	a = \dfrac{0.42747R^2T_c^2}{P_c}, \quad
	b = \dfrac{0.08664RT_c}{P_c}, \nonumber \\
	\alpha = \left( 1 + \left( 0.48508 + 1.55171 \omega - 0.15613 \omega^2 \right) \nonumber 
    \times \left( 1 - T_r^{0.5} \right) \right)^2,  \nonumber \\
	T_r = \dfrac{T}{T_c},
\end{eqnarray}
where $R$ is the ideal gas constant (Erg/(K $\cdot$ mol)), $T$ is the absolute temperature, $V_m$ is the molar volume of the species (cm$^3$/mol), $T_c$ and $P_c$ are the critical temperature and pressure of the species, respectively, and $\omega$ is the acentric factor of the species. Additionally, $a$, $b$, and $\alpha$ are all species-specific parameters calculated from the given species thermodynamic properties as listed above. All cases are run with a single species, that being \gls{co2}. This equation is used in conjunction with approximations to ideal gas behavior provided by NASA Polynomial fits \cite{NASAPoly}, where the \gls{srk} then contributes the behavior attributed to real gas effects. The \gls{srk} and Chung high pressure corrections are commonly used models for thermodynamic and transport properties in simulations of turbulent \gls{sco2} systems related to advanced cycles for power generation \cite{osti_2228671, kasuya_flameletann_2022}. Though higher order \gls{eos} exist and provide better accuracy, as a cubic \gls{eos}, the \gls{srk} provides a good balance between accuracy while incorporating real gas effects and computational efficiency. It has been shown to outperform other cubic \gls{eos} and would serve as a better baseline for further \gls{eos} development \cite{GHANBARI201713}. Multiple works have compared the use of different \gls{eos} in \gls{sco2} simulation applications \cite{rasmussen_how_2021, mecheri_supercritical_2016, manikantachari_thermal_2017}. Additionally, work has gone into improving the computational efficiency of these high-cost \gls{eos} regimes, though the extension to multi-species mixtures is still underway \cite{ZHANG2024112752}.

While many of the investigations referenced earlier note that \gls{sgs} models may need modification to deal with supercritical flows \cite{LES_N, PETIT201361, doi:10.1063/1.4937948, doi:10.1080/00102200500287613}, it is noted by Muller \emph{et al.} that the impact of \gls{sgs} modeling and numerical flux discretization is primarily limited to second-order moments when using a sufficiently refined grid while the choice in thermodynamic modeling is a key component in capturing first-order moments \cite{doi:10.1063/1.4937948}. With that, they also note that the choice in \gls{sgs} model and numerical flux discretization had a larger than expected effect on resolved Reynolds stress profiles \cite{doi:10.1063/1.4937948}. Specifically, the constant Smagorinsky model yielded decaying fluctuation magnitudes early in the jet evolution, resulting in the transition to a fully turbulent mixing zone beginning from lower turbulence levels. However, their results did agree with the jet break-up location they obtained from mean density profiles, which were shifted slightly downstream by comparison to other \gls{sgs} models \cite{doi:10.1063/1.4937948}. We will be using the compressible version of the dynamic Smagorinksy \gls{sgs} closures for our investigation, with further consideration of any influence of \gls{sgs} model on our quantities of interest being noted later on. 
\subsection{Transport properties}
Transport coefficients include real gas behavior through Chung's high pressure corrections \cite{chung:1988}:
\begin{equation} \label{trans_decomp}
  q = q_k + q_p,
\end{equation}
where $q$ is the transport coefficient of interest (e.g., $\lambda$), $q_k$ is the low-pressure component relating to ideal gas behavior, and $q_p$ is the high-pressure deviation from ideal behavior. Ideal gas behavior $q_0$ is approximated with EGLib functions \cite{ERN1995105} for use in $q_k$:
\begin{equation} \label{EGLIB}
  \ln{q_0} = \sum_{n=1}^{4}a_{q,n}\left(\ln(T)\right)^{n-1},
\end{equation}
where $a_{q,n}$ are pre-calculated polynomial fit coefficients for transport coefficient $q$. Additional details regarding this formulation can be found in Appendix A
\subsection{Discretization}

\begin{figure}[H]
\begin{center}
\includegraphics[scale=0.5]{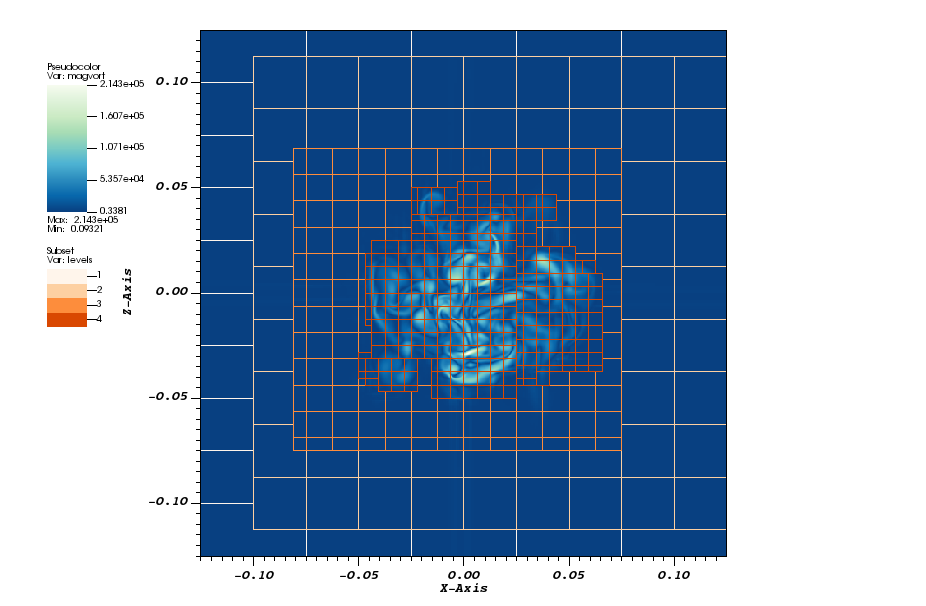}
\end{center}
\caption{Slice normal to the jet axis at $y=2d$ demonstrating \gls{amr} block structure, where internal faces are displayed for ease of visualization. Refinement criterion is based on vorticity.}
\label{fig:mesh}
\end{figure}

To discretize and evolve the system of partial differential equations \eqref{filtered_NSE_FINAL}-\eqref{SRK_EOS}, we use \textit{PeleC} \cite{PeleC1,PeleC2,PeleSoftware, 10.1145/3581784.3607065, PhysRevFluids.8.110511, PhysRevFluids.6.110504, PERRY2022112286, HassanalyQSSA}, a highly scalable compressible hydrodynamics code for reacting flows developed as part of the \gls{ecp} through the \gls{doe}. While full details regarding \textit{PeleC} can be found in the works cited, a brief summary of key features is mentioned here for convenience. \textit{PeleC} uses a second-order, conservative finite volume formulation with data stored at cell centers. Cell-centered data is interpolated to cell faces with a second-order \gls{ppm} for use with an approximate Riemann solver. A standard second-order predictor-corrector approach is used for temporal discretization. The time step is dynamically limited using a Courant number of 0.9 for scaling and minimizing between advective and diffusive restrictions. \textit{PeleC} leverages \textit{AMReX} \cite{amrex1, amrex2, amrex3} for \gls{amr}, using linear cell conservative interpolation and sub-cycling between refinement levels. An example of the adaptively-generated grid at a slice in the domain is presented in Figure \ref{fig:mesh}. Modeling of transport terms and thermodynamic quantities as described above are done through \textit{PelePhysics} \cite{PelePhysics}, a library for complex physics, including chemical reactions, non-ideal \gls{eos} options, and higher fidelity transport models.

\section{Computational Implementation}
In this section, we discuss key components of the computational cases, including discretization and \gls{amr} considerations, handling of the jet inflow and boundary conditions, operating values for parameters of interest, and post-processing data slicing and averaging procedures. 

\subsection{Adaptive Mesh Refinement Specification}
\begin{figure}[h!]
\centering
\begin{tikzpicture} [scale = .6]
	\draw [draw=c3med, fill=c3med!60, thick] (.5, 0) -- (4.5,0) -- (4.5, 12) -- (.5, 12) -- (.5, 0); 
	\draw [draw=c2med, fill=c2med!60, thick] (2.1, 0) -- (2.1, .4) -- (2.9, .4) -- (2.9, 0) -- (2.1, 0); 
	\draw [thick] (0,0) -- (5,0) -- (5,12.5) -- (0,12.5) -- (0,0);
	\draw [ultra thick] (2.4,0) -- (2.6, 0);
	\node [below] at (2.6,0) {\scriptsize{Jet Inlet}};
	\node [below] at (2.6,-.5) {\scriptsize{$v_{in}$, $T_{in}$}};
	\node [right] at (0.8, 10.4) {\scriptsize{Ambient Fluid}};
	\node [right] at (1.6, 9.9) {\scriptsize{$v_{\infty} = 0$}};
	\node [right] at (1.6, 9.4) {\scriptsize{$T_{\infty}$, $p_{\infty}$}};
	\node [right] at (5, 6.25) {\footnotesize{$62.5d$}};
	\node [above] at (2.5, 12.5) {\footnotesize{$25d$}};
	\node [below] at (2.5, 12.6) {\scriptsize{Outflow BC}};
\end{tikzpicture}
\hspace{.1in}
\begin{tikzpicture} [scale = .75]
	\draw [<->] (0,2.15) -- (0,1.15) -- (1,1.15);
	\draw[->] (0,1.15) -- (.5,1.4);
	\node [right] at (.85,1.175) {\tiny{$x$}};
	\node [right] at (.3, 1.5) {\tiny{$z$}};
	\node [above] at (0,2.025) {\tiny{$y$}};
\end{tikzpicture}
\caption{Two dimensional slice schematic of jet setup. Four levels of refinement are enforced within the green box based on proximity to jet inlet. Refinement based on vorticity criterion then occurs within the blue region. Outside the blue region, \gls{amr} is explicitly turned off to allow flow structures to be dissipated numerically and allowed to leave the domain without incurring spurious reflections.}
\label{fig:domain}
\end{figure}
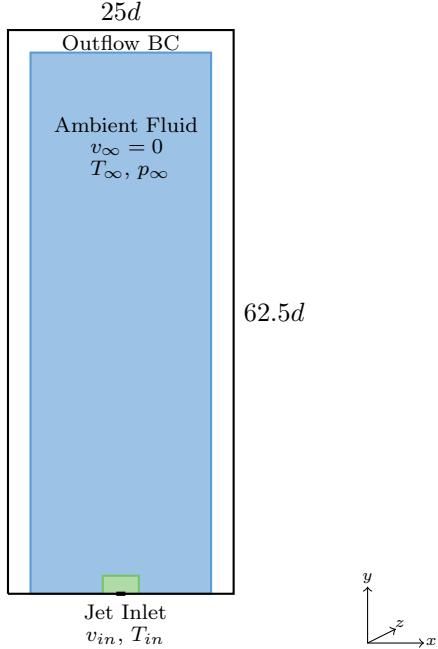

Our target simulation is a 3D \gls{les} with a grid size that is 100 times larger than the smallest scale of the turbulent flow. The smallest scale of these turbulent flows, known as the Kolmogorov scale \cite{kolmogorov}, can be approximated as:
\begin{equation} \label{Kolmogorov}
	\eta = \left( \dfrac{\nu^3}{\varepsilon} \right)^{1/4},
\end{equation}
where $\nu = \nicefrac{\mu}{\rho}$ is the kinematic viscosity and $\varepsilon = \nicefrac{v_{in}^3}{d}$ approximates the average rate of dissipation of turbulent kinetic energy per unit mass, where $v_{in}$ is the reference inflow velocity in the axial direction $y$ and $d=\SI{.01}{cm}$ is the jet diameter. For these turbulent jets, $\eta = \SI{5.37e-6}{cm}$. Figure \ref{fig:domain} provides a 2D schematic of the computational domain. The length in each direction is $x = 25d$, $y = 62.5d$, and $z = 25d$. To keep the calculation tractable and achieve an adequate \gls{les} grid size, we implement four levels of refinement, with a refinement ratio of two, leading to 80, 200, and 80 cells on the coarsest level in the $x$, $y$, and $z$ directions, respectively. This results in an initial mesh size of $\Delta x_{0}=\Delta y_{0}=\Delta z_{0}=0.3125 d$, leading to $\Delta x_{3}=\Delta y_{3}=\Delta z_{3}=\SI{3.9062e-4}{}d$, where the subscript denotes the \gls{amr} level. The full four levels of refinement are implemented in a box around the inlet of lengths $x=z=4d$ and $y=2d$ to ensure high refinement at the inflow. These four levels are then adaptively refined based on the given refinement criterion in the region outward from the jet inlet up to a distance of $20d$ in the $x$ and $z$ direction and $60d$ in the $y$ direction. 

The grid refinement criterion is given by the vorticity, specifically with $\omega \geq 5000^{2l}$, where $\omega$ is the magnitude of the vorticity and $l$ is the \gls{amr} level. For the first ten flow-throughs of the simulation, mesh refinement only occurs up to one level within the refinement region to establish the flow pattern. Thereafter, the simulation proceeds with the four levels of mesh refinement for ten more flow-throughs. After that, statistics are collected over the next two flow-throughs for analysis, where a steady state is assumed to have been reached.

\subsection{Initial and Boundary Conditions}
Our inlet consists of an opening centered in the $xz$-plane with diameter $d=\SI{.01}{cm}$ through which the \gls{sco2} jet is initialized. The pressure in the jet at the inlet is the same as that of the quiescent background fluid and it is given by $p_{in}=p_{\infty}=\SI{1.01325e+8}{Barye}$. The ambient fluid remains at rest while the jet is initialized with an inflow velocity of $v_{ref} = \SI{1800}{cm.s^{-1}}$, leading to a Reynolds number of the initialized jet of $Re_{ref} = 22910$, with $v_{ref}$ and $d$ being the reference velocity and length scale respectively. For the jet temperature and pressure conditions given, the inflow density is $\rho_{in}=$ \SI{3.019e-1}{g.cm^{-3}}, as calculated via the \gls{srk} in \textit{PeleC}. To implement a turbulent inflow, we formulate our mean velocity and \gls{rms} values by scaling, interpolating, and adding noise to a predetermined velocity profile calculated via \gls{dns} \cite{DNS}, with jet diameter $D$, axial velocity $v_{\text{\tiny{DNS}}}$, and \gls{rms} values given by $v^{\prime}_{\text{\tiny{DNS}}} =  \langle v_{\text{\tiny{DNS}}}^2 \rangle ^{1/2}$. For \gls{dns} quantities, $(u,v,w)$ values are in the $(r,v,\theta)$ direction. Finalized values are converted to Cartesian values for the simulation. We begin by scaling the \gls{dns} values with our reference jet velocity and length values:
\begin{subequations} \label{Jet_Inflow}
	\begin{align}
		r_{\text{\tiny{DNS}},\ scaled} &= d \cdot \left( \dfrac{r}{D} \right)_{\text{\tiny{DNS}}}, \\
		v_{\text{\tiny{DNS}},\ scaled} &=  v_{ref} \cdot v_{\text{\tiny{DNS}}}, \\
		v^{\prime}_{\text{\tiny{DNS}},\ scaled} &=  v_{ref} \cdot v^{\prime}_{\text{\tiny{DNS}}}, \\
		u^{\prime}_{\text{\tiny{DNS}},\ scaled} &=   v_{ref} \cdot u^{\prime}_{\text{\tiny{DNS}}}, \\
		w^{\prime}_{\text{\tiny{DNS}},\ scaled} &=   v_{ref} \cdot w^{\prime}_{\text{\tiny{DNS}}}. 
	\end{align}
\end{subequations}  
These values are then linearly interpolated onto our grid values $r(i)$ with: 
\begin{subequations} \label{interpolation}
\begin{eqnarray}
f_1 = \tfrac{r(i) - r_{\text{\tiny{DNS}},\ scaled}(i) }{r_{\text{\tiny{DNS}},\ scaled}(i+1) - r_{\text{\tiny{DNS}},\ scaled}(i)}, 
\end{eqnarray}
\begin{eqnarray}
f_2 = \tfrac{r(i) - r_{\text{\tiny{DNS}},\ scaled}(i+1) }{r_{\text{\tiny{DNS}},\ scaled}(i) - r_{\text{\tiny{DNS}},\ scaled}(i+1)}, 
\end{eqnarray}
\begin{eqnarray}
\phi_{\text{\tiny{DNS}},\ inter} = f_1\ \phi(i) + f_2\ \phi(i+1), 
\end{eqnarray}
\end{subequations}
where $\phi$ is each of the velocity components mentioned in Equations \eqref{Jet_Inflow}. Finally, noise is added to each cylindrical component of the velocity as follows before conversion back to Cartesian coordinates for use in the boundary inflow:
\begin{subequations} \label{Jet_Inflow_noise}
	\begin{eqnarray}
		&&v_{in,\ cyl} = \langle v_{\text{\tiny{DNS}},\ inter} \rangle + \nonumber \\ 
        &&\big(  v^{\prime}_{\text{\tiny{DNS}}, inter}  + \beta  v^{\prime}_{\text{\tiny{DNS}}, inter}    	\psi_1\sin{\theta_1}  \big) \cdot \psi_2 \sin{\theta_2}, \label{y_flow}
        \end{eqnarray}
        \begin{eqnarray}
		u_{in,\ cyl} =   u^{\prime}_{\text{\tiny{DNS}}, inter}  + \beta  u^{\prime}_{\text{\tiny{DNS}}, inter}  \psi_3\sin{\theta_3}, \label{t_flow} 
        \end{eqnarray}
        \begin{eqnarray}
		w_{in,\ cyl} =   w^{\prime}_{\text{\tiny{DNS}}, inter}  + \beta  w^{\prime}_{\text{\tiny{DNS}}, inter}  \psi_4\sin{\theta_4},  \label{theta_flow}
	\end{eqnarray}
\end{subequations}  
where $\beta = 0.1$ and each $\psi_i$ and $\theta_k$ value is randomly generated as follows:
\begin{subequations} \label{Random_Variables}
	\begin{align}
		r_i &= \sqrt{-2.0 \log{(X_i)}} \label{Random_r}, \\
		\theta_k &= 2 \pi X_k \label{Radom_theta},
	\end{align}
\end{subequations}
where $X_n$ are random uniformly distributed numbers between 0 and 1. The inflow parameters are finalized as $\phi_{in}$ after being converted to Cartesian coordinates $(r,y,\theta) \to (x,y,z)$. 

We implement zero-gradient boundary conditions for all boundaries not involving the jet inflow with first order extrapolations. Additionally, \gls{amr} is halted at a distance of $2.5d$ from the boundary in the $x$ and $z$ directions, and that of $5d$ in the axial direction. This low-refinement perimeter is implemented to act in a similar fashion to a sponge with the goal of dissipating waves and thus reducing the chance of spurious reflections from the boundaries.

\subsection{Operating Conditions} \label{OpCon}
Three cases are investigated in this study. The first case pertains to an isothermal jet, in which the jet conditions match that of the ambient fluid for all quantities aside from velocity. The other two cases consider non-isothermal jets, in which the ambient fluid is adjusted to no longer match the temperature of the jet inflow. All three cases consider the same inflow conditions, with ambient fluid conditions varying for each case. Temperatures for the ambient fluid in the non-isothermal cases are chosen such that the ambient fluid is supercritical but the temperature for each case falls on either side of the peak in specific heat seen near the critical point. This peak is associated with the pseudo-boiling phenomenon seen in supercritical fluids, whereupon crossing the Widom line, the supercritical fluid experiences a shift from being more gas-like to more liquid-like in nature, although distinct phase separation between the two is not present. A summary of the conditions for each case can be seen in Table \ref{Case-Params}, including values for the sound speed $C_s$, compressibility factor $Z$, and reference Mach number $M_{ref}$; subscripts denote values at the inlet ($in$) vs the ambient fluid ($\infty$).
\begin{table}
\caption{Summary of jet parameters for each case.}
\label{Case-Params}
\begin{center}
\begin{tabular}{ p{3.5cm} | p{3cm} p{3cm} p{3cm}  }
Parameter & Case 1 & Case 2 & Case 3 \\
\hline
$T_{\infty}$ (K)& 330& 350 & 314  \\
$T_{in}$ (K)& 330 & 330 & 330 \\
$p_{\infty}$ (Barye)& \num{1.01325e+08} & \num{1.01325e+08} & \num{1.01325e+08} \\
$p_{in}$ (Barye) & \num{1.01325e+08} & \num{1.01325e+08} & \num{1.01325e+08} \\
$\rho_{\infty}$ (g/cm$^3$)& \num{3.0192e-01} & \num{2.2567e-01} & \num{5.1274e-01} \\
$\rho_{in}$ (g/cm$^3$) & \num{3.0192e-01} &\num{3.0192e-01}  & \num{3.0192e-01}  \\
$c_{p_{\infty}}$ (Erg/g*K) &\num{3.389e+07} & \num{1.927e+07} & \num{5.703e+07} \\
$c_{p_{in}}$ (Erg/g*K) & \num{3.389e+07} & \num{3.389e+07} & \num{3.389e+07} \\
$C_{s_{\infty}}$ (cm/s) & \num{2.593e+04} & \num{2.661e+04} & \num{3.131e+04} \\
$C_{s_{in}}$ (cm/s) & \num{2.593e+04} & \num{2.593e+04}&\num{2.593e+04} \\
$\mu_{\infty}$ (P) & \num{2.748e-04} & \num{2.445e-04} & \num{4.110e-04} \\
$\mu_{in}$ (P) & \num{2.748e-04} & \num{2.748e-04} & \num{2.748e-04} \\
$\lambda_{\infty}$ (Erg/(cm*s*K))  & \num{3.966e+03}  & \num{3.443e+03} & \num{6.361e+03}  \\
$\lambda_{in}$ (Erg/(cm*s*K))  & \num{3.966e+03} & \num{3.966e+03} & \num{3.966e+03} \\
$Z_{\infty}$ & \num{5.329e-01} & \num{6.691e-01} & \num{3.331e-01} \\
$Z_{in}$ & \num{5.329e-01}& \num{5.329e-01}& \num{5.329e-01}\\
$v_{ref}$ (cm/s)& 1800 & 1800 & 1800  \\
$R_{ref}$ & 22911 & 22911 & 22911  \\
$M_{ref}$ & 0.08 & 0.08 & 0.08  \\
\end{tabular}
\end{center}
\end{table}

\begin{figure}[hbtp!]
\begin{center}
\vspace{-28pt}
\subfloat[\label{rho_vs_cases}] { 
	\includegraphics[scale=.25, width=0.45\columnwidth]{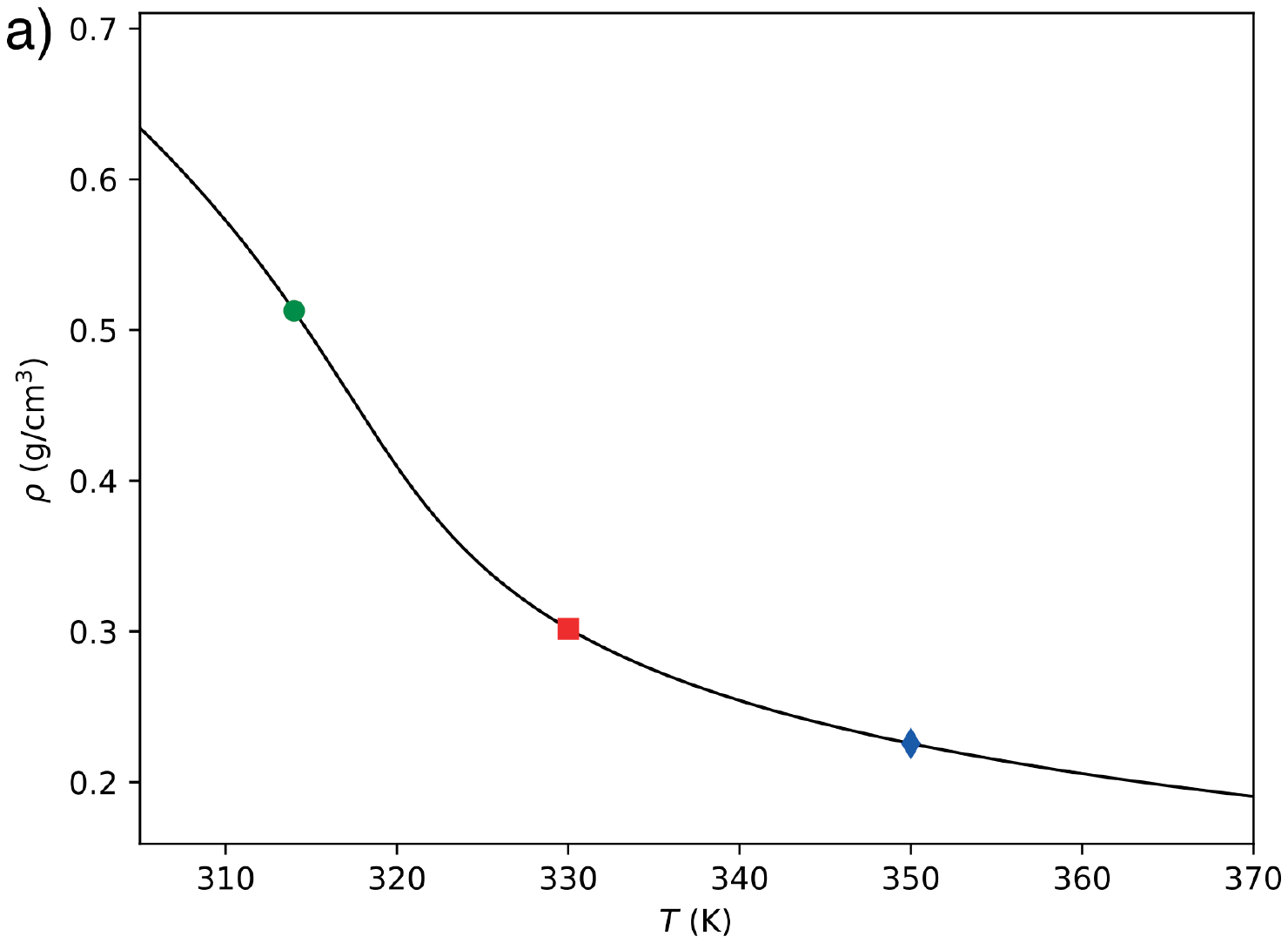}
}
\subfloat[\label{cp_vs_cases}] { 
	\includegraphics[scale=.25, width=0.45\columnwidth]{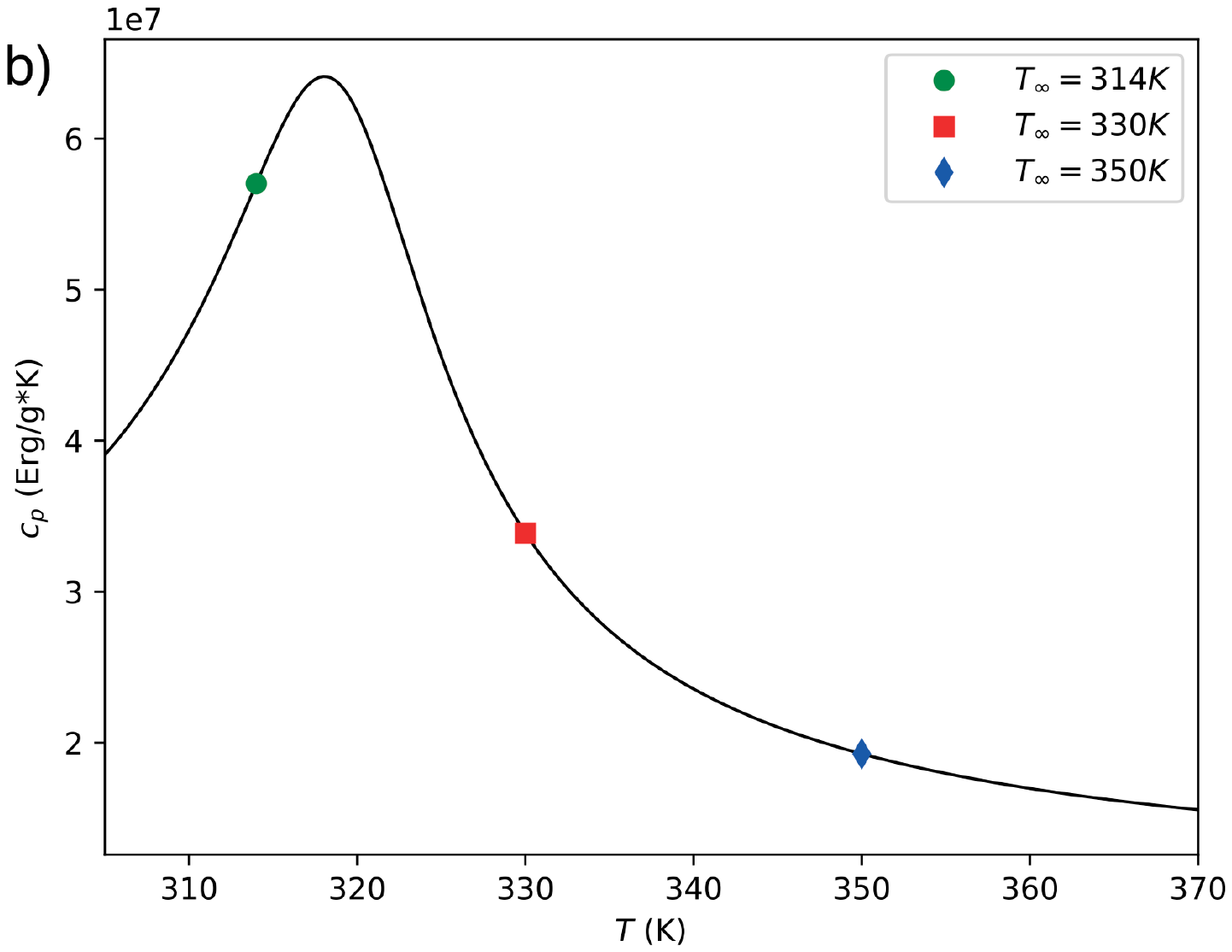}
} \\
\vspace{-28pt}
\subfloat[    \label{mu_vs_cases}] { 
	\includegraphics[scale=.25, width=0.45\columnwidth]{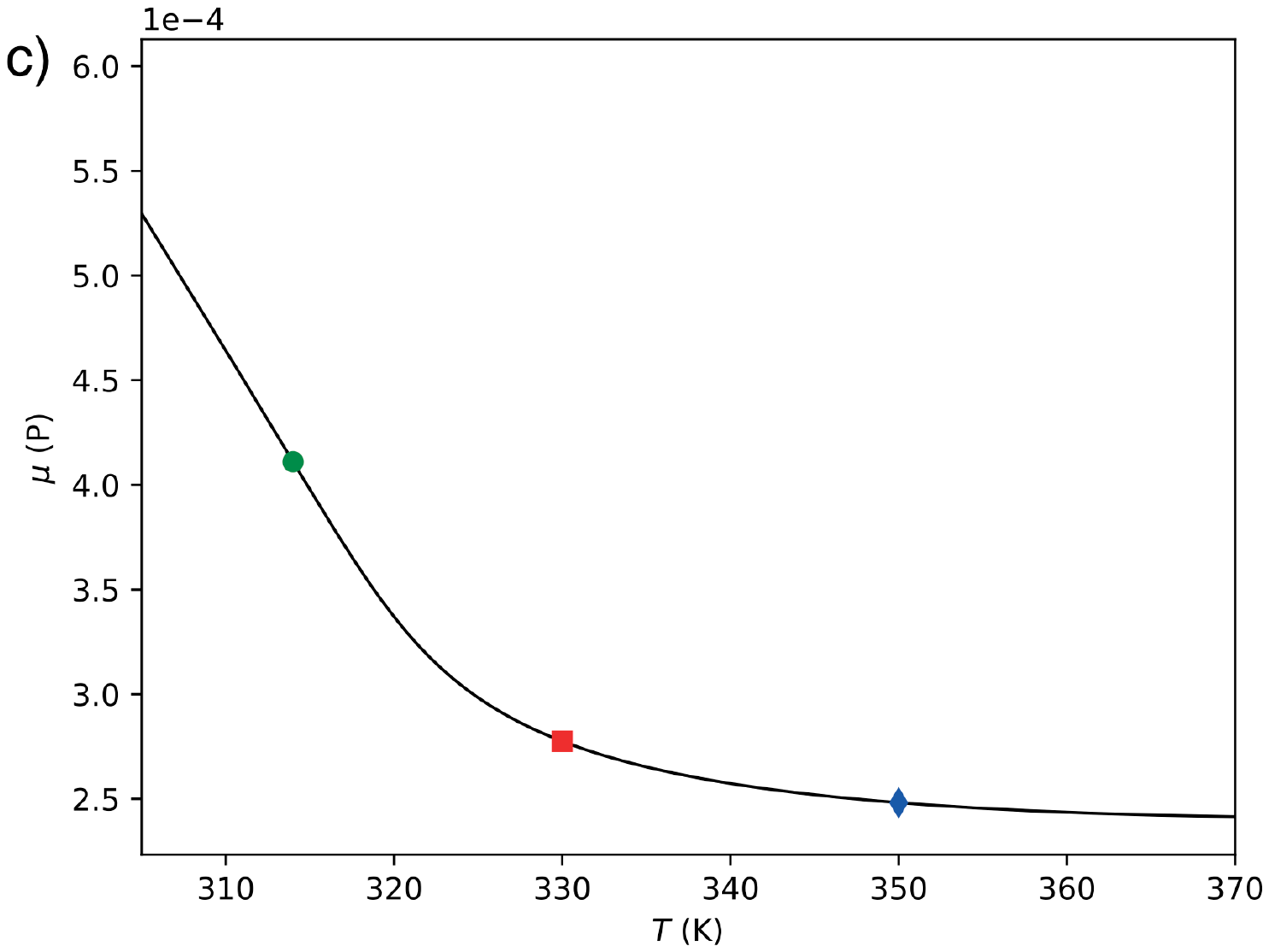}
}
\subfloat[\label{lam_vs_cases}] { 
	\includegraphics[scale=.25, width = 0.45\columnwidth]{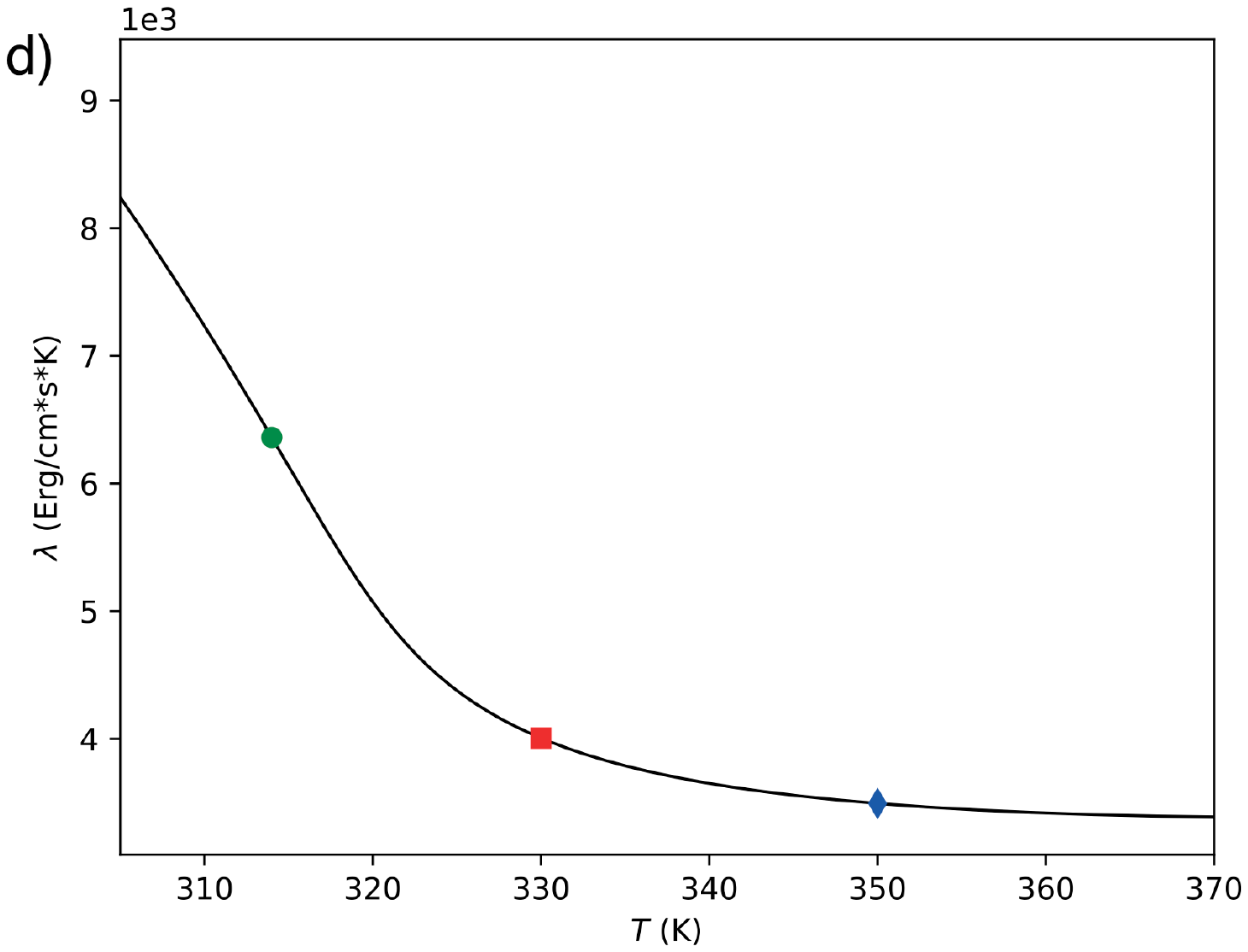}
}
\caption{Extreme variation seen near critical point for (a) density, (b) constant-pressure specific heat, (c) viscosity, and (d) thermal conductivity. Ambient temperature conditions are highlighted at 314~K (green circle), 330~K (red square), and 350~K (blue diamond).}
\label{thermo_cases}
\end{center}
\end{figure}

The goal in choosing the ambient temperature in this way is to investigate the impact of large thermodynamic changes on the flow field of the supercritical jet. An example of this can be seen with the sharp peak in constant pressure specific heat in Figure \ref{cp_vs_cases}. This choice differs from similar studies, which typically involve transcritical injection. 

\clearpage

\subsection{Post-Processing Procedures}
Post-processing procedures were written in Python and primarily leveraged the yt \cite{YT} and pandas \cite{pandas1,pandas2} libraries. Original 3D data at each data output time were sliced in three different ways using yt: along the centerline to create 1D data, normal to the jet axis at various points down stream to create 2D data, and along the axial direction at $z=0$ to create 2D data. A graphic depicting these different slicing regimes can be seen below:

\begin{figure}[H]
\centering
\vspace{12pt}
\begin{tikzpicture} [scale = .4]
\pgfmathsetmacro{\x}{3}
\pgfmathsetmacro{\y}{3}
\pgfmathsetmacro{\z}{6}
\path (0,0,\y) coordinate (A) (\x,0,\y) coordinate (B) (\x,0,0)
coordinate (C) (0,0,0) coordinate (D) 
(0,\z,\y) coordinate (E) (\x,\z,\y) coordinate (F) (\x,\z,0) coordinate (G) (0,\z,0) coordinate (H)
(\x/2, 0 ,\y/2) coordinate (I) (\x/2,\z,\y/2) coordinate (J);
\draw (A)--(B)--(C)--(G)--(F)--(B) (A)--(E)--(F)--(G)--(H)--(E);
\draw (A)--(D)--(C) (D)--(H);
\draw[red,thick] (I)--(J);
\node[align=left] at (-2.25,-0.5,0) {a)};
\end{tikzpicture}
\hspace{32pt}
\begin{tikzpicture} [scale = .4]
\pgfmathsetmacro{\x}{3}
\pgfmathsetmacro{\y}{3}
\pgfmathsetmacro{\z}{6}
\path (0,0,\y) coordinate (A) (\x,0,\y) coordinate (B) (\x,0,0)
coordinate (C) (0,0,0) coordinate (D) 
(0,\z,\y) coordinate (E) (\x,\z,\y) coordinate (F) (\x,\z,0) coordinate (G) (0,\z,0) coordinate (H)
(\x/2, 0 ,\y/2) coordinate (I) (\x/2,\z,\y/2) coordinate (J);
\draw (A)--(B)--(C)--(G)--(F) (A)--(E)--(F)--(G)--(H)--(E);
\draw (A)--(D)--(C) (D)--(H);
\filldraw[draw=red,
         fill=red!20]          
         (0,2,\y)--(\x,2,\y)--(\x,2,0)--(0,2,0)--cycle;
\draw (B)--(F);
\node[align=left] at (-2.25,-0.5,0) {b)};
\end{tikzpicture}
\hspace{32pt}
\begin{tikzpicture} [scale = .4]
\pgfmathsetmacro{\x}{3}
\pgfmathsetmacro{\y}{3}
\pgfmathsetmacro{\z}{6}
\path (0,0,\y) coordinate (A) (\x,0,\y) coordinate (B) (\x,0,0) coordinate (C) (0,0,0)
coordinate (D) (0,\z,\y) coordinate (E) (\x,\z,\y) coordinate (F) (\x,\z,0) coordinate (G)
(0,\z,0) coordinate (H);
\draw (A)--(B)--(C)--(G)--(F) (A)--(E)--(F)--(G)--(H)--(E);
\draw (A)--(D)--(C) (D)--(H);
\filldraw[draw=red,
         fill=red!20]          
         (0,0,\y/2)--(\x,0,\y/2)--(\x,\z,\y/2)--(0,\z,\y/2)--cycle;
\draw (B)--(F) (E)--(F)--(G);
\node[align=left] at (-2.25,-0.5,0) {c)};
\end{tikzpicture}
\hspace{32pt}
\begin{tikzpicture} [scale = .5]
	\draw [<->] (0,2.15) -- (0,1.15) -- (1,1.15);
	\draw[->] (0,1.15) -- (.5,1.4);
	\node [right] at (.85,1.175) {\tiny{$x$}};
	\node [right] at (.3, 1.5) {\tiny{$z$}};
	\node [above] at (0,2.025) {\tiny{$y$}};
\end{tikzpicture}
\caption{Data point collection for (a) centerline, (b) normal, and (c) axial slicing methods.}
\label{fig:slicing}
\end{figure}
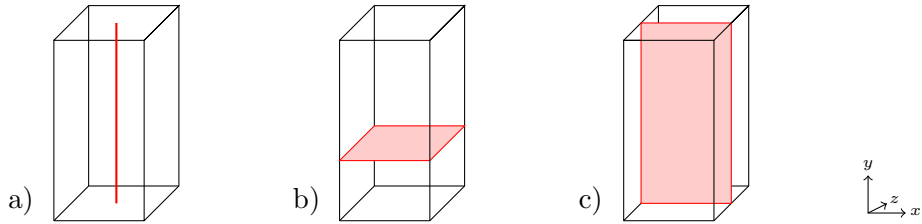

For time averaging, a basic discrete averaging procedure was implemented over $i$ saved plot files from the final 2 flow-throughs of the simulation, with 100 plots saved per flow-through:
\begin{equation} \label{discrete_time_avg}
\phi_{avg,\ slice} = \dfrac{1}{200} \sum\limits_{i=1}^{200} \phi_{i,\ slice}.
\end{equation}
For each slice normal to the jet axis, radial averages are approximated with $r = 0$ lying in the center of the $xz-$plane with $r = \sqrt{x^2 + z^2}$. The $N$ points nearest to each fixed $r$ distance away from the center were then averaged together:
\begin{equation} \label{discrete_time_avg}
\phi_{avg,\ r} = \dfrac{1}{N} \sum\limits_{k=1}^{N} \phi_{k,\ r}.
\end{equation}
For quantities that were averaged both temporally and radially, temporal averaging was performed first, with the radial average then taken of the time averaged slice data. Fluctuating quantities are calculated on 2D slices as follows:
\begin{equation} \label{discrete_rey_decomp}
\phi'_{i,\ slice} = \phi_{avg,\ slice} - \phi_{i,\ slice}.
\end{equation}
These perturbations can then be used to calculate average resolved Reynolds stresses for each $u_i$ component of velocity, $\langle u'_i u'_j \rangle$, and \gls{tke} components \cite{iso_comp_2} as follows: 
\begin{equation} \label{TKE}
\langle TKE \rangle = \dfrac{1}{2} \left( \langle u' u' \rangle + \langle v' v' \rangle + \langle w' w' \rangle  \right).
\end{equation}
In the results section, when these quantities are introduced, they are presented without angle brackets for the sake of streamlining notation, but note that all of the results presented from here on are averaged either temporally or temporally and radially as was presented in this section, unless explicitly stated otherwise.

\section{Results and Discussion}

Simulations were run on the \gls{nrel} Eagle supercomputer using Intel Skylake processors with the Intel suite of compilers \cite{10920476}. Simulations used 576 \gls{mpi} ranks in total, with 36 ranks spread across 16 nodes. Roughly 17,000 cells were handled per \gls{mpi} rank. Dynamic load balancing of \gls{amr} levels allowed for maximum utilization of the system. In total, over 100 TB of data were stored from the original 3D simulations run over the course of the 10 flow-throughs of the domain needed for averaging procedures. To help manage data storage issues, slicing procedures were implemented to reduce full 3D simulation data down to 2D slice data, where averaging procedures were then performed as outlined in the previous section. 

Here, we present the results of the three cases outlined in section \ref{OpCon}. In addition to visualizing some qualitative aspects of jet slices along the axial direction, a variety of quantities of interest are analyzed. Quantities of interest we consider include axial velocity decay (in both the radial and axial directions), centerline velocity decay, resolved Reynolds stresses, and resolved \gls{tke}, among others. These quantities are commonly explored in turbulent jet investigations and have been chosen to highlight the differences in flow field structure between our three cases. In the graphs presented throughout this chapter, all quantities are averaged over time unless specifically denoted otherwise. Additionally, for graphs plotted in the radial direction away from the centerline, quantities are radially-averaged. 

These results are organized as follows: first, the isothermal jet case is presented and quantities of interest are related to those of comparable subcritical turbulent jets studies from relevant literature \cite{iso_comp_2, iso_comp_1, iso_comp_1_ref_1}. This establishes a baseline for comparison with the non-isothermal jet cases and is used to validate that the numerical jet setup is behaving in a reasonable manner. The two non-isothermal cases are then presented together, with the 350~K ambient case compared against the isothermal case. Variations between these two cases are related to similar non-isothermal turbulent jets from the subcritical regime. Our main case of interest is the 314~K ambient case as it contains the transition over the pseudo-boiling point. We then compare this case with the 350~K ambient case, which is additionally related to similar supercritical investigations regarding the work of Mayer \emph{et al.}, where transcritical injection involves a phase change \cite{mayer2003raman}. Through the comparisons of our 314~K  cases, we specifically investigate the impact of the transition through the pseudo-boiling point on the supercritical flow field. 

\subsection{Isothermal Jet}
Our first case involves the isothermal jet where both jet and ambient fluid are at 330~K. Quantities of interest for this case are compared against an incompressible round turbulent jet \cite{Pope} and apt compressible jet cases from literature \cite{iso_comp_1, iso_comp_1_ref_1, iso_comp_2}. This case serves as a baseline for comparison with the other two cases involving non-isothermal jets.

\subsubsection{Flow Field Features}
All images here depict a 2D slice of the 3D flow field at $z=0$. Each figure contains an instantaneous snapshot and a time-averaged snapshot of the entire slice domain, plus an additional zoomed-in image of the time-averaged quantity of interest near the inlet. All instantaneous images are taken from the final data point of the simulation.

\begin{figure}[hbtp!]
\centering
\hspace{28pt}
\subfloat[ \label{330_v_1}] { 
	\includegraphics[clip, trim={0 0 6.5cm 0}, scale=.3]{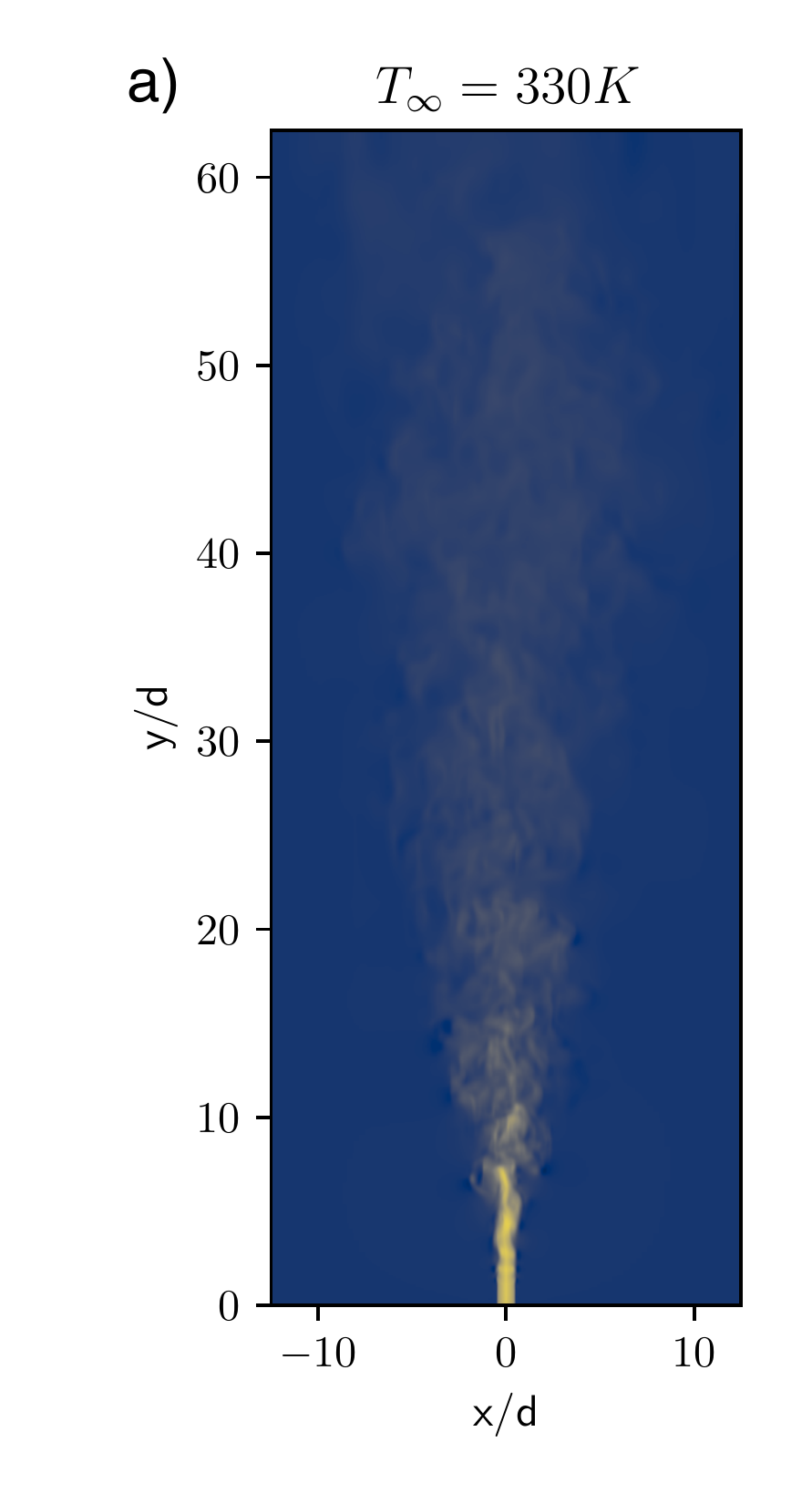}
 } 
\subfloat[ \label{330_v_3}] { 
	\includegraphics[trim={0 0 6.5cm 0},clip, scale=.3]{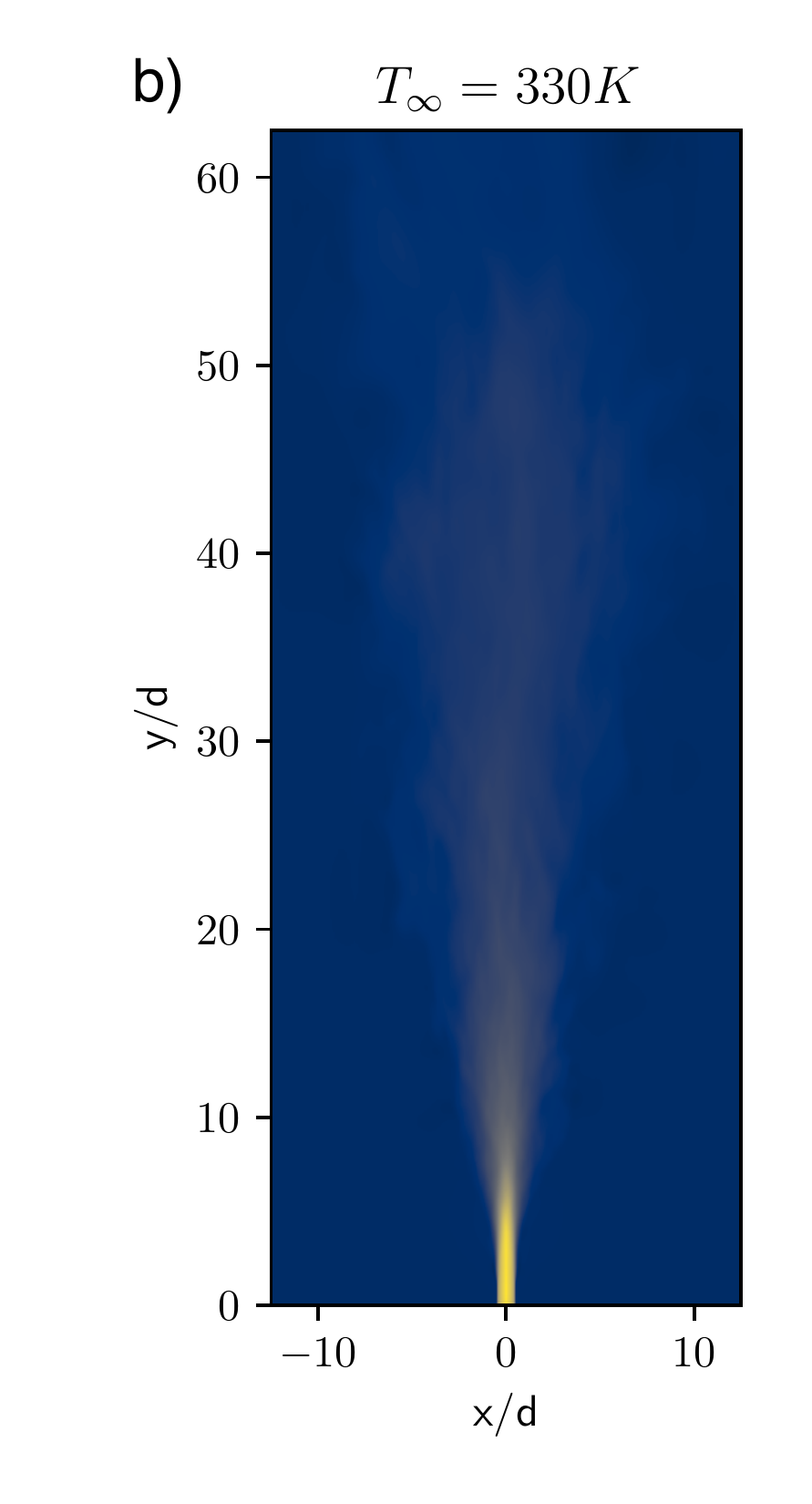} 
} 
\subfloat[\label{330_v_4}] { 
	\includegraphics[clip, scale=0.3]{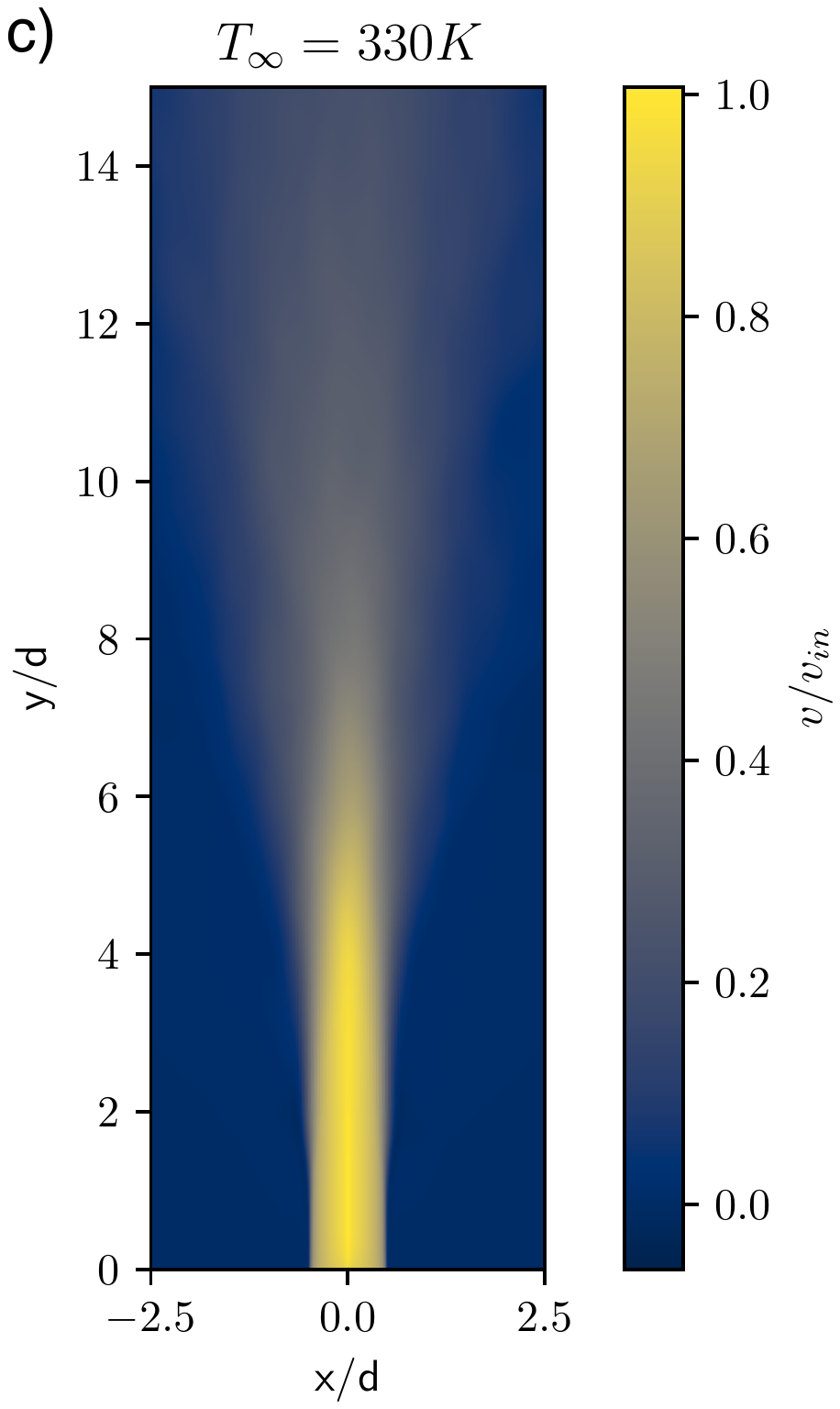} 
} \hfill
\vspace{-28pt}
\caption{(a) Instantaneous and (b) time-averaged axial velocity features of the isothermal jet, with a closer look at averaged values (c) near the inlet.} \label{330_v_features}
\end{figure}

Figure \ref{330_v_features} shows the axial velocity component of the flow field. The general spreading rate and decay of the velocity field can be seen from the instantaneous snapshot in Figure \ref{330_v_1}. From these images, the flow field can be distinguished into three main regions, as noted in \cite{iso_comp_1_ref_1}: the potential core region, the transition region, and the fully developed region. The flow is laminar up until about $\nicefrac{y}{d}=2$ before perturbations begin. The stream mostly stays together through these initial perturbations up until $\nicefrac{y}{d}=7$ where spreading then begins, marking the end of the potential core. The averaged axial velocity fields in Figures \ref{330_v_3} and \ref{330_v_4} shed more light on the development of the jet. The transition region occurs approximately from $7 \leq \nicefrac{y}{d} \leq 15$, where the laminar state evolves into the turbulent state as flow structures begin to interact with one another. Thereafter the jet appears fully developed and nonlinear interactions give way to eddies across a range of large and small scales. Centerline analysis of the turbulent kinetic energy later on will provide more information for these boundaries.  

\begin{figure}[hbtp!]
\centering
\subfloat[] { 
	\includegraphics[clip, trim={0 0 7cm 0}, scale=.3]{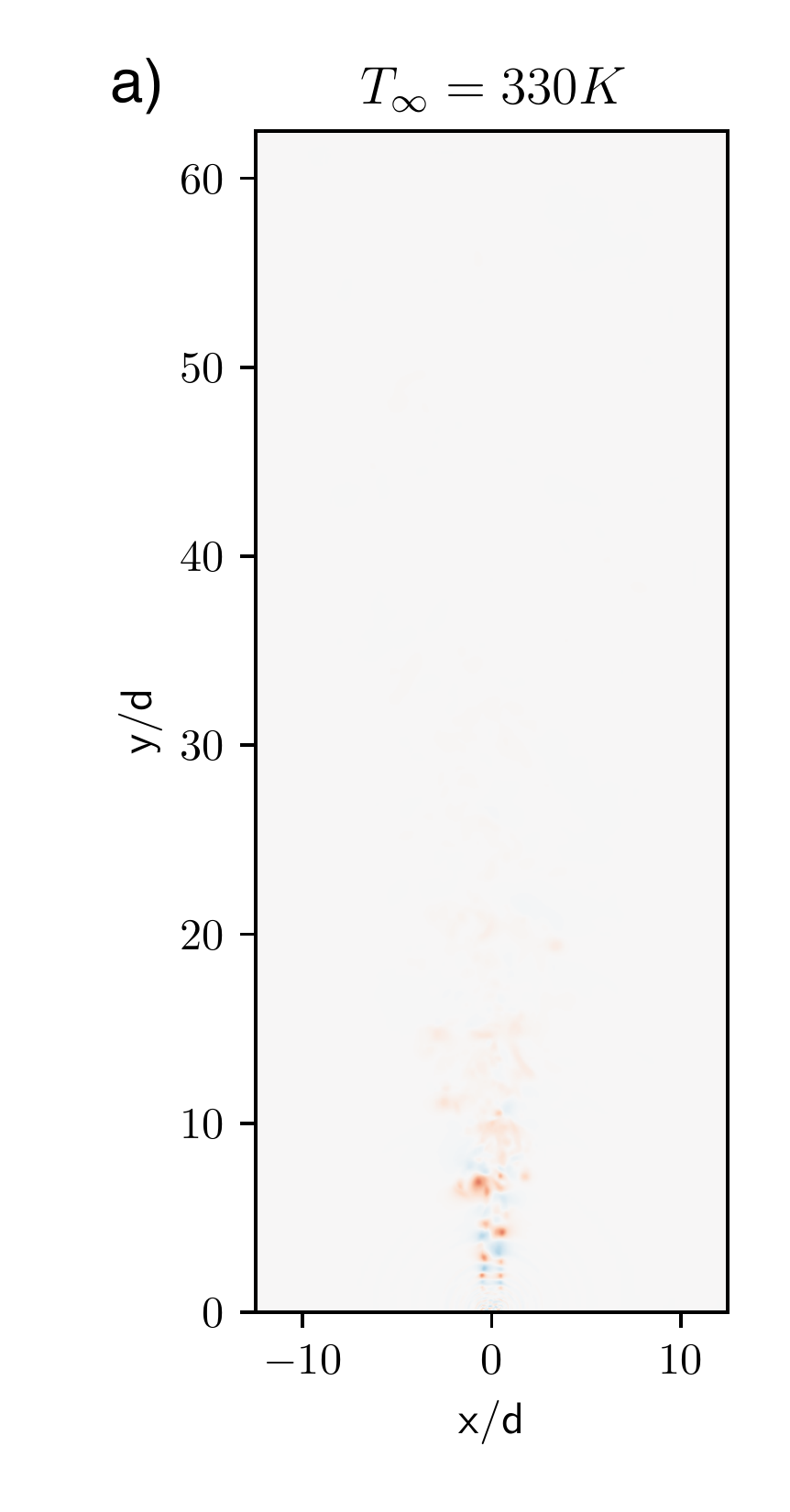} 
    \label{330_pressure_1}
}
\subfloat[] { 
	\includegraphics[trim={0 0 7cm 0},clip, scale=.3]{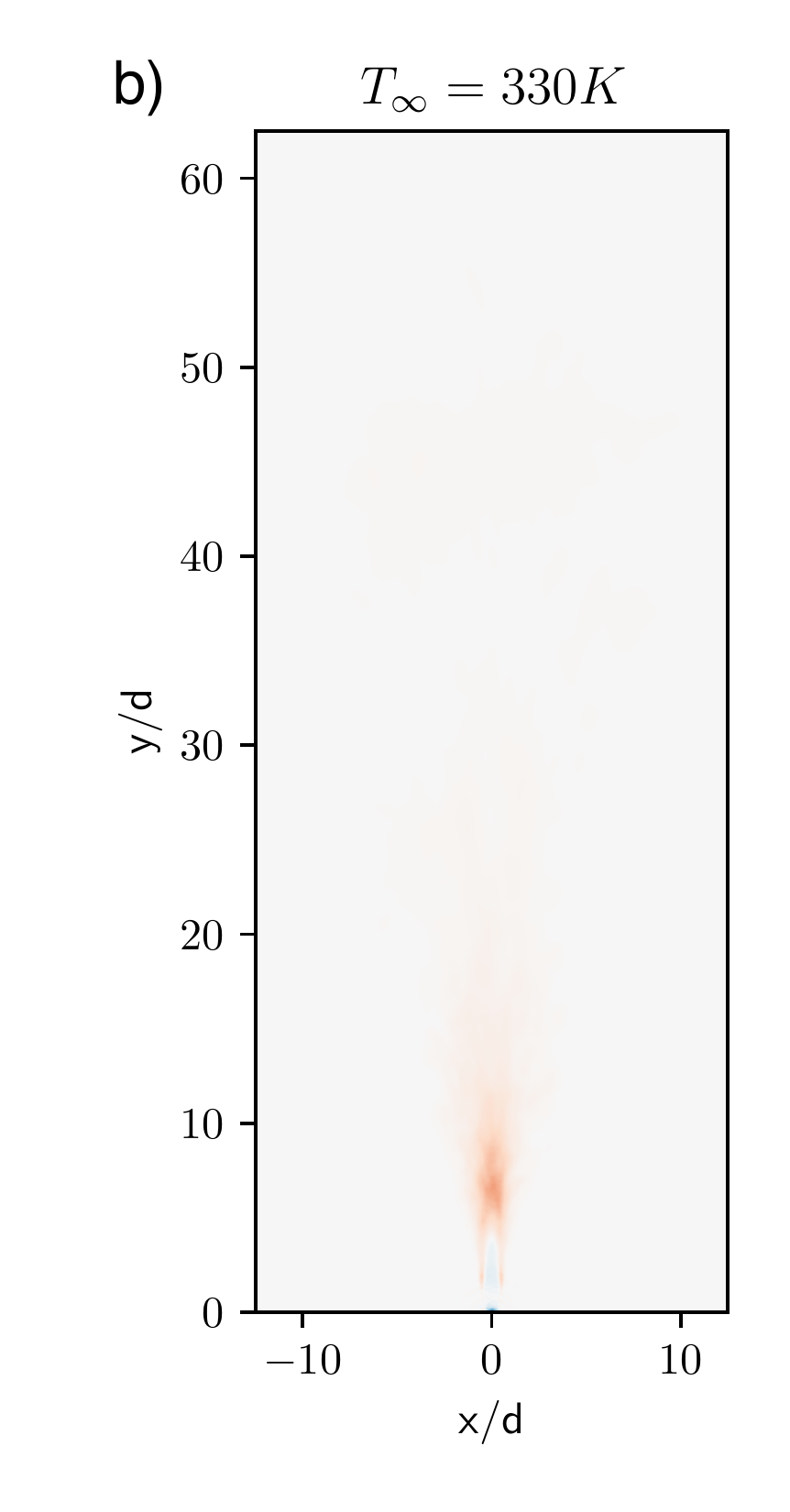} 
    \label{330_pressure_3}
}
\subfloat[] { 
	\includegraphics[clip, scale=.3]{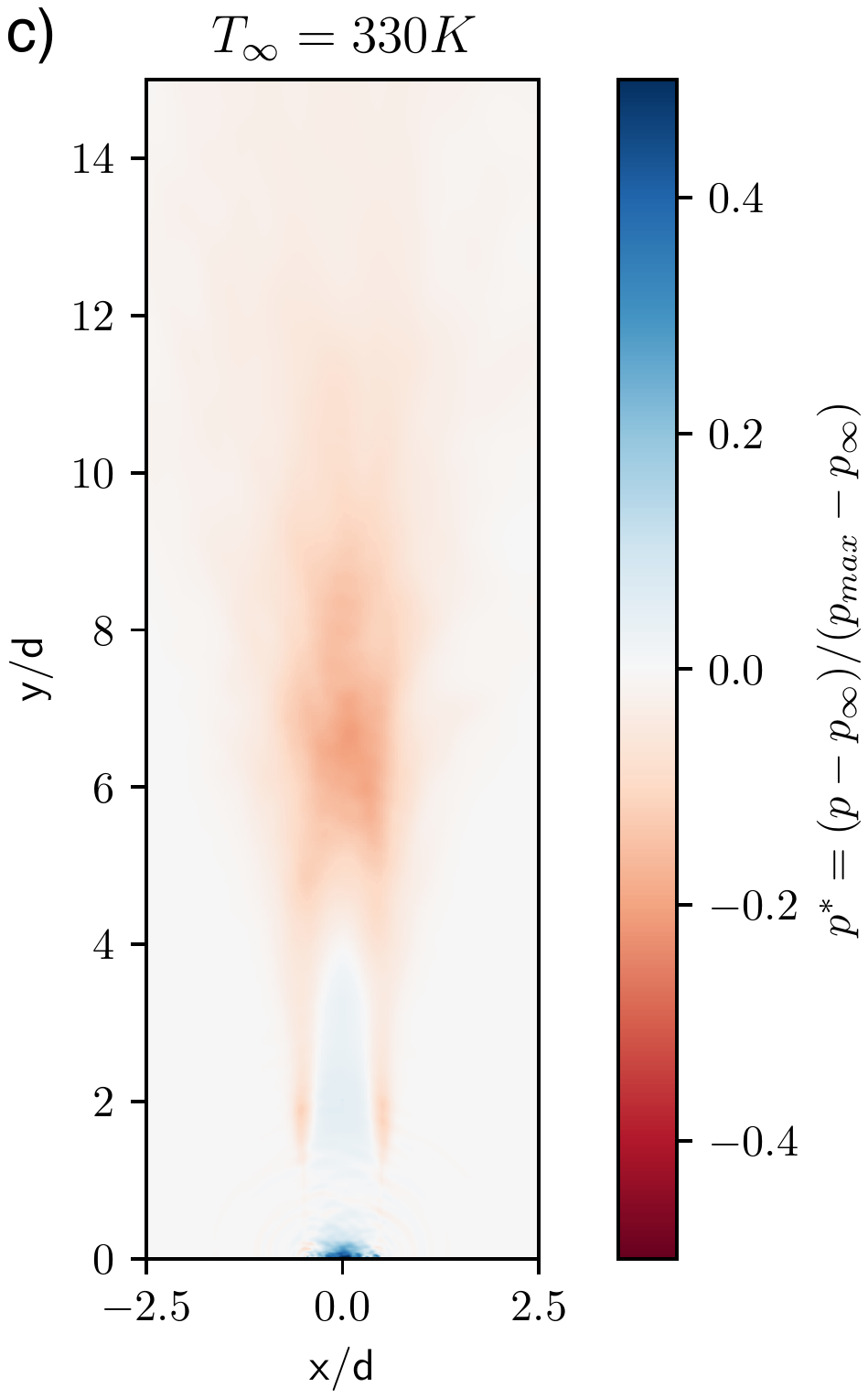} 
    \label{330_pressure_4}
}
\vspace{-28pt}
\caption{(a) Instantaneous and (b) time-averaged pressure features of the isothermal jet, with a closer look at averaged values (c) near the inlet.}
\label{330_pressure_features}
\end{figure}

Figure \ref{330_pressure_features} shows minor pressure fluctuations in the flow, scaled against the maximum pressure achieved above the ambient pressure. Figure \ref{330_pressure_1} shows minor pressure oscillations mirrored on each side of the jet edge in the same zone as the initial velocity fluctuations see in Figure \ref{330_v_1}. Thereafter, the oscillations become asymmetric, correlating to the beginning of the jet disintegration as seen in the velocity field. Pressure fluctuations are concentrated near the inlet and decrease in amplitude past the transition zone. This can be seen more clearly in Figure \ref{330_pressure_3}. On average, pressure fluctuations yield a minor increase within the potential core and decrease on the jet perimeter, as can be seen in Figure \ref{330_pressure_4}. Fluctuations also lead to a minor pressure drop on average in the transition region.

\begin{figure}[hbtp!]
\centering
\subfloat[] { 
	\includegraphics[clip, trim={0 0 7cm 0}, scale=.3]{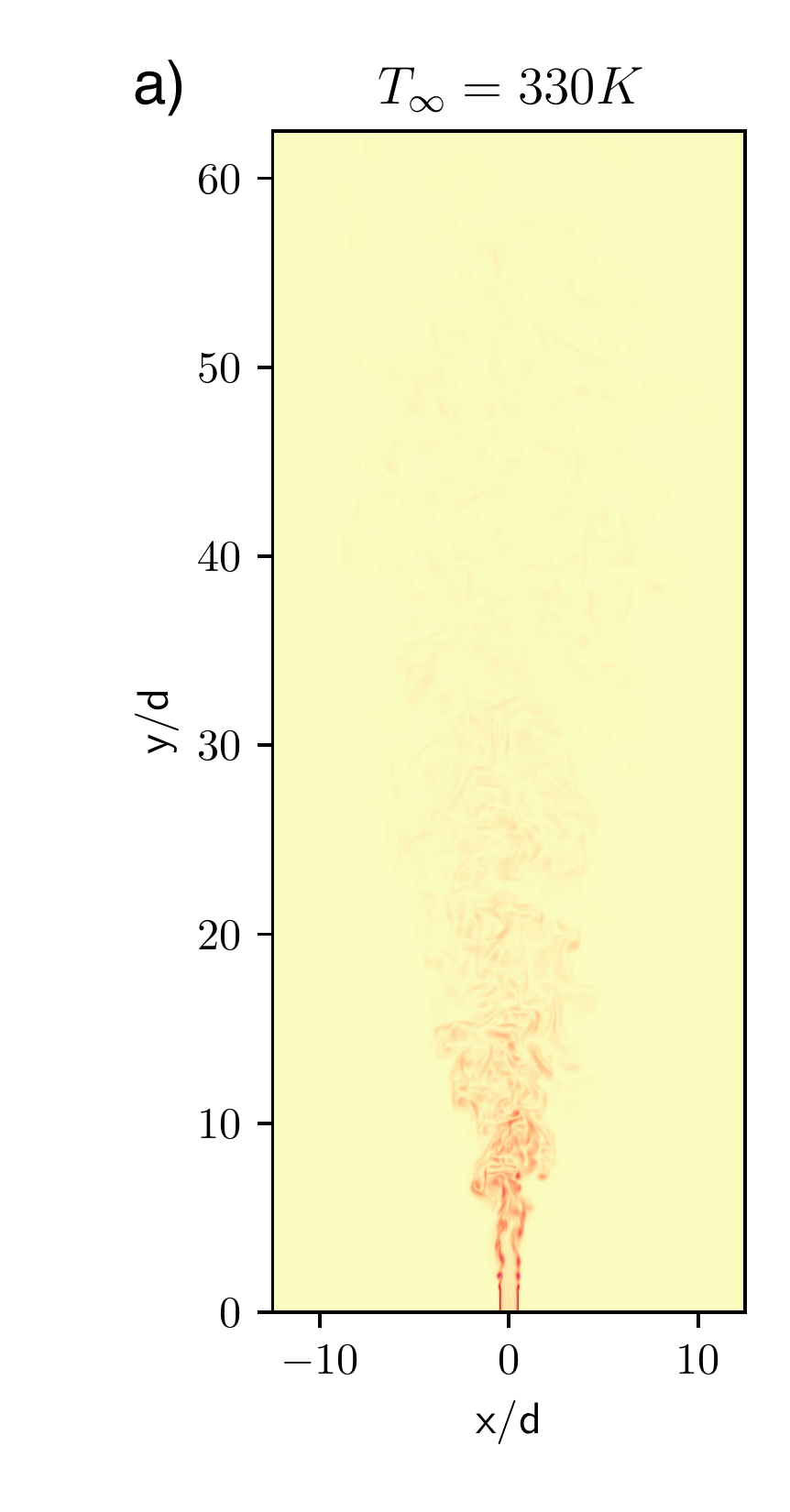}
    \label{330_magvort_1}
    }
\subfloat[] { 
	\includegraphics[trim={0 0 7cm 0},clip, scale=.3]{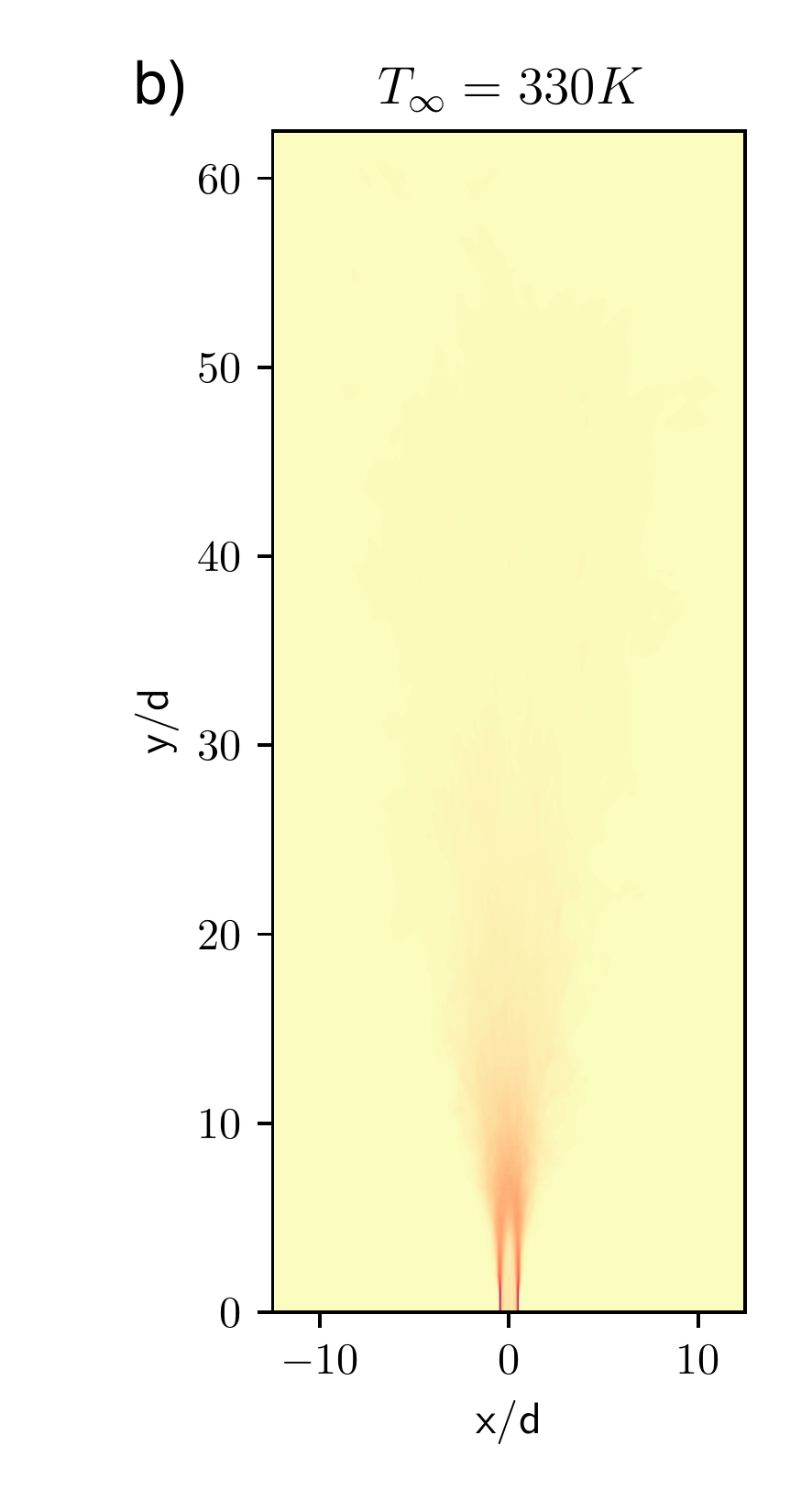}
    \label{330_magvort_3}
    }
\subfloat[] { 
	\includegraphics[clip, scale=.3]{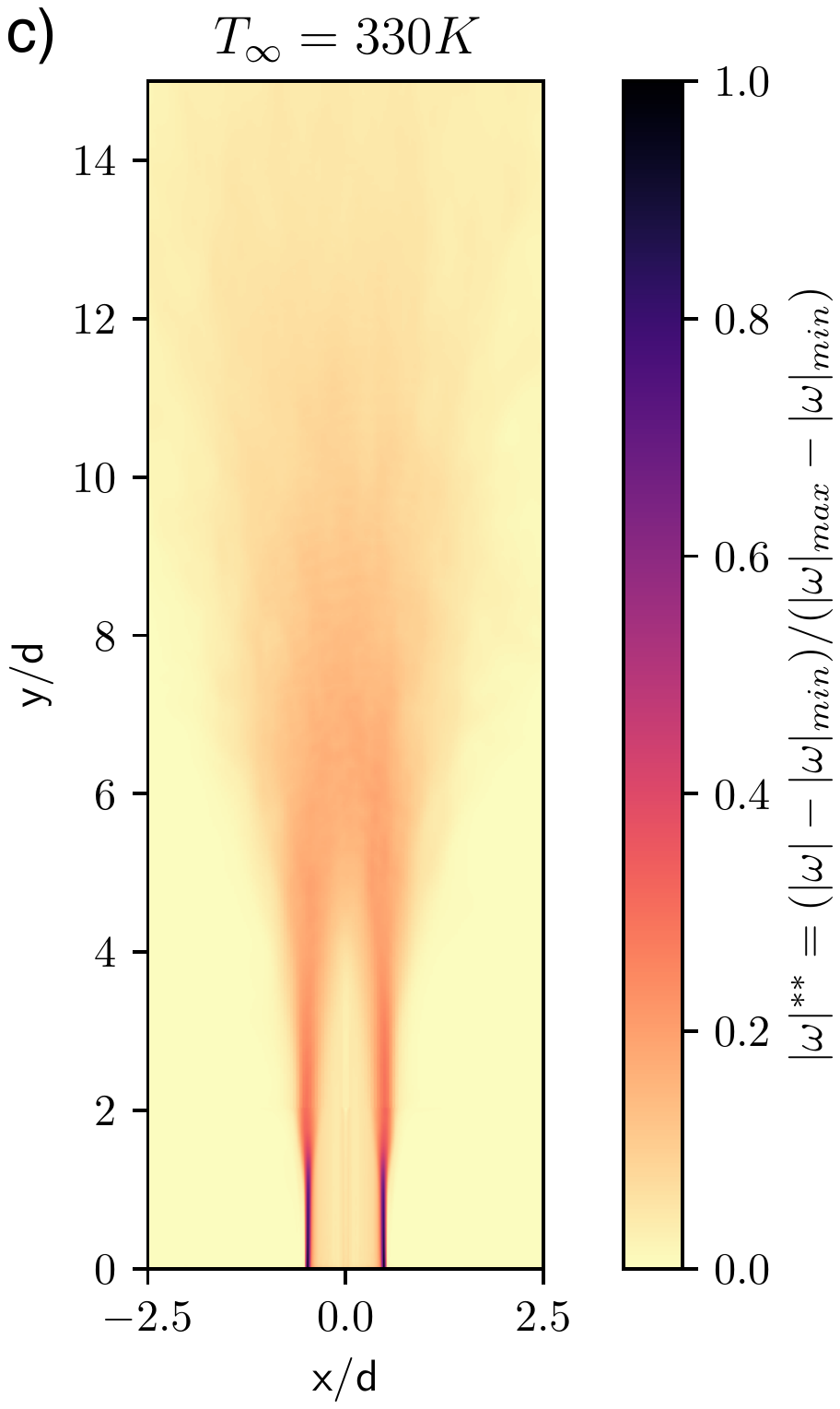}
    \label{330_magvort_4}
    }
\vspace{-28pt}
\caption{(a) Instantaneous and (b) time-averaged vorticity magnitude features of the isothermal jet, with a closer look at averaged values (c) near the inlet.}
\label{330_magvort_features}
\end{figure}

Figure \ref{330_magvort_features} shows the magnitude of the vorticity scaled between the maximal and minimal values. Figure \ref{330_magvort_1} shows the strongest vorticity occurring at the jet interface with the ambient fluid near the inlet. Past this initial stage, a Kelvin-Helmholtz instability causes a shear rollup at the outer edge of the jet around $\nicefrac{y}{d} = 2$, forming small but coherent vortical structures. These structures then become unstable and lead to more complex vortices beginning around $\nicefrac{y}{d}= 5$. The averages in Figures \ref{330_magvort_3} and \ref{330_magvort_4} both show again that the most intense vorticity occurs at the inlet along the outer edge of the jet. This high intensity remains constant until about $\nicefrac{y}{d}=1$ before more mixing with the ambient fluid occurs as the jet spreads and the vorticity lessens in intensity. Vortices are still restricted to the jet edge until around $\nicefrac{y}{d} = 5$ where the transition to fully developed turbulence enables vortical motions to extend across the fully spread of the jet. Generally, vorticity then dissipates as the jet spreads farther downstream.

\subsubsection{Mean Flow Properties}
\begin{figure}[hbtp!]
	\includegraphics[scale=.6]{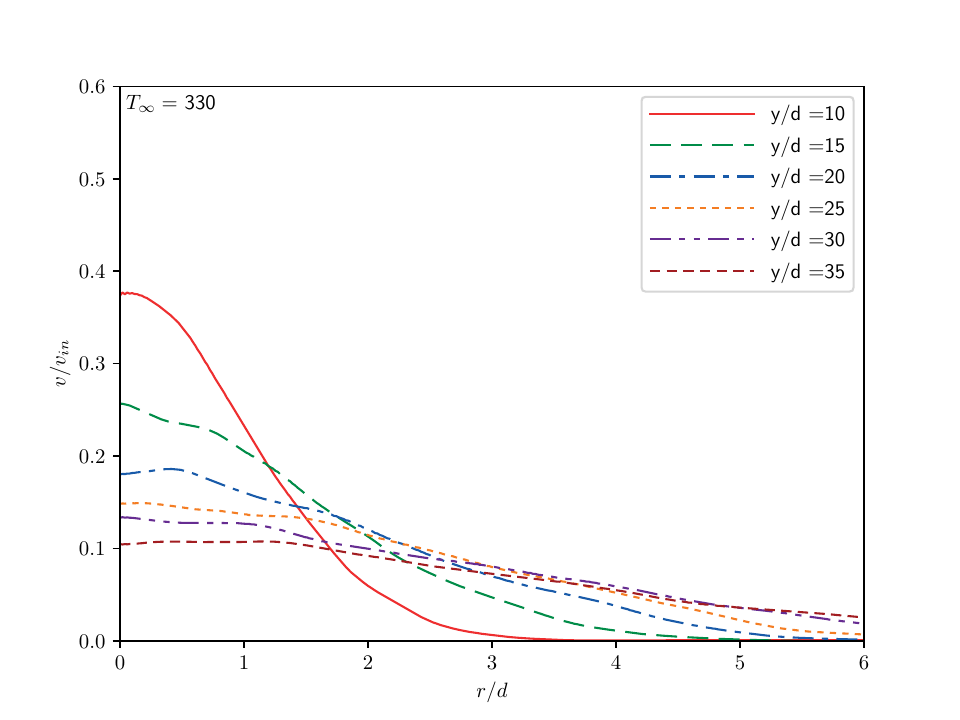}
	\caption{Average (both in time and radially) axial velocity scaled by inlet value plotted along radial distance from centerline. Profile decay follows similar trajectory to what is expected in incompressible round jet theory \cite{Pope}.} \label{330_centerline_decay}
\end{figure}

Figure \ref{330_centerline_decay} depicts the time and radially averaged scaled axial velocity component plotted against radial distance from the centerline at multiple normal slices downstream from the inlet. The velocity is scaled by the average axially velocity value at the inflow while the radial direction is scaled by the jet diameter. These plots demonstrate the axial velocity decay as the flow progresses farther downstream. As the velocity value along the centerline decreases, the \gls{hmhw} increases, both flattening and widening the velocity profile. Typically, for the round turbulent jet, this expansion would occur in such a way that upon specially selected scaling, these profiles would collapse into one profile after a certain point. This potential is explored farther in the next figure. This decay is a common feature of how the axial velocity of the jet develops as it leaves the inlet \cite{Pope}. 

\begin{figure}[htbp!]
    \centering
    \subfloat[] { 
	\includegraphics[trim={0 5cm 0 7cm}, scale=.37, clip=tight]{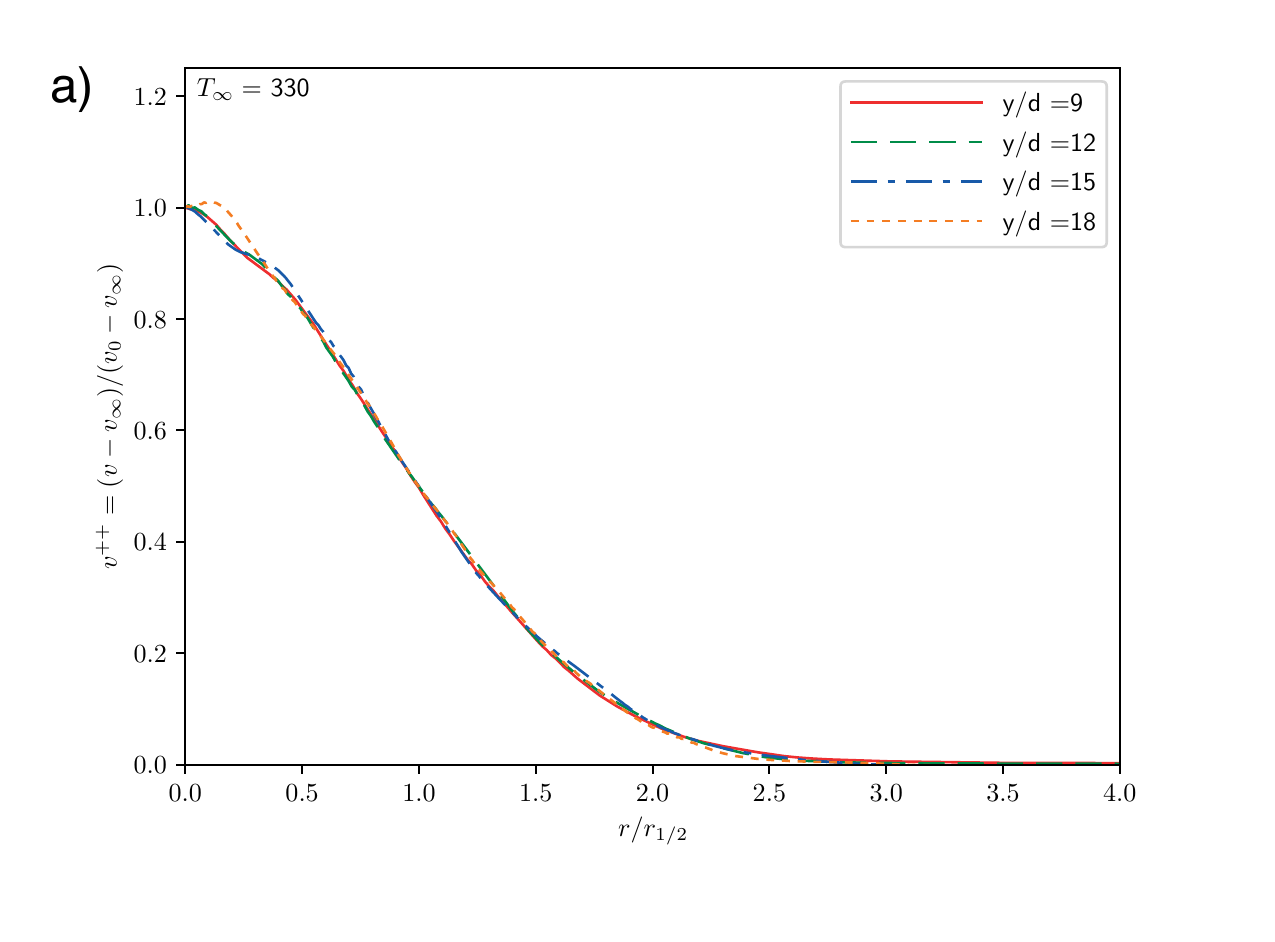}
	\label{330_r_vs_v_1}
    } 
    \subfloat[] { 
	\includegraphics[trim={0 5cm 0 7cm}, scale=.37, clip=tight]{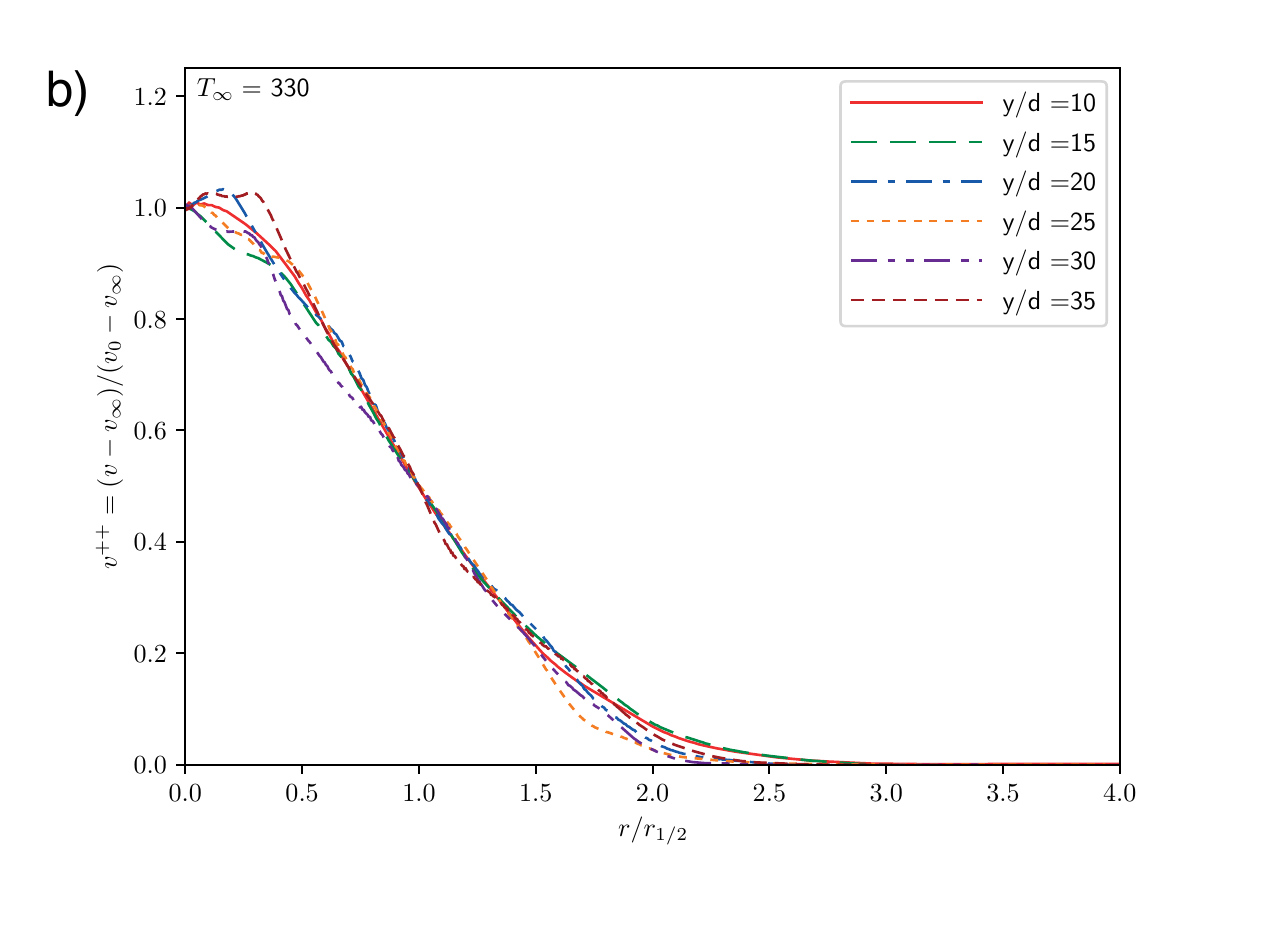}
	\label{330_r_vs_v_2}
    }
\vspace{-28pt}
\caption{Normal slices of scaled axial velocity, averaged in both time and the radial direction, plotted in the radial direction scaled by $r_{1/2}$. Both (a) near- and (b) far-field regions demonstrate the self-similarity within the round turbulent jet.}
\label{330_r_v_features}
\end{figure}

Figure \ref{330_r_v_features} depicts time and radially averaged scaled axial velocity components plotted against the radial distance from the centerline. Each curve is made at a slice normal to the axial flow direction at different points downstream. Figure \ref{330_r_vs_v_1} depicts velocity curves every $3d$ downstream from the inlet near where the transition region begins while Figure \ref{330_r_vs_v_2} contains plots taken every $5d$. The axial velocity is scaled by the centerline value $v_c$ while the radial distance is scaled by the \gls{hmhw}, where the velocity component is equal to half the value on the centerline $v(r_{1/2}) = \nicefrac{v_c}{2}$.

Figure \ref{330_r_vs_v_1} shows axial velocity profiles in the transition region of the jet. They exhibit self-similarity collapse into one profile which, is a common feature of round turbulent jets \cite{Pope, iso_comp_1, iso_comp_2}. farther downstream as depicted in Figure \ref{330_r_vs_v_2}, self-similarity is fairly well maintained with minor fluctuations in the center and edge of the jet. These fluctuations are most likely the result of low resolution in the time averaging of available data. Overall though, this general collapse is in agreement to that which is seen in comparable low-Mach isothermal round jet simulations and experiments, as are summarized in \cite{iso_comp_1}. 

\begin{figure}[hbtp!]
\begin{center}
	\includegraphics[scale=.6]{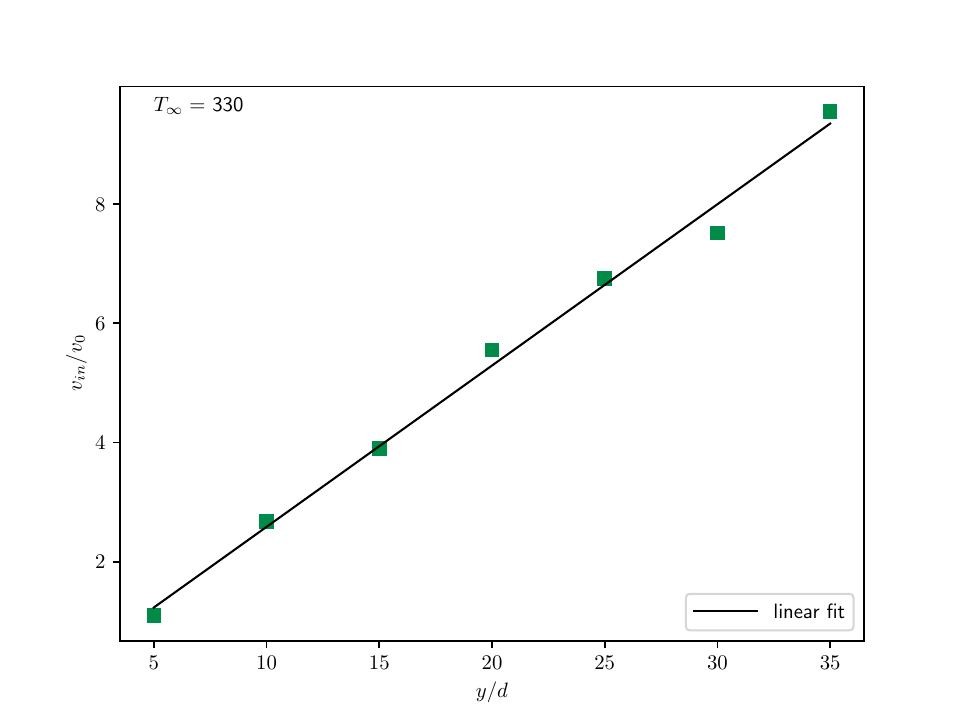}
	\caption{Axial inlet velocity scaled by centerline values along the axial direction. When the distance downstream is scaled by the jet diameter, linear decay of the centerline axial velocity is observed.} \label{330_centerline_scaling}
\end{center}
\end{figure}

Another common feature of round turbulent jets is the development of a linear relationship between the jet centerline value and the distance downstream. This comparison for the transition region of the jet is depicted in Figure \ref{330_centerline_scaling}. Here, the centerline value of the axial velocity $v_0$ is inversely linearly proportional to the distance downstream. The decay rate can be characterized by the following relationship \cite{iso_comp_1}:
\begin{equation} \label{decay_rate}
\dfrac{v_{in}}{v_0} = \dfrac{1}{B_v}\left[ \dfrac{y}{d} - \dfrac{y_0}{d} \right],
\end{equation}
where $y_0$ is the virtual origin of the jet \cite{Pope} and $B_v$ is the decay rate. For this case, the decay rate is given by $B_v = 3.69$.  The decay rate here is smaller than those of comparable low-Mach jets in the subcritical regime, as summarized in \cite{iso_comp_1_ref_1}. 

\begin{figure}[htbp!]
\begin{center}
	\includegraphics[scale=.6]{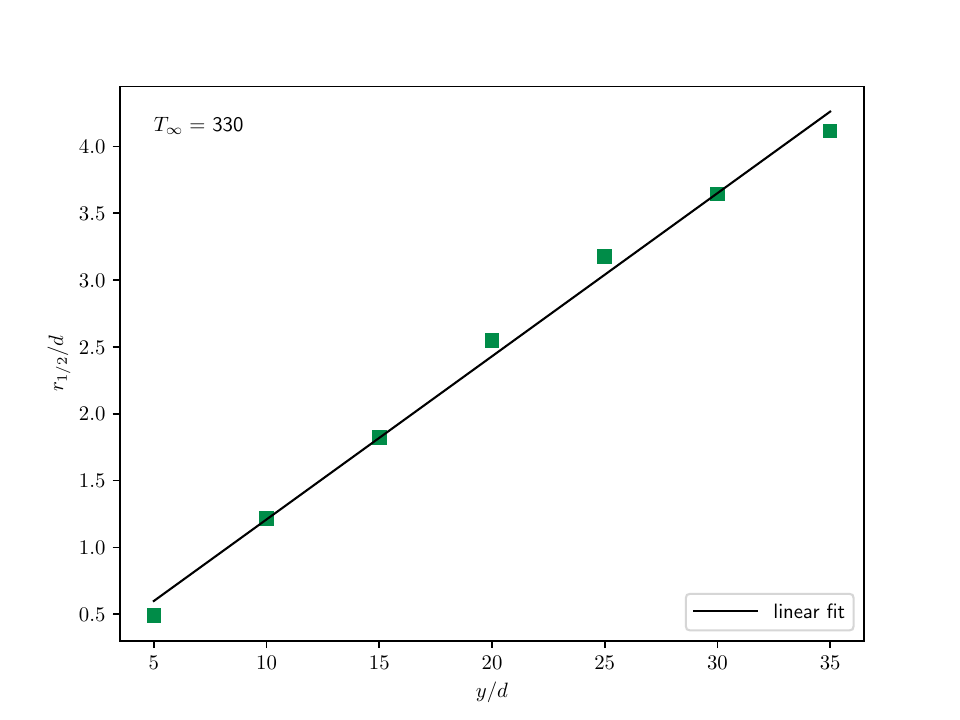}
	\caption{\gls{hmhw} of axial velocity along the axial direction. When the distance downstream is scaled by the jet diameter, linear growth of the axial velocity \gls{hmhw} is observed.} \label{330_hwhm_scaling}
\end{center}
\end{figure}

As can be seen in Figure ~\ref{330_hwhm_scaling}, the spreading rate of the jet can also be characterized by linear development downstream in a similar fashion \cite{iso_comp_1_ref_1}:
\begin{equation} \label{spread_rate}
\dfrac{r_{1/2}}{d} = C_v\left[ \dfrac{y}{d} - \dfrac{y_0}{d} \right],
\end{equation}
where $C_v$ is the spreading rate of the jet. The spreading rate here is given by $C_v=0.122$. The spreading rate here is larger than those of comparable low-Mach jets in the subcritical regime, as summarized in \cite{iso_comp_1_ref_1}. 

\subsubsection{Turbulence Dynamics}
\begin{figure}[htbp!]
\centering
    \subfloat[] { 
 	\includegraphics[trim={0 5cm 0 7cm}, scale=.37, clip=tight]{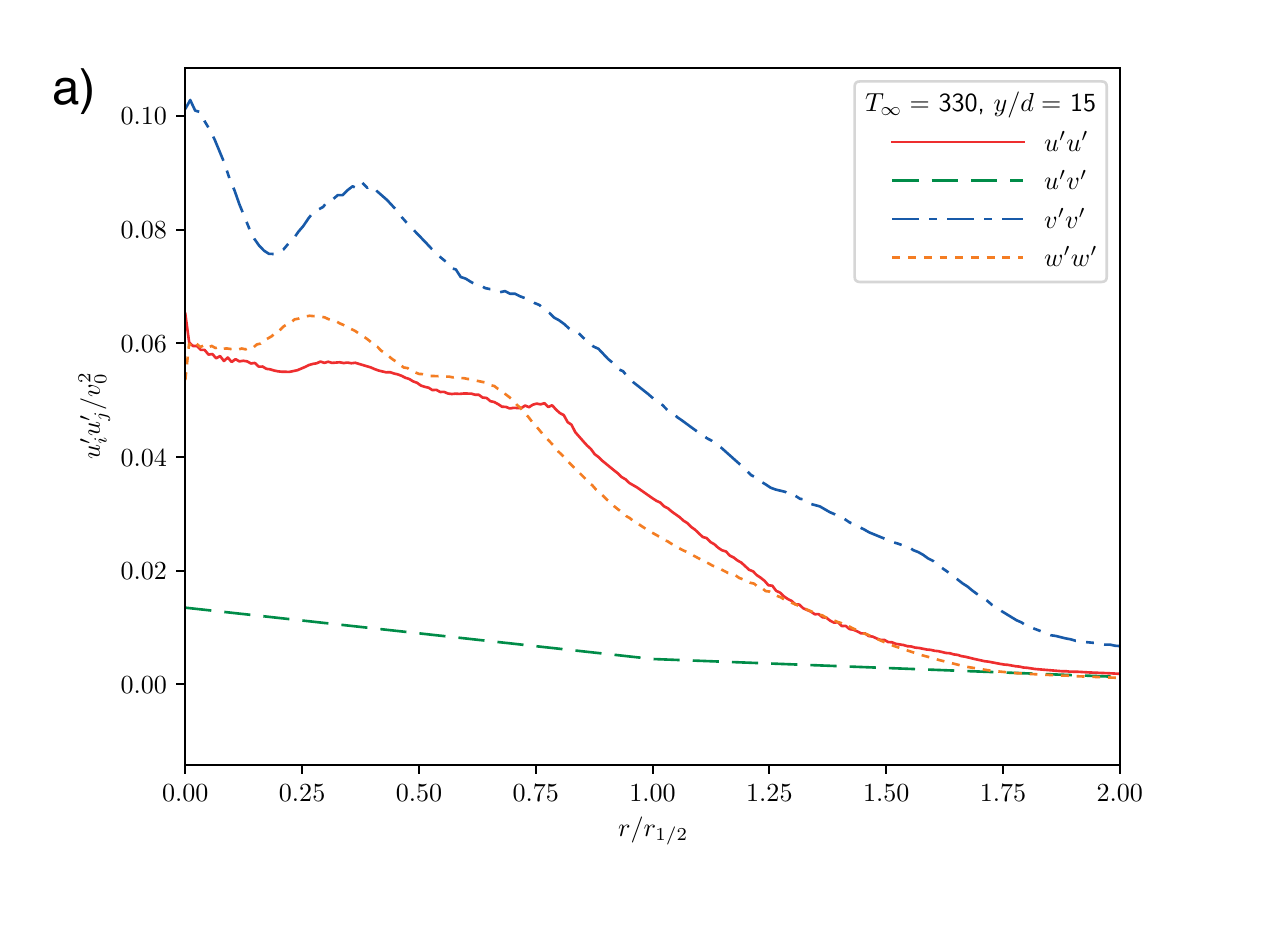} 
    \label{330_rey_15}
    } 
    \subfloat[] { 
 \includegraphics[trim={0 5cm 0 7cm}, scale=.37, clip=tight]{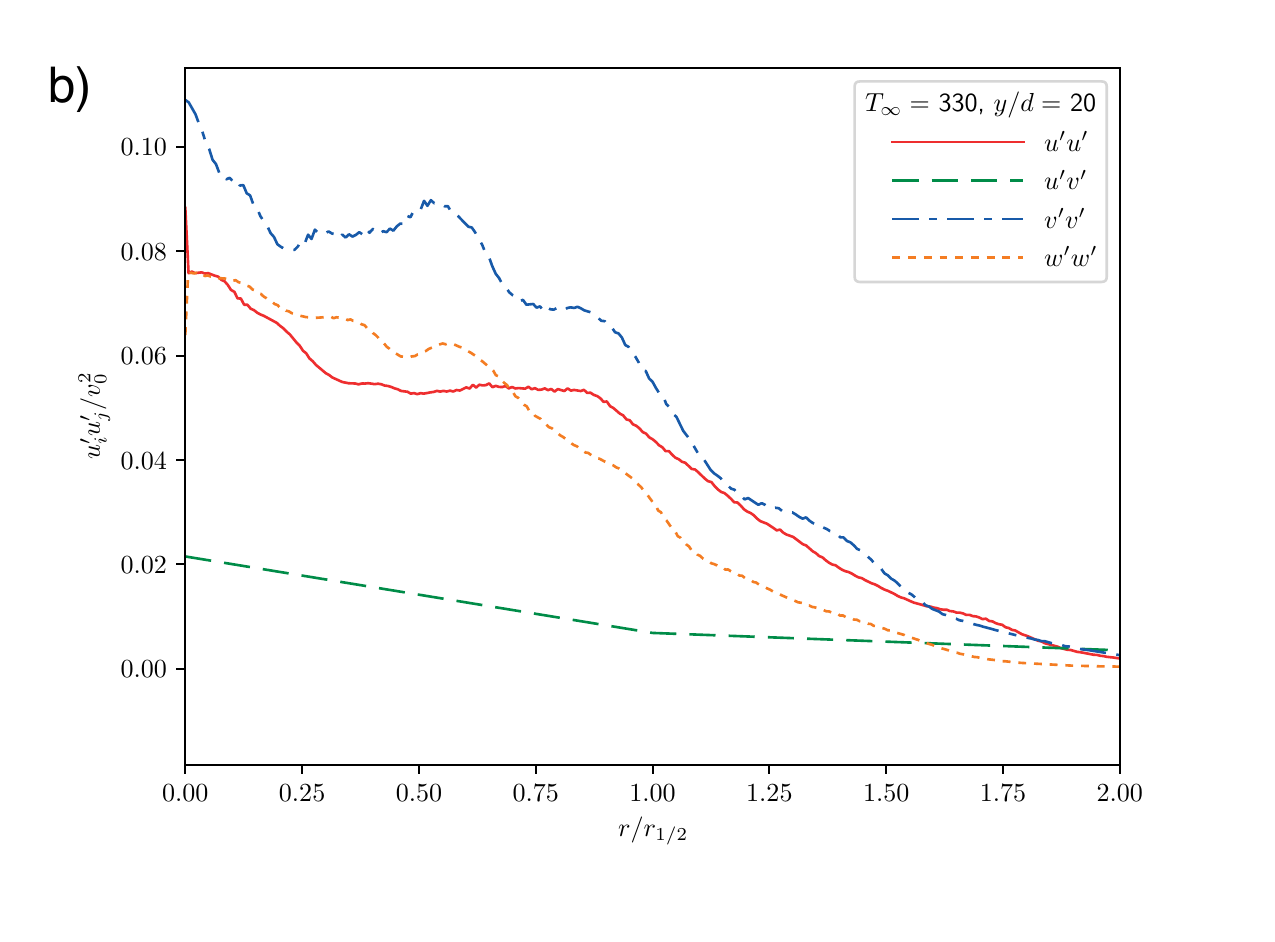}
    \label{330_rey_20}
    }
\vspace{-28pt}
\caption{Time and radially averaged Reynolds stresses for the isothermal jet at two locations downstream: a) $y/d=15$ and b) $y/d=20$. Both slices follow similar Reynolds stress relations seen in incompressible round jets \cite{Pope}.}
\label{330_reynolds_features}
\end{figure}

Figure \ref{330_reynolds_features} shows the time and radially averaged resolved Reynolds stresses at two points downstream from the inlet. Here, velocity components $(u,v,w)$ correspond to the $(r,y,\theta)$ directions, respectively. Each Reynolds stress component follows general trends associated with round turbulent jets \cite{Pope, iso_comp_1_ref_1}, with the axial component providing the leading contribution, followed by the other two directional components closely with these two being of roughly the same magnitude, and the cross-directional component contributing the least. Centerline values are slightly higher than the typical values seen in other works as is summarized in \cite{iso_comp_1_ref_1}, but are of the same order of magnitude. This discrepancy could be due to error induced by the \gls{sgs} model accuracy for second order moments as noted in \cite{doi:10.1063/1.4937948}. Note also that self-similarity is not exhibited, as each component exhibits an increase in magnitude at the center of the jet as distance downstream is increased. 

\begin{figure}[hbtp!]
\begin{center}
	\includegraphics[scale=.6]{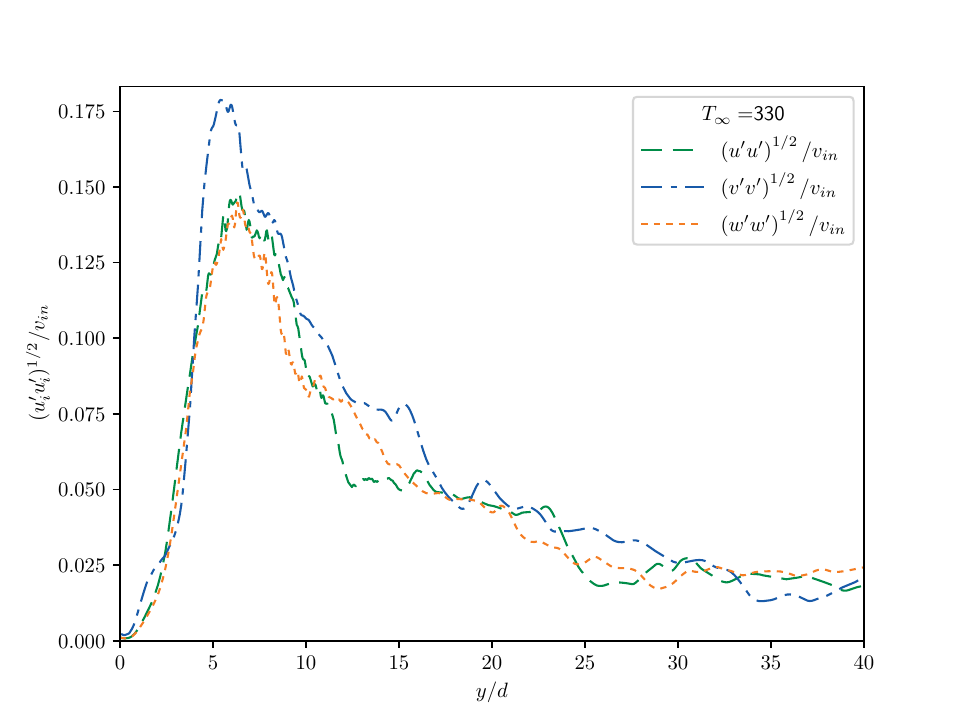}
	\caption{Average resolved turbulent kinetic energy components along centerline. Note: averaging as described in Chapter 4 occurs prior to square rooting in the given quantity description.} \label{330_TKE_features}
\end{center}
\end{figure}

Figure \ref{330_TKE_features} shows the time average \gls{tke} components along the centerline of the jet. Each component grows through the potential core region of the jet up until all components reach a peak in energy around $\nicefrac{y}{d}=6$, with the axial component's peak coming slightly before the other two directions. Rapid decay is then observed up until around $\nicefrac{y}{d} = 15$ before a slower decay sets in up until a leveling off is achieved around $\nicefrac{y}{d} = 30$. This rapid decay and then farther progression correspond to the transition and fully developed jet regions, respectively. Compared to similar \gls{tke} analysis regarding compressible turbulent jet flows, this case displays anisotropic separation more akin to higher mach flows \cite{iso_comp_2}.

\subsubsection{Discussion}
Overall, the isothermal \gls{sco2} jet appears to behave in a similar fashion to low-Mach compressible and incompressible jets within similar parameter regimes \cite{iso_comp_1, iso_comp_1_ref_1, iso_comp_2, Pope}. Qualitative features such as discernible regions of development and self-similarity are in agreement with ranges provided in the literature for subcritical jets. Decay rate and spreading rate are slightly below and above the typical ranges seen for similar subcritical cases, respectively. The supercritical isothermal jet appears to persist and spread more than its subcritical comparable counterpart, at least in the transition region. Resolved \gls{tke} components also exhibit slight anisotropy between the axial and span-wise components. 

\subsection{Non-Isothermal Jets}
The two non-isothermal jet cases as described in the operating conditions are presented here. Some features are compared to the isothermal jet case while others are used for direct comparison between the two non-isothermal cases. Again, the 350~K ambient case depicts a cooler jet entering a warmer ambient fluid, moving farther away from the supercritical point while the 314~K ambient case depicts a warmer jet entering a cooler ambient fluid with a transition over the pseudo-boiling point. It is anticipated that the 314~K ambient case will vary much more prominently from the other two cases across various quantities of interest examined. 

\subsubsection{Flow Field Features}
\begin{figure}[hbtp!]
\centering 
\subfloat[] { 
	\hspace{-24pt}
	\includegraphics[trim={0 6cm 3.6cm 6cm},clip, scale=.45]{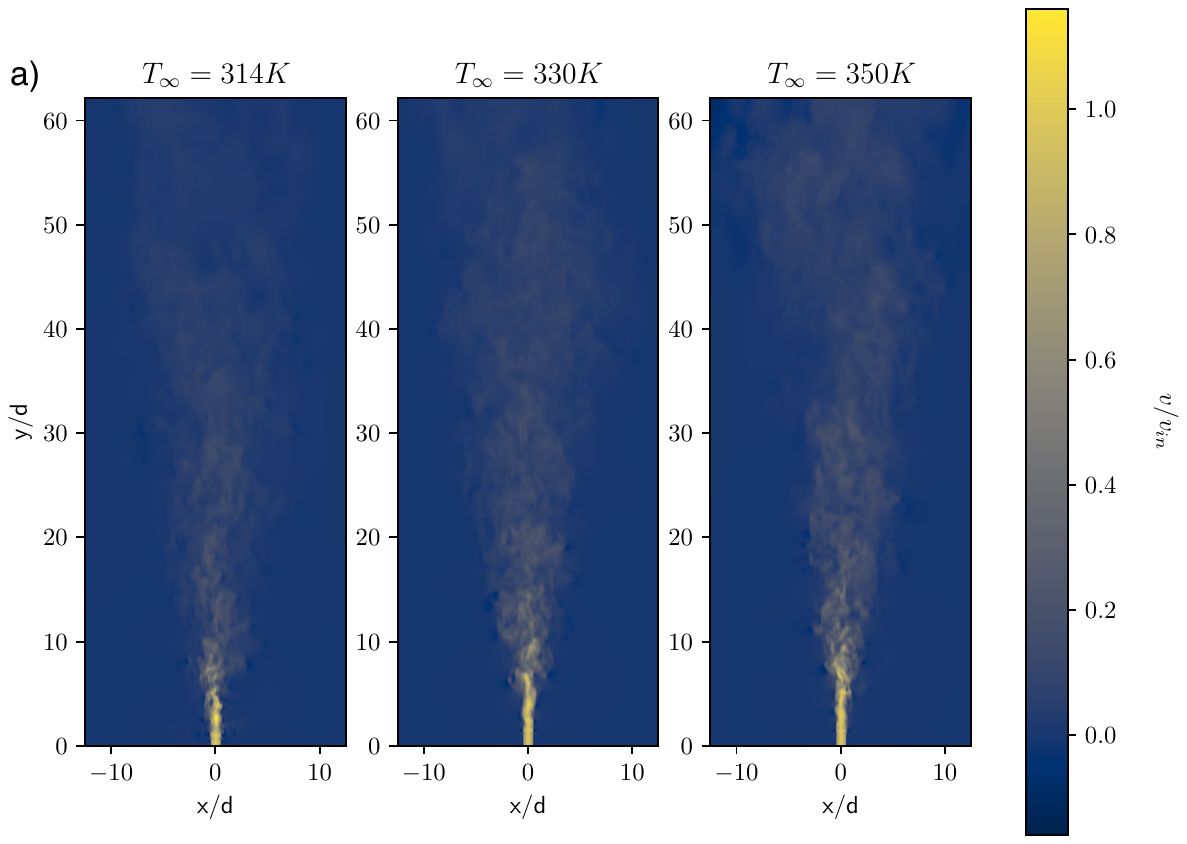}
    \label{all_v_1}
} 
\subfloat[] { 
	\includegraphics[trim={0 6cm 3.6cm 2cm},clip,scale=.45]{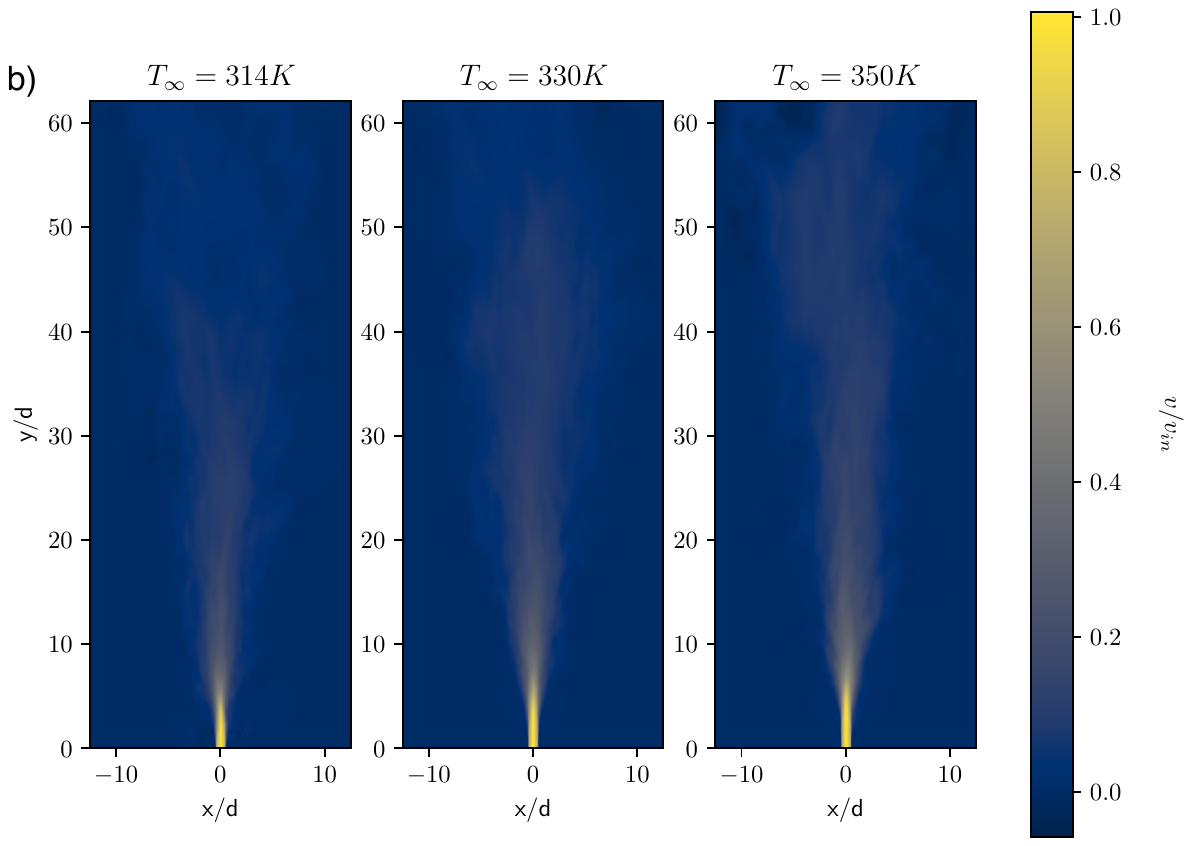}
    \label{all_v_2}
} \\
\vspace{-96pt}
\subfloat[] { 
    \centering
	\includegraphics[trim={0 5cm 0 2cm},clip, scale=.45]{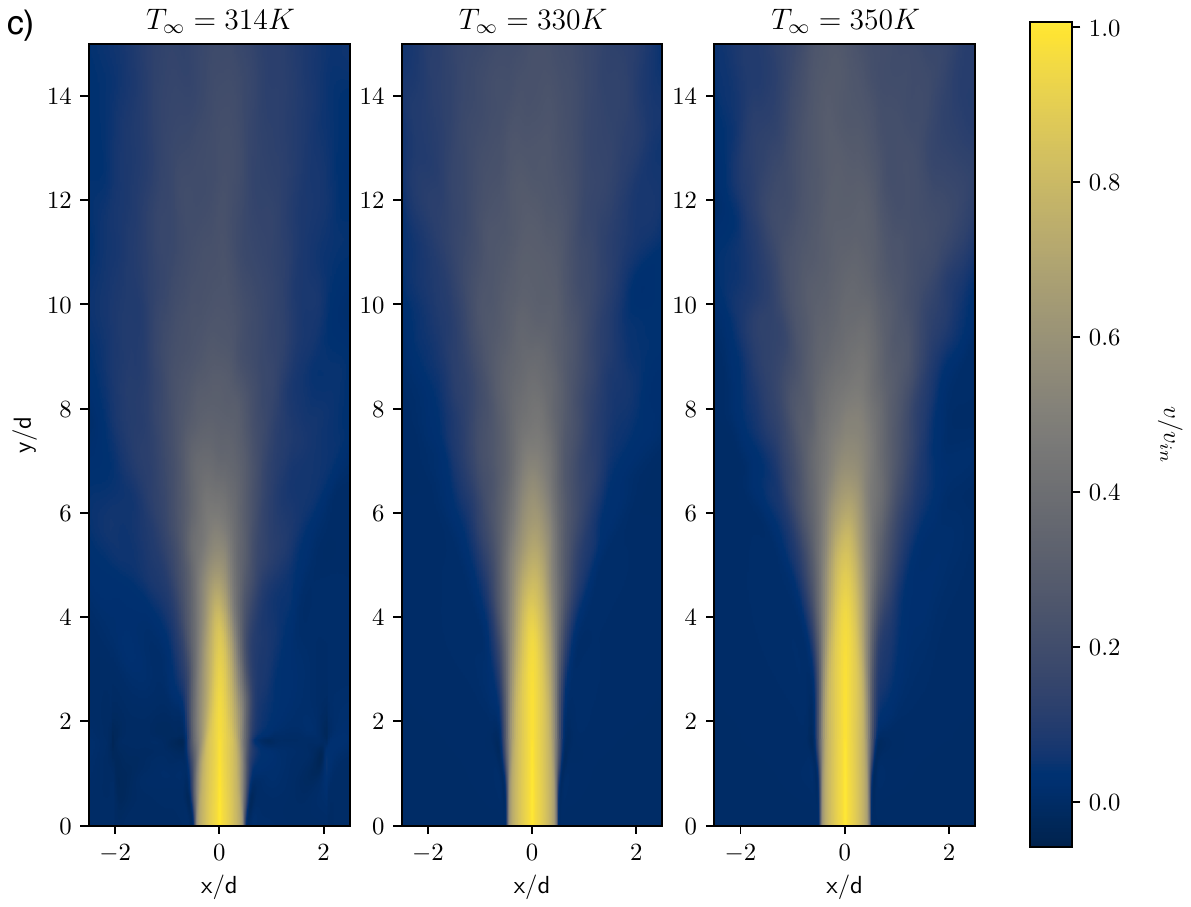} 
    \label{all_v_3}
}
\vspace{-32pt}
\caption{(a) Instantaneous and (b) time-averaged axial velocity field comparison between 314~K ambient case (left), 330~K ambient case (middle), and 350~K ambient case (right), with a closer look at averaged values (c) near the inlet.}
\label{all_v_features}
\end{figure}

Figure \ref{all_v_features} shows various axial velocity field comparisons between the isothermal and non-isothermal jets. The instantaneous axial velocity over the whole domain is depicted in Figure \ref{all_v_1}. The 350~K ambient case has a laminar flow region of similar length to the 330~K ambient case, with the jet core staying laminar up until $\nicefrac{y}{d} = 2$ before fluctuations begin to take effect. The developing region after this within $7 \leq \nicefrac{y}{d} \leq 15$ appears to be slightly more volatile in the 350~K ambient case compared to the 330~K ambient case, with the region containing finer-scale fluctuations. The spreading rate of the 350~K ambient case is smaller than that of the base case while overall intensity levels seem to be comparable in the downstream direction. The 314~K ambient case has a much shorter laminar region compared to the base case, with fluctuations beginning to become apparent as early as $\nicefrac{y}{d} = 1$. Perturbations after that appear to be more symmetric in structure, with vortices along the jet edge staying in tandem instead of the asymmetric shedding apparent in the other two cases. The 314~K ambient case also appears to have a spreading rate comparable to the 350~K ambient case, but with a faster decay in intensity in the downstream direction. Figure \ref{all_v_2} helps further illuminate the difference in spread of the three jets. Here the 350~K ambient core intensity is more persistent downstream with a more distinct and thinner outer edge flow present than in the 330~K ambient case, while the 314~K case has a lower intensity central flow region but a comparably sized edge flow region compared to the 330~K ambient case. Finally, Figure \ref{all_v_3} depicts the potential core and transition regions of the three cases more clearly. Here, the similarity between the potential cores of the 330~K and 350~K ambient cases is clear while the 314~K ambient case has a much shorter core length. 

\begin{figure}[hbtp!]
\centering
\subfloat[] { 
	\hspace{-24pt}
	\includegraphics[trim={0 6cm 4cm 5cm},clip,scale=.45]{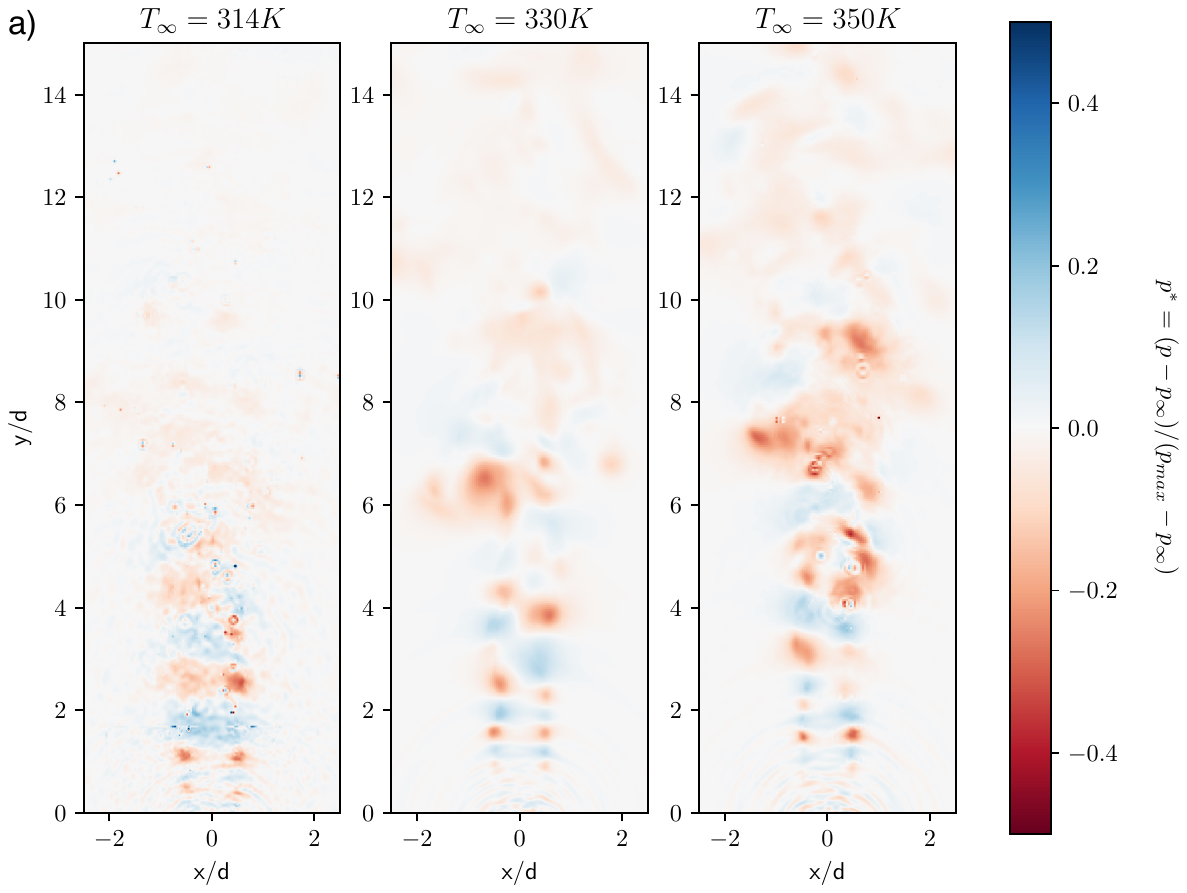}
    \label{all_pressure_1}
} 
\subfloat[] { 
	\includegraphics[trim={0 6cm 0 2cm},clip, scale=.45]{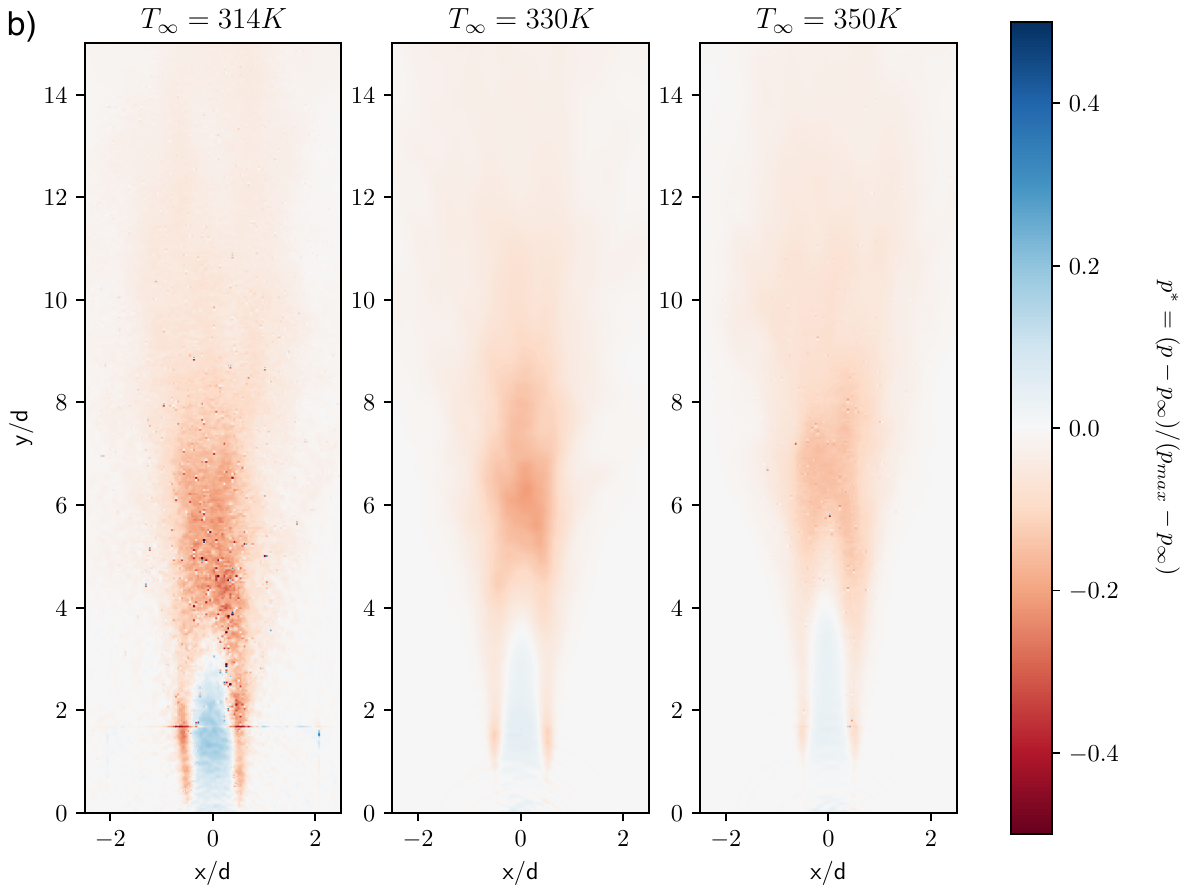}
    \label{all_pressure_2}
}
\vspace{-28pt}
\caption{(a) Instantaneous and (b) time-averaged pressure fluctuation field comparison between 314~K ambient case (left), 330~K ambient case (middle), and 350~K ambient case (right) near the jet inlet.}
\label{all_pressure_features}
\end{figure}

Figure~\ref{all_pressure_features} shows pressure fluctuations near the inlet for all three cases. Figure~\ref{all_pressure_1} contains an instantaneous snapshot of these fluctuations near the inlet. The 350~K ambient case is comparable to the 330~K ambient case, with similar large coherent structures forming in the pressure oscillations. However, the initial pressure waves emanating from the inlet are stronger in the 350~K ambient case, resulting in slightly more intense fluctuations downstream. Minor spurious perturbations are also seen downstream in the 350~K ambient case. The 314~K ambient case is qualitatively much different from the other two cases. Larger coherent structures are much less defined, with smaller fluctuations appearing much more prominently throughout. There is also a band of cells slightly before $\nicefrac{y}{d}=2$ where perturbations are more concentrated. Average pressure fluctuation can be seen in Figure~\ref{all_pressure_2}. All cases have a slight increase in pressure isolated within the potential core surrounded by and followed with a pressure drop. The difference between these low and high pressure pockets is most prominent in the 314~K ambient case, followed by the 330~K ambient case. Numerical artifacts can be seen in the 314~K ambient case with some similar artifacts present in the 350~K ambient case as well. These artifacts are most likely a combined result of the \gls{amr} and \gls{srk} interplay in this region, \emph{i.e.}, slight inconsistencies of states between coarse cells may be exacerbated by the \gls{amr} at this point in the domain. For conditions nearer the pseudo-boiling point, where density gradients are steeper, the fluctuations are more pronounced, hence our suspicion that some aspect of the \gls{srk} implementation plays a part in this phenomenon. Further investigation is needed to discern the specifics involved with this feature. Pressure in general is a sensitive field, and since these fluctuations are not apparent in other plots, we do not expect them to impact the conclusions drawn from other quantities investigated here.  

\begin{figure}[hbtp!]
\centering
\subfloat[] { 
	\hspace{-24pt} 
	\includegraphics[trim={0 6cm 3.6cm 6cm},clip,scale=.45]{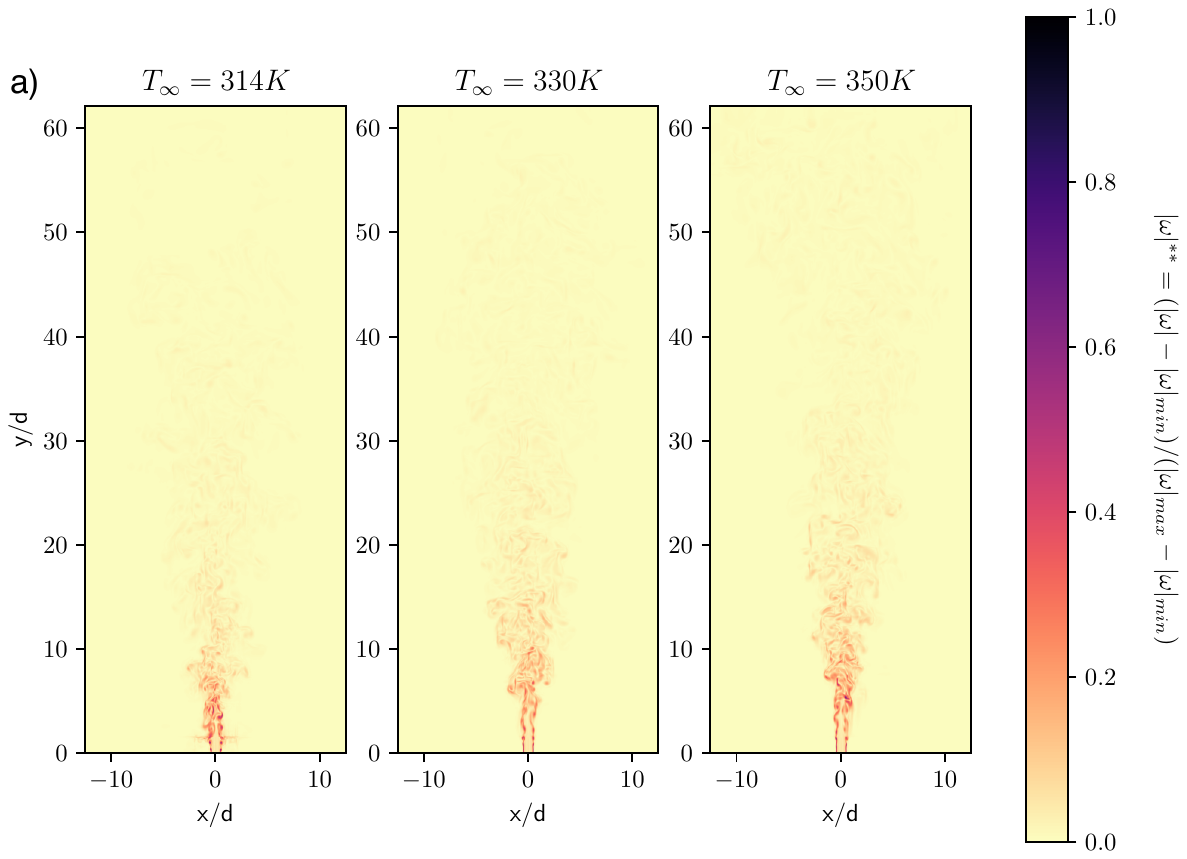}
    \label{all_magvort_1}
} 
\subfloat[] { 
	\includegraphics[trim={0 6cm 3.6cm 2cm},clip,scale=.45]{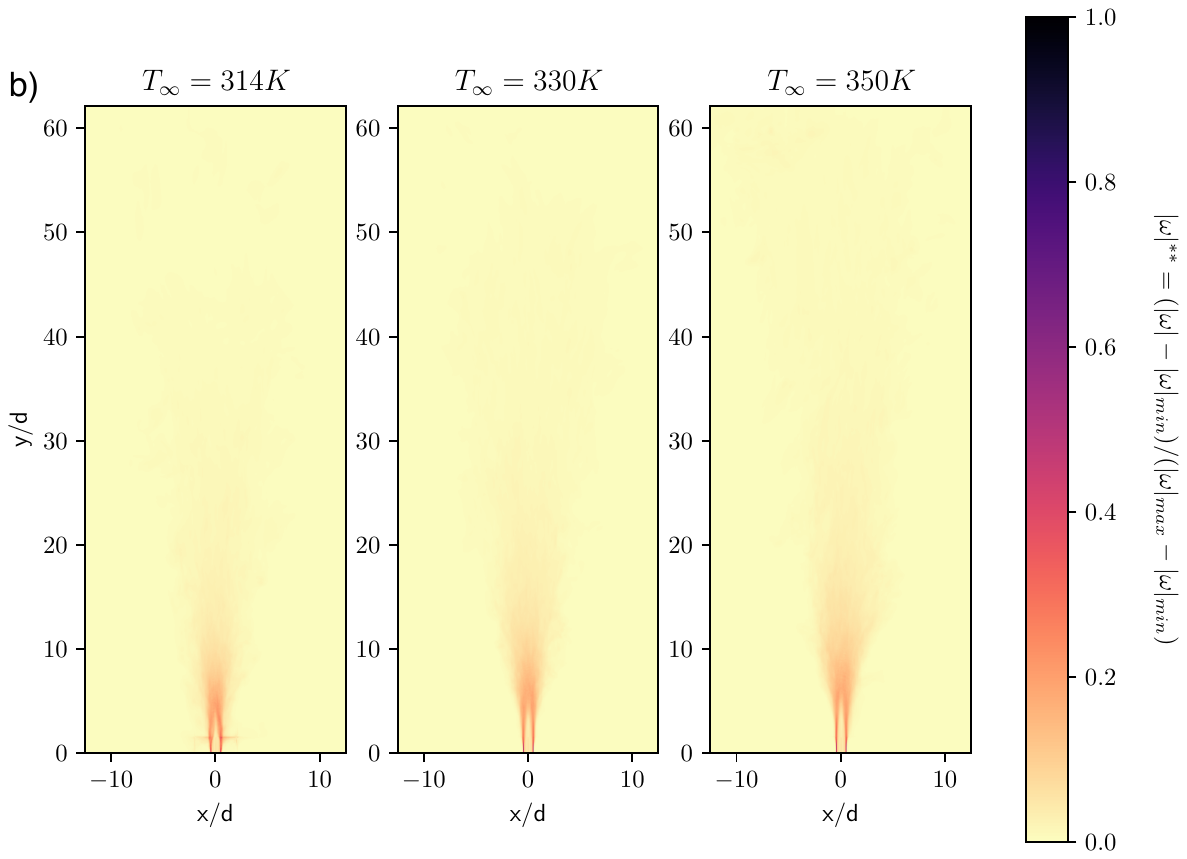}
    \label{all_magvort_2}
} \\
\vspace{-96pt}
\subfloat[] { 
	\includegraphics[trim={0 5cm 0 2cm},clip, scale=.45]{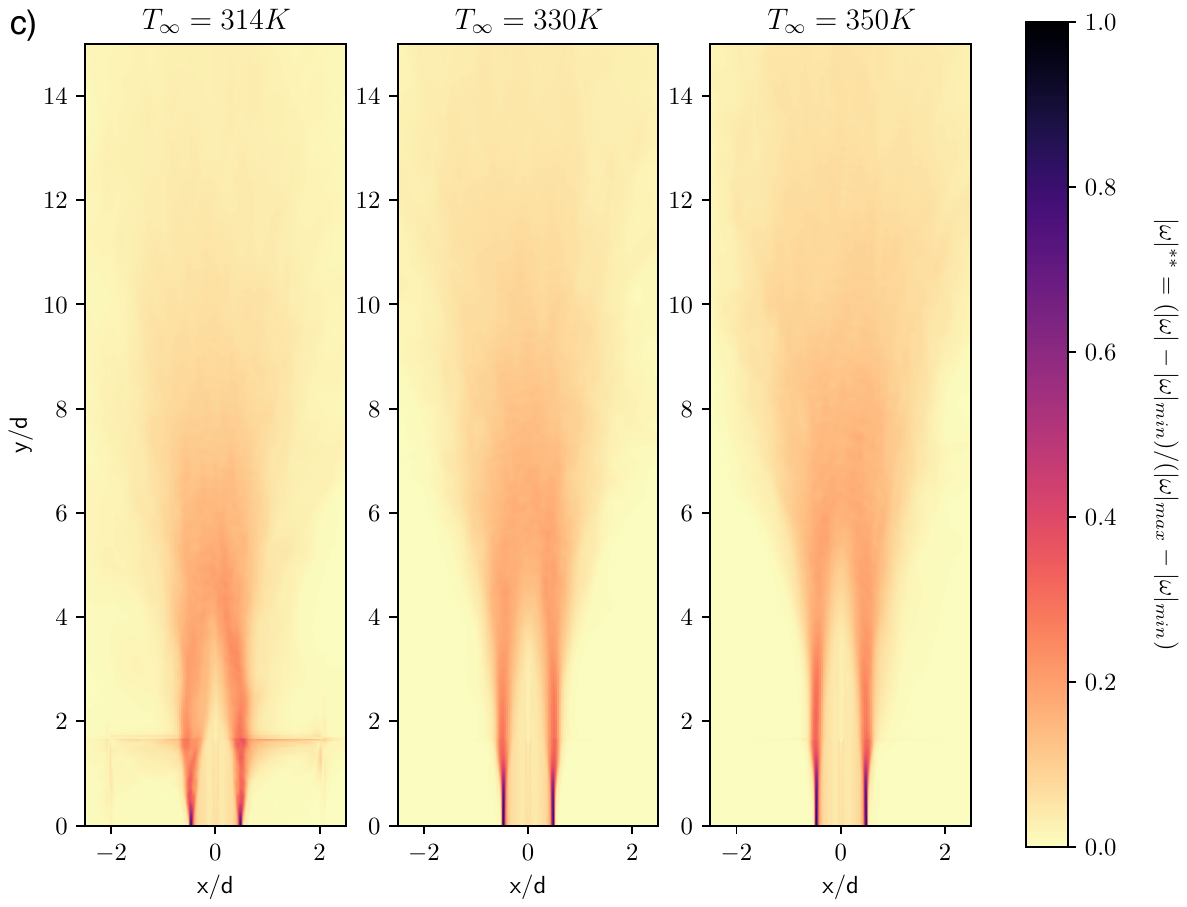}
    \label{all_magvort_3}
}
\vspace{-32pt}
\caption{(a) Instantaneous and (b) time-averaged vorticity magnitude field comparison between 314~K ambient case (left), 330~K ambient case (middle), and 350~K ambient case (right), with a closer look at averaged values (c) near the inlet.}
\label{all_magvort_features}
\end{figure}

Figure \ref{all_magvort_features} shows both instantaneous and time-averaged flow fields of the vorticity magnitude for the three cases. The vorticity magnitude features seen in Figure \ref{all_magvort_1} follow similar trends as the axial velocity. In all three cases, the strongest vorticity occurs along the jet edge at the inflow, leading to vortex formation and subsequent fluctuations. The 314~K ambient case has a shorter region of steady flow in and also features larger, more distinct vortex formation along the edge of the jet. These vortices roll along the edge in tandem before mixing to lead into the transition region around approximately $\nicefrac{y}{d} = 3$ for the 314~K ambient case and $\nicefrac{y}{d} = 5$ for the 350~K. Immediately following this initial vortex formation, jet behavior begins to diverge between the two cases. The 314~K ambient case maintains coherent large vortex structures farther along downstream compared to the 350~K case. As these larger structures disintegrate, it appears that positive interference between edge vortices leads to finer, more compact filament structures with higher pockets of vorticity as compared to the 350~K ambient case. These finer structures then more rapidly decay in the downstream direction. The 350~K ambient case maintains higher vorticity intensity downstream with more prolonged pockets of higher vorticity seen in the transition region of the jet compared to that of the 330~K ambient case. Similar to what was seen with the 314~K ambient pressure field, there are some vorticity artifacts present in the mesh at the edge of the initial high-refinement zone surrounding the jet inlet. Similar to the 350~K ambient case, the 314~K ambient case also maintains some pockets of high vorticity, but then decays more rapidly overall compared to the other two cases. The averaged vorticity magnitude fields in Figures \ref{all_magvort_2} and \ref{all_magvort_3} showcases the different jet regions and jet spreading. From Figure \ref{all_magvort_2} one can see that the 350~K ambient case has more widespread vorticity in the transition region from around $5 \leq \nicefrac{y}{d} \leq 15$ and then maintains the intensity formed here for farther along downstream compared to the 330~K ambient case. The 314~K ambient case has a much narrower vorticity range in the same region with intensity decaying much more rapidly downstream. Near the inlet, as depicted in Figure \ref{all_magvort_3}, it is easier to see that the 314~K ambient case begins spreading much earlier than the other two cases and has much less uniform vorticity along the jet edges. This earlier breakup also impacts the way the high-vorticity outer edges of the jet merge around the low-vorticity center at the end of the potential core, where the merging of the two jet edges across the middle meet in an almost concave point as compared to the more convex central region seen in the other two cases. The vorticity fluctuations before the mesh refinement change may contribute to the vorticity artifacts seen here. We believe these artifacts are due to a post-processing issue from a previous iteration of the code that has since been resolved. Again, since the velocity field appears relatively smooth by comparison, we do not expect the artifacts seen here to significantly impact the conclusions discussed later on. 

\begin{figure*}[hbtp!]
	\centering
    \subfloat[] { 
    \centering
	\includegraphics[trim={0 1cm 0 5cm}, scale=.35]{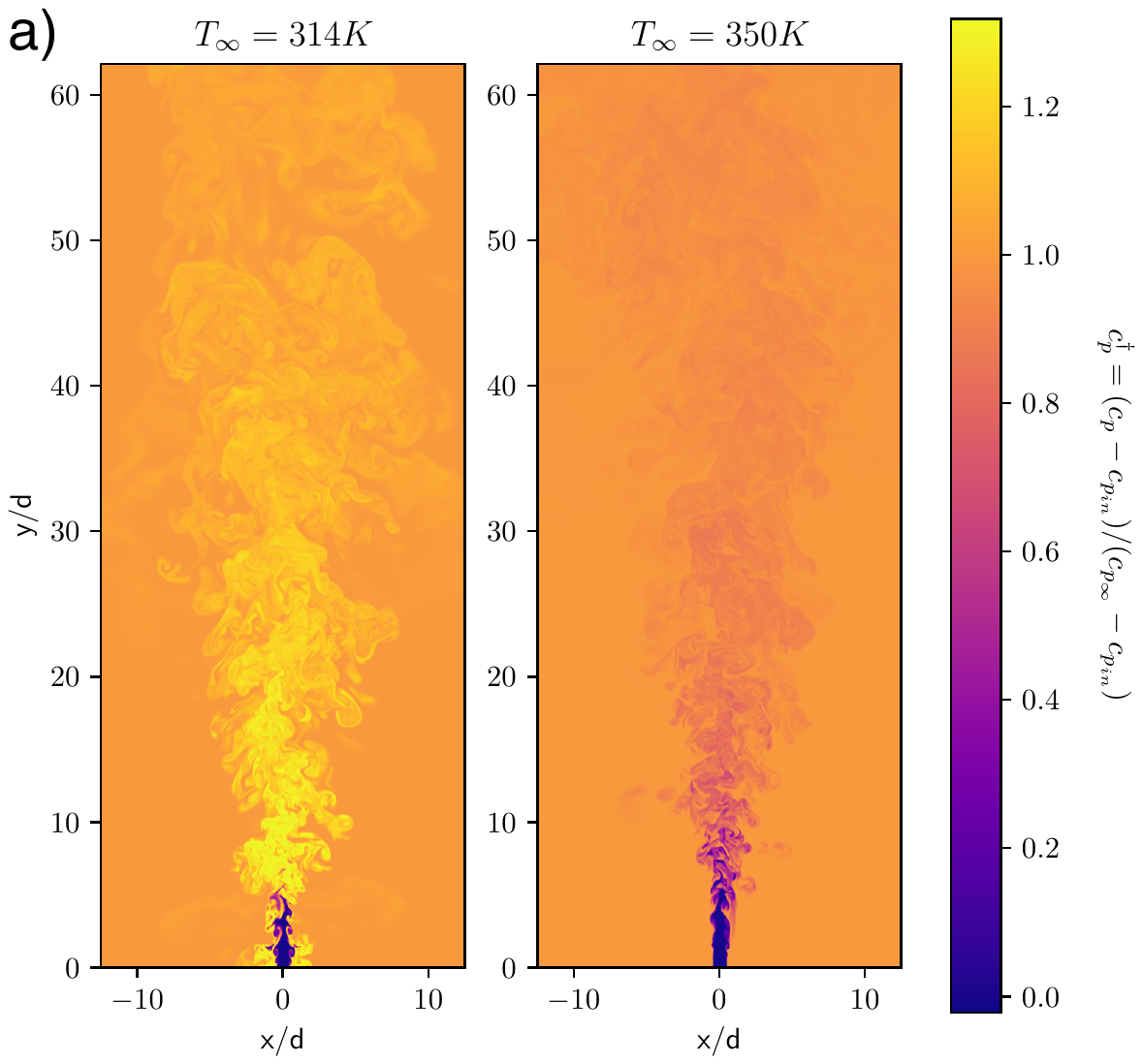} 
    \label{noniso_cp_1} 
    }
    \subfloat[] { 
    \centering
	\includegraphics[trim={0 1cm 0 5cm}, scale=.35]{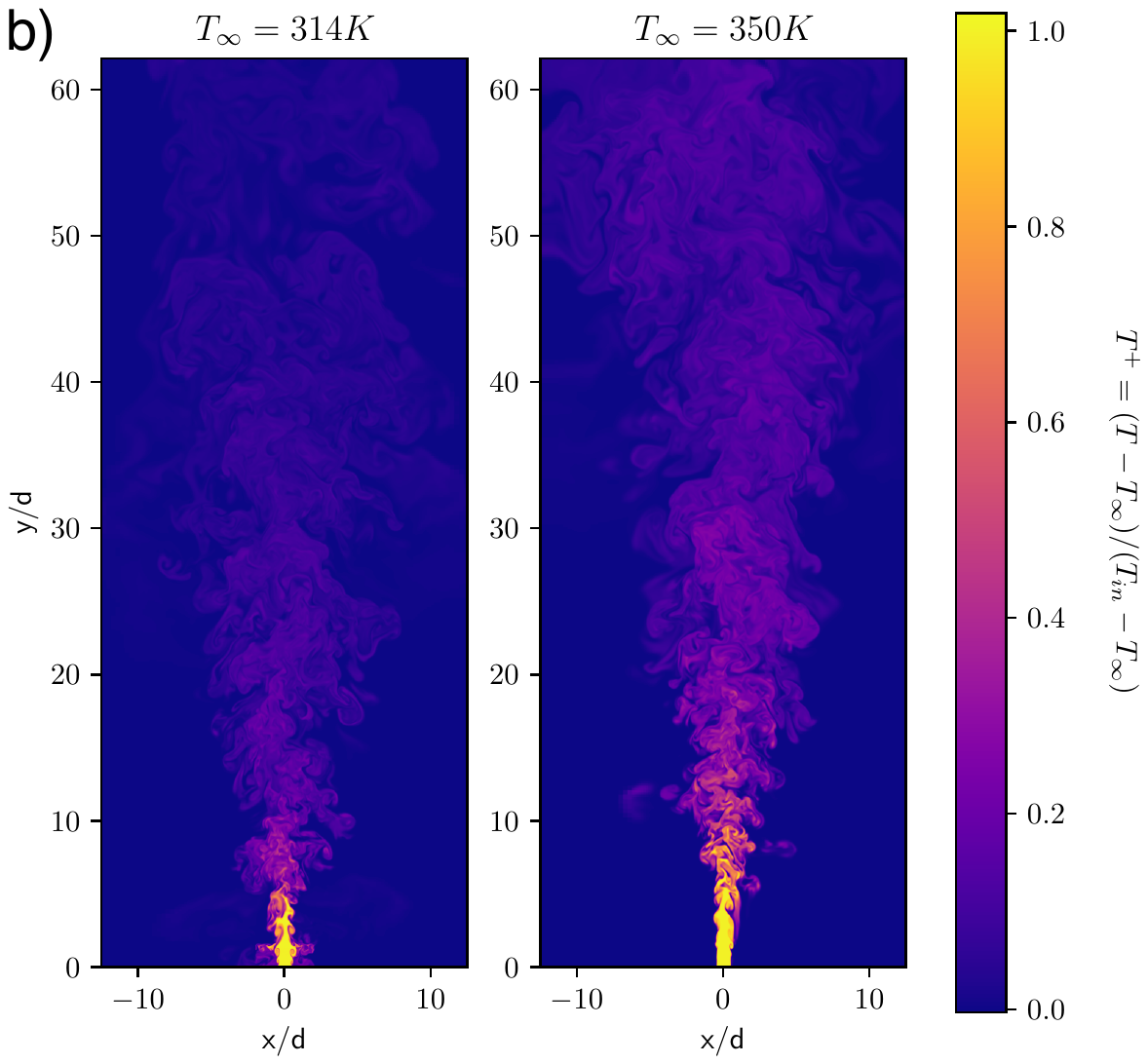} 
    \label{noniso_temp_1}
    } \\
    \vspace{-64pt}
    \subfloat[] { 
    \centering
	\includegraphics[trim={0 1cm 0 5cm}, scale=.35]{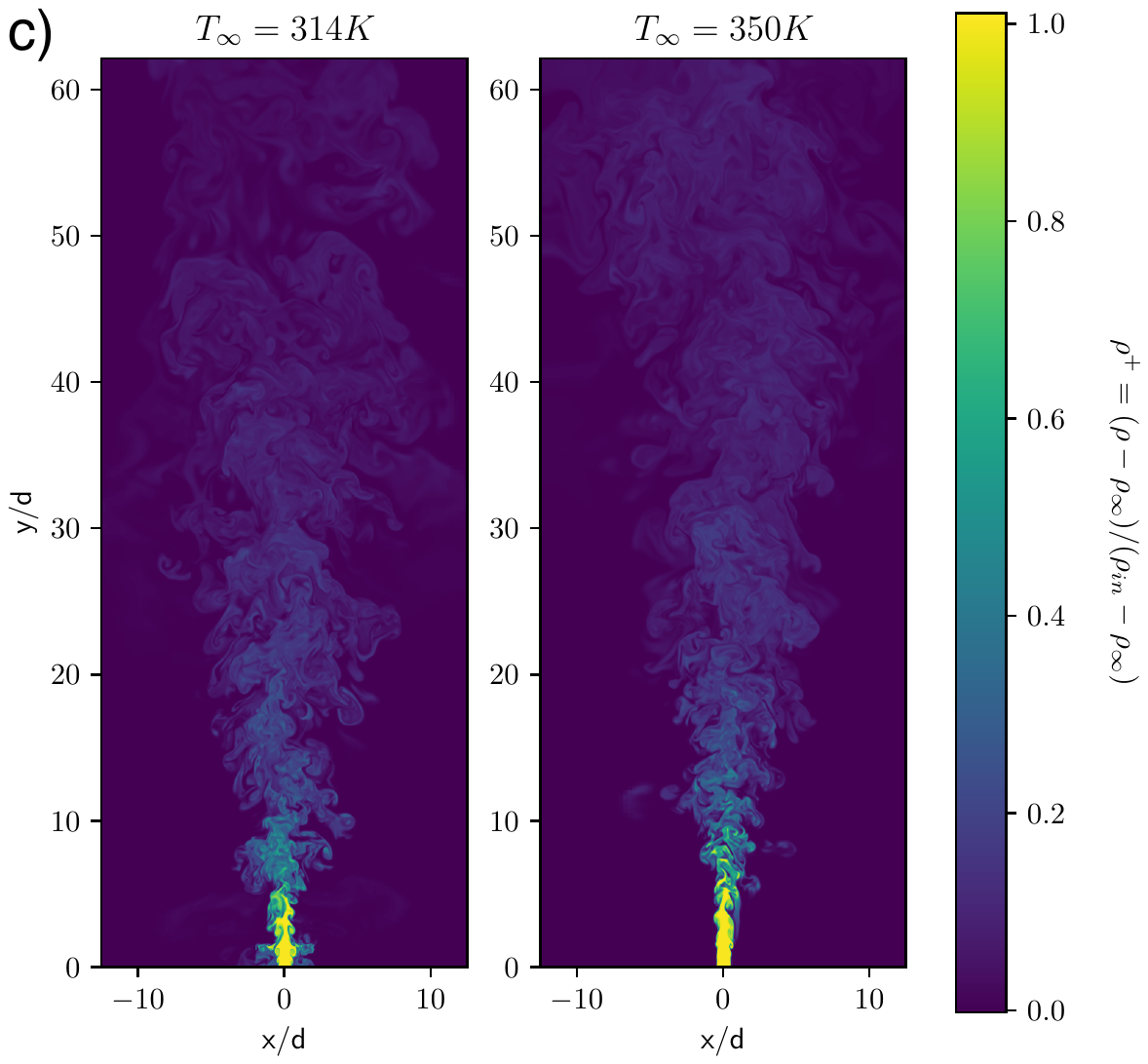} 
    \label{noniso_rho_1}
    }
    \subfloat[] { 
    \centering
	\includegraphics[trim={0 1cm 0 4cm}, scale=.35]{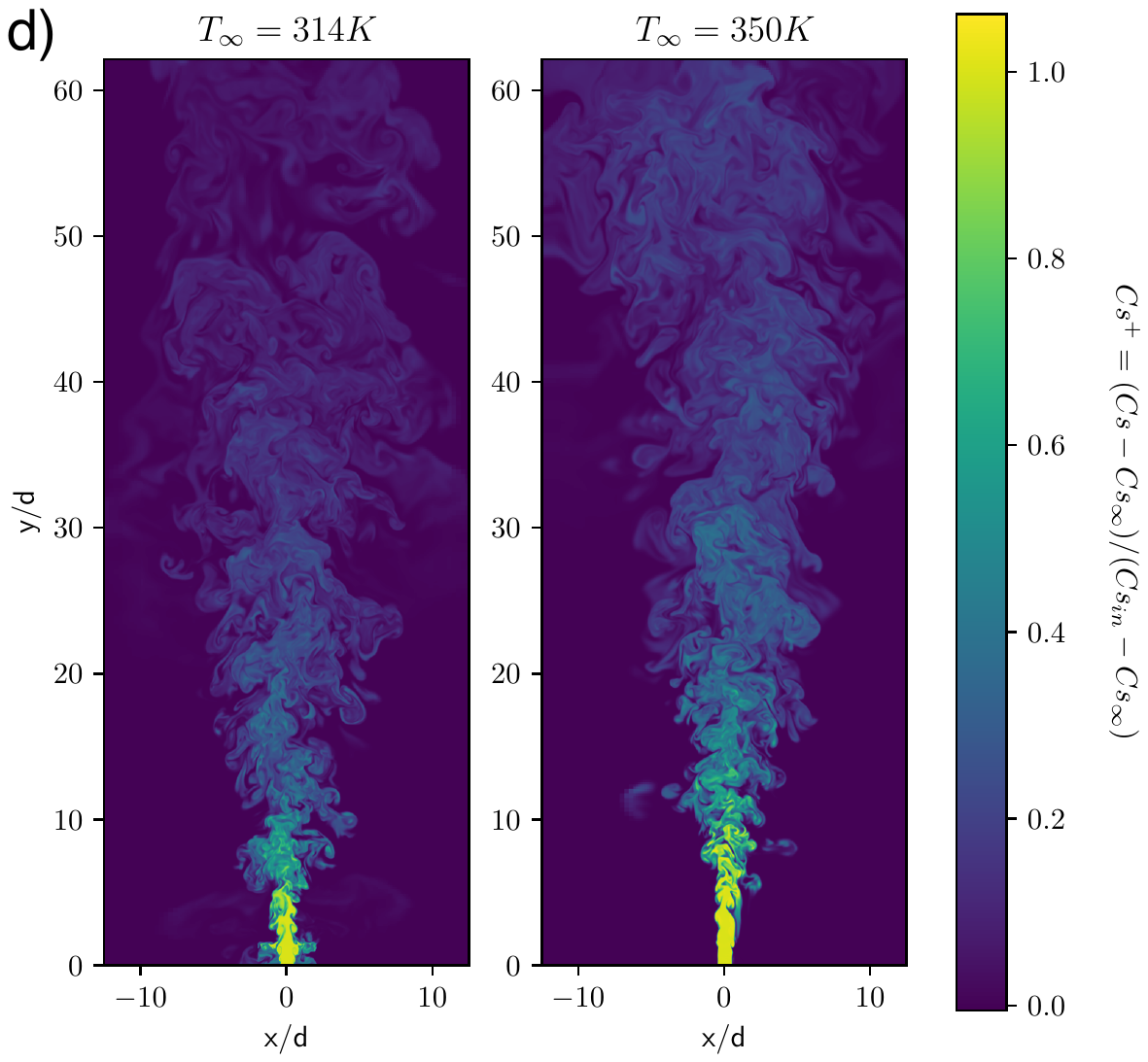} 
    \label{noniso_Cs_1}
    } \\
    \vspace{-64pt}
    \subfloat[] { 
    \centering
	\includegraphics[trim={0 1cm 0 4cm}, scale=.35]{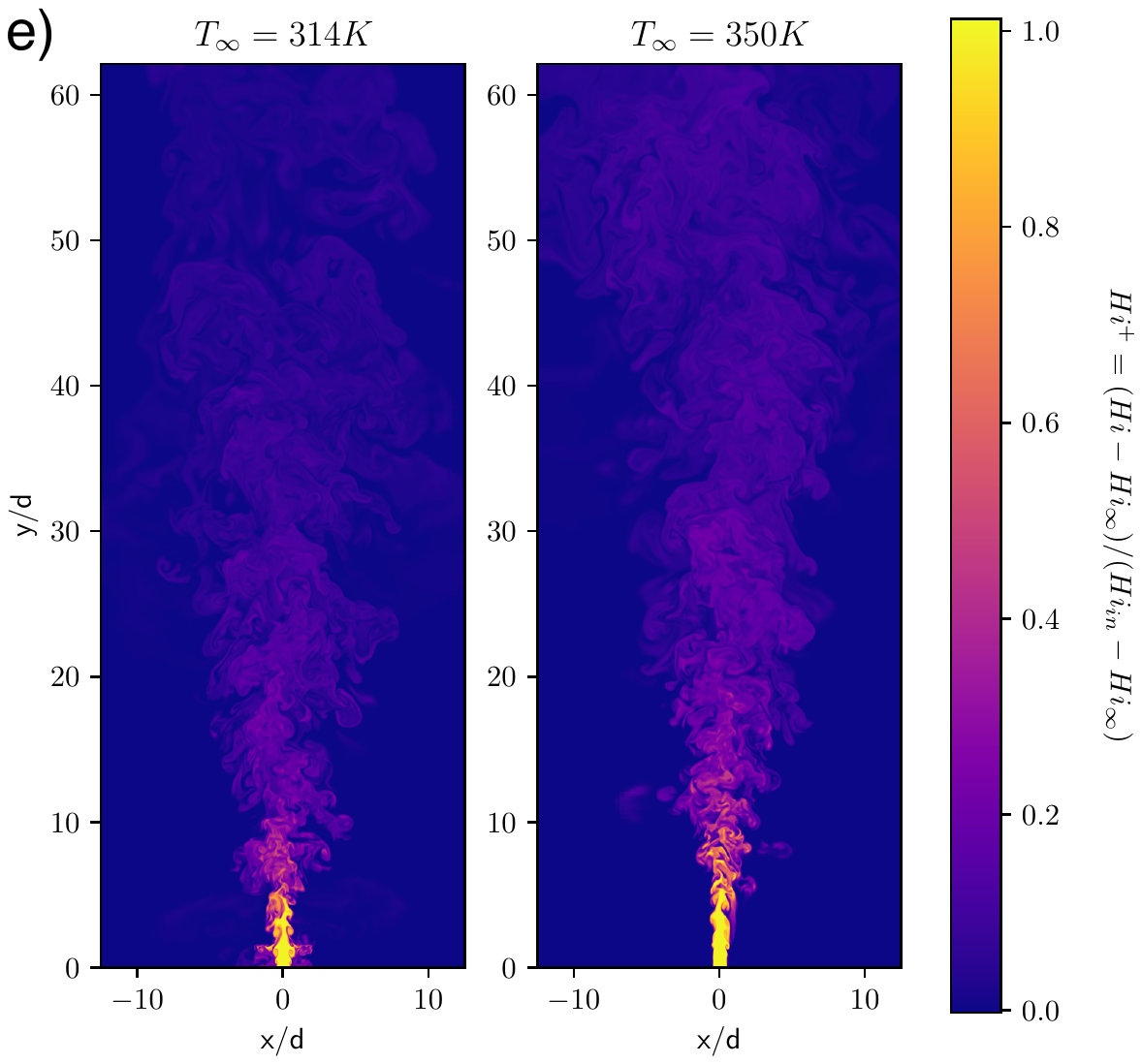} 
    \label{noniso_Hi_1}
    }
    \subfloat[] { 
    \centering
	\includegraphics[trim={0 1cm 0 4cm}, scale=.35]{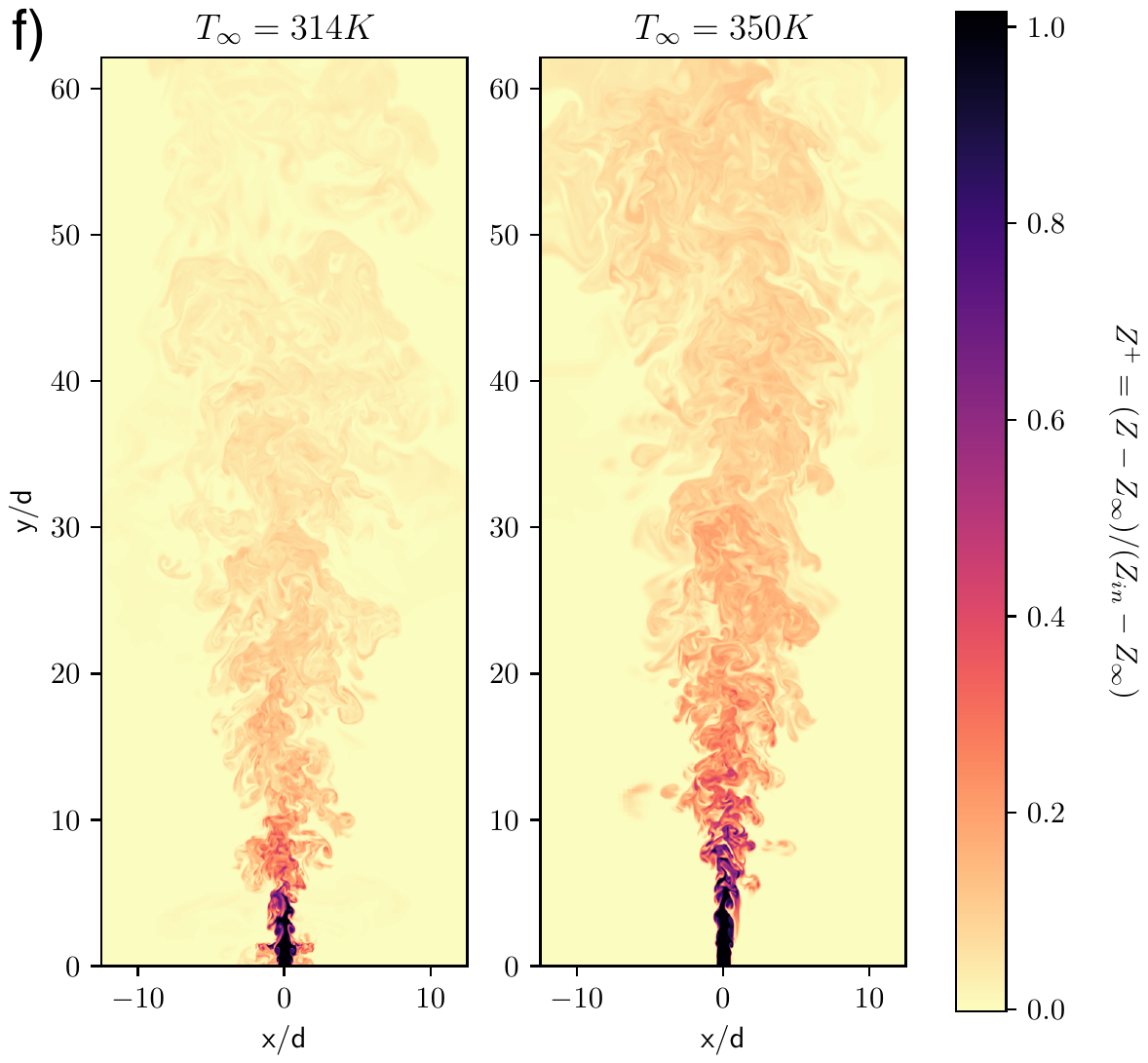} 
    \label{noniso_Z_1}
    }
    \vspace{-56pt}
\caption{Direct comparison between 314~K ambient case (left) and 350~K ambient case (right) for various instantaneous fluid quantities of interest: a) constant pressure specific heat, b) temperature, c) density, d) sound speed, e) enthalpy, and f) compressibility factor.}
\label{noniso_various_features}
\end{figure*}

The final figure in this section, Figure \ref{noniso_various_features}, contains direct comparisons between the two non-isothermal cases for various quantities of interest. Figure \ref{noniso_cp_1} shows the constant-pressure specific heat variation between the two cases. The added peak in specific heat can be seen in the 314~K ambient case as the fluid transitions through the pseudo-boiling point. This peak occurs at the jet edge in the potential core and is additionally maintained after the transition region, showing a slower transition to the ambient specific heat as compared to the 350~K ambient case. This layer of peak specific heat between the jet and ambient fluid creates a thermal shield around the jet, where most energy goes toward expanding the volume of the fluid as opposed to raising the temperature \cite{10.1063/1.5054797}. This effect can be seen in comparing Figures \ref{noniso_temp_1} and \ref{noniso_rho_1}. In regions of heightened constant-pressure specific heat, density change in the 314~K ambient case is comparable to the 330~K ambient case, even though the former involves gas-like injection into a liquid-like background and the latter involves gas-like supercritical fluids for both injection and background flow. In the transition region of the 314~K ambient jet, temperature drops dramatically in an area of high mixing just past the potential core. Thereafter, temperature decays much more slowly as the specific heat remains at peak levels for the remainder of the jet propagation. 

The last group of images, Figures \ref{noniso_Cs_1}, \ref{noniso_Hi_1}, and \ref{noniso_Z_1}, show the sound speed, enthalpy, and compressibility factor of the fluid, respectively. All three quantities exhibit similar trends, with the 350~K ambient case showing a more rapid transition to the ambient conditions as compared to the 314~K ambient case. These follow the same trend as seen in the temperature, but to varying degrees of intensity. For example, enthalpy has an overall faster downstream decay for both cases as compared to the temperature decay, while sound speed and compressibility factor have comparable decay rates. 

\subsubsection{Mean Flow Properties}
\begin{figure}[hbtp!]
    \centering
    \subfloat[] { 
	\includegraphics[trim={0 5cm 0 7cm}, scale=.37, clip=tight]{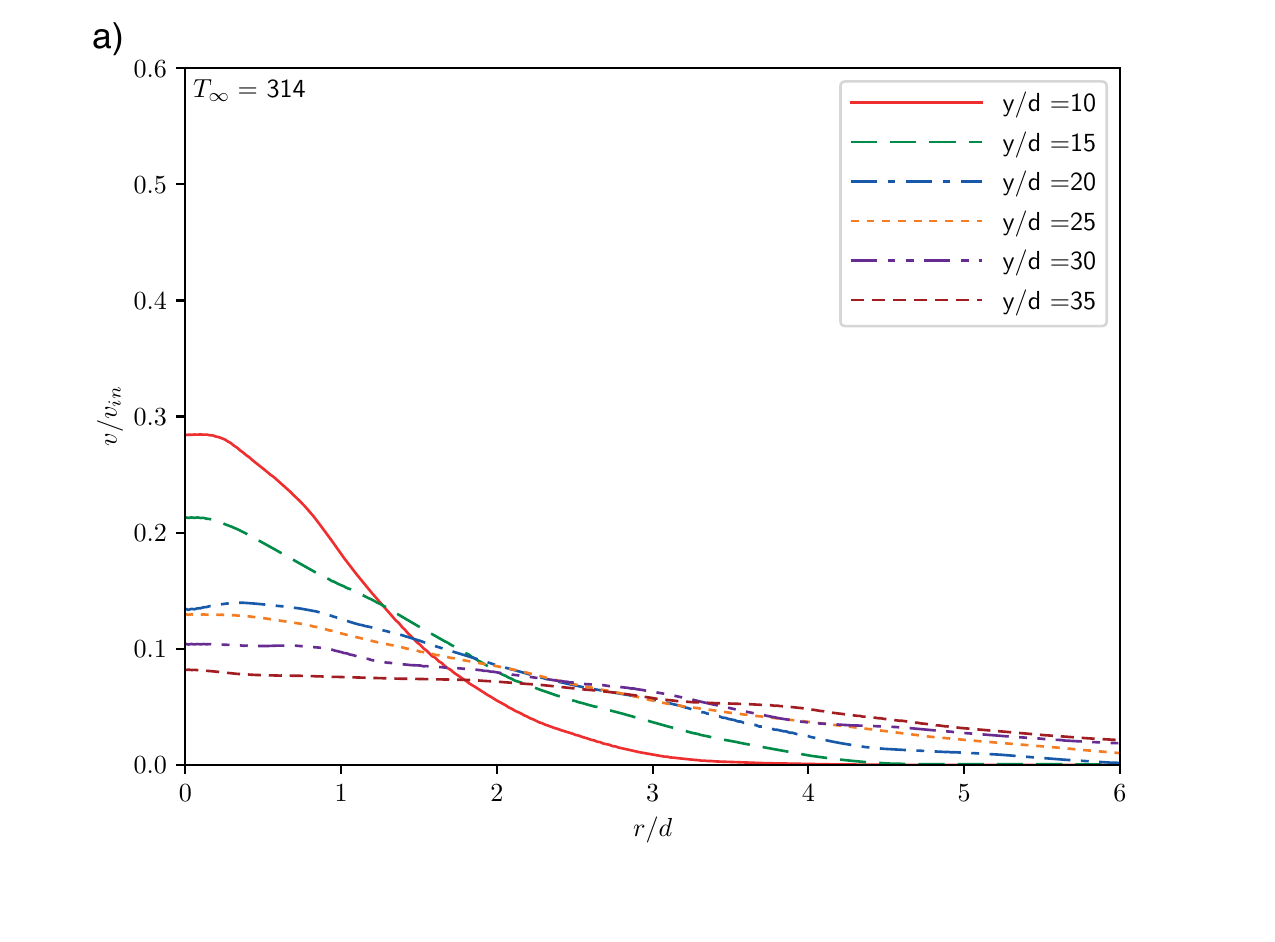}
    \label{noniso_v_vin_r_d_1}
} 
    \subfloat[] { 
	\includegraphics[trim={0 5cm 0 7cm}, scale=.37, clip=tight]{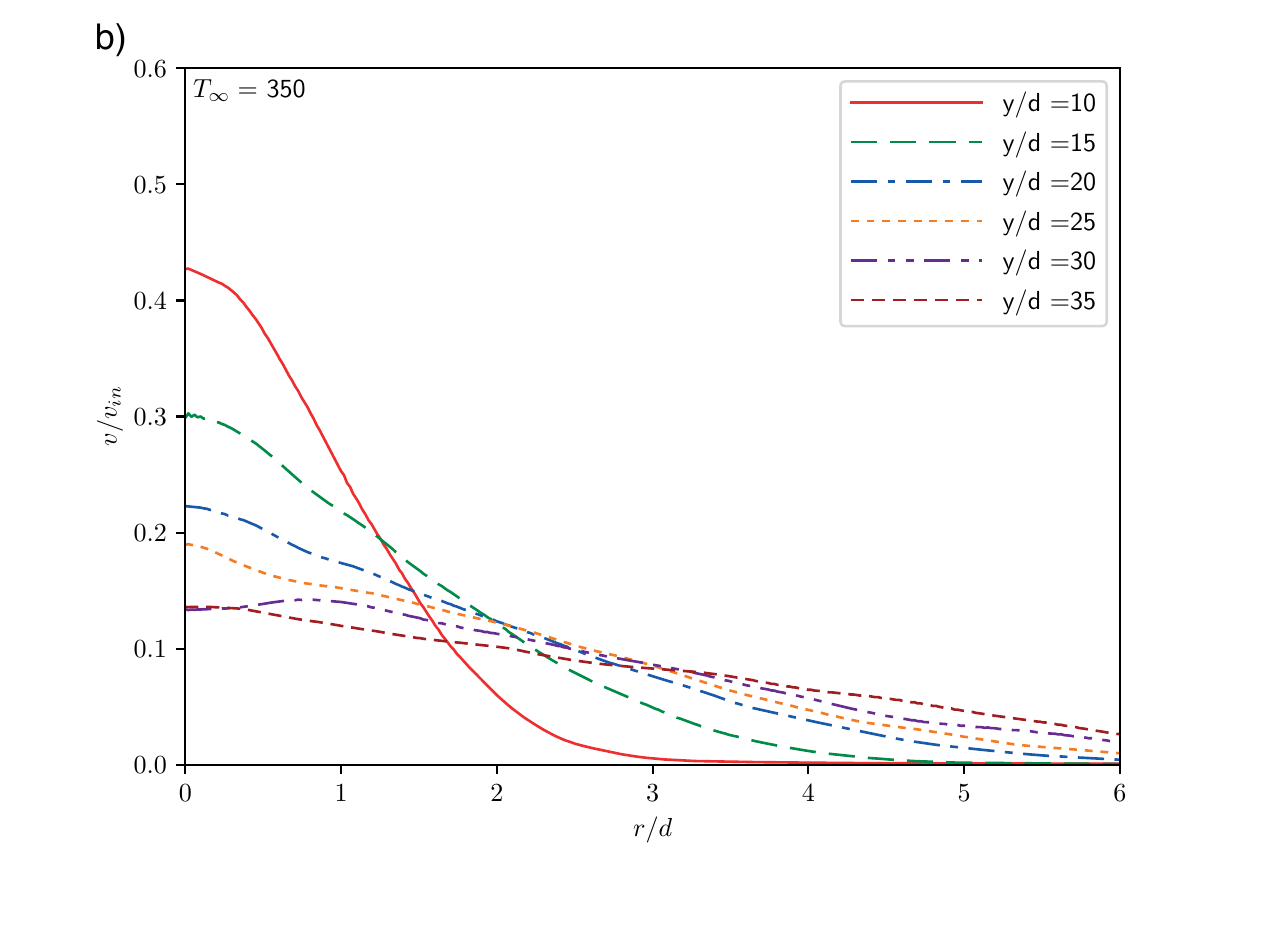}
    \label{noniso_v_vin_r_d_2}
}
\vspace{-28pt}
\caption{Average (both in time and radially) axial velocity scaled by inlet value plotted along radial distance from centerline for (a) 314~K and (b) 350~K ambient cases.}
\label{noniso_v_vin_r_d_features}
\end{figure}

Figure \ref{noniso_v_vin_r_d_features} depicts the time and radially averaged scaled axial velocity component plotted against radial distance from the centerline at multiple normal slices downstream from the inlet. The velocity is scaled by the average axial velocity value at the inflow while the radial direction is scaled by the jet diameter. In both cases, velocity profiles flatten as they get farther downstream. Compared to the 330~K ambient case, the 350~K ambient case has a slower axial velocity decay downstream while the 314~K ambient case has a much faster decay. This agrees with the trends seen in the vertical slice images. 

\begin{figure}[hbtp!]
\begin{center}
\vspace{-28pt}
\subfloat[] { 
	\includegraphics[scale=.27]{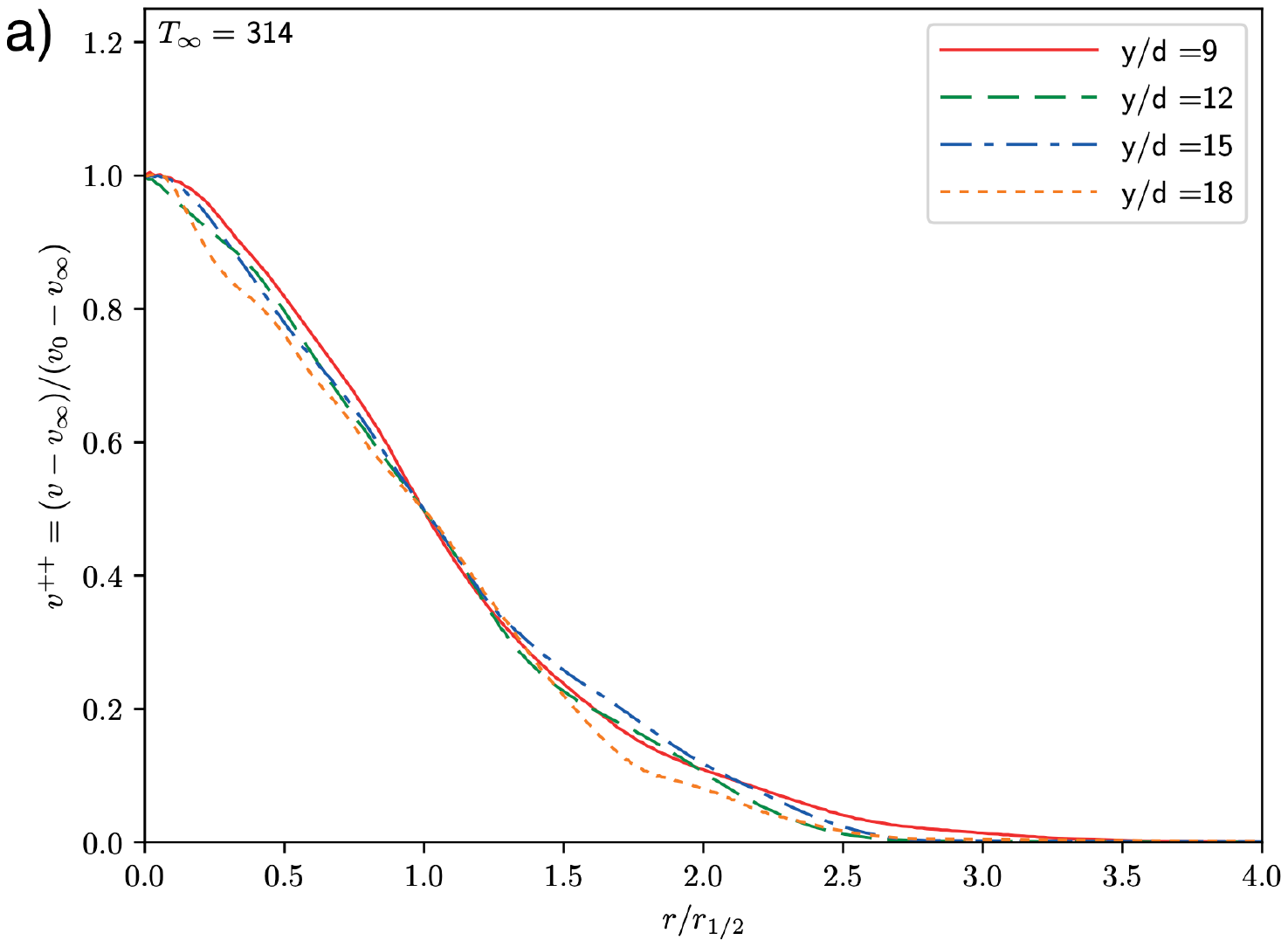}
 \label{noniso_near_r_vs_v_1}
}
\subfloat[] { 
	\includegraphics[scale=.27]{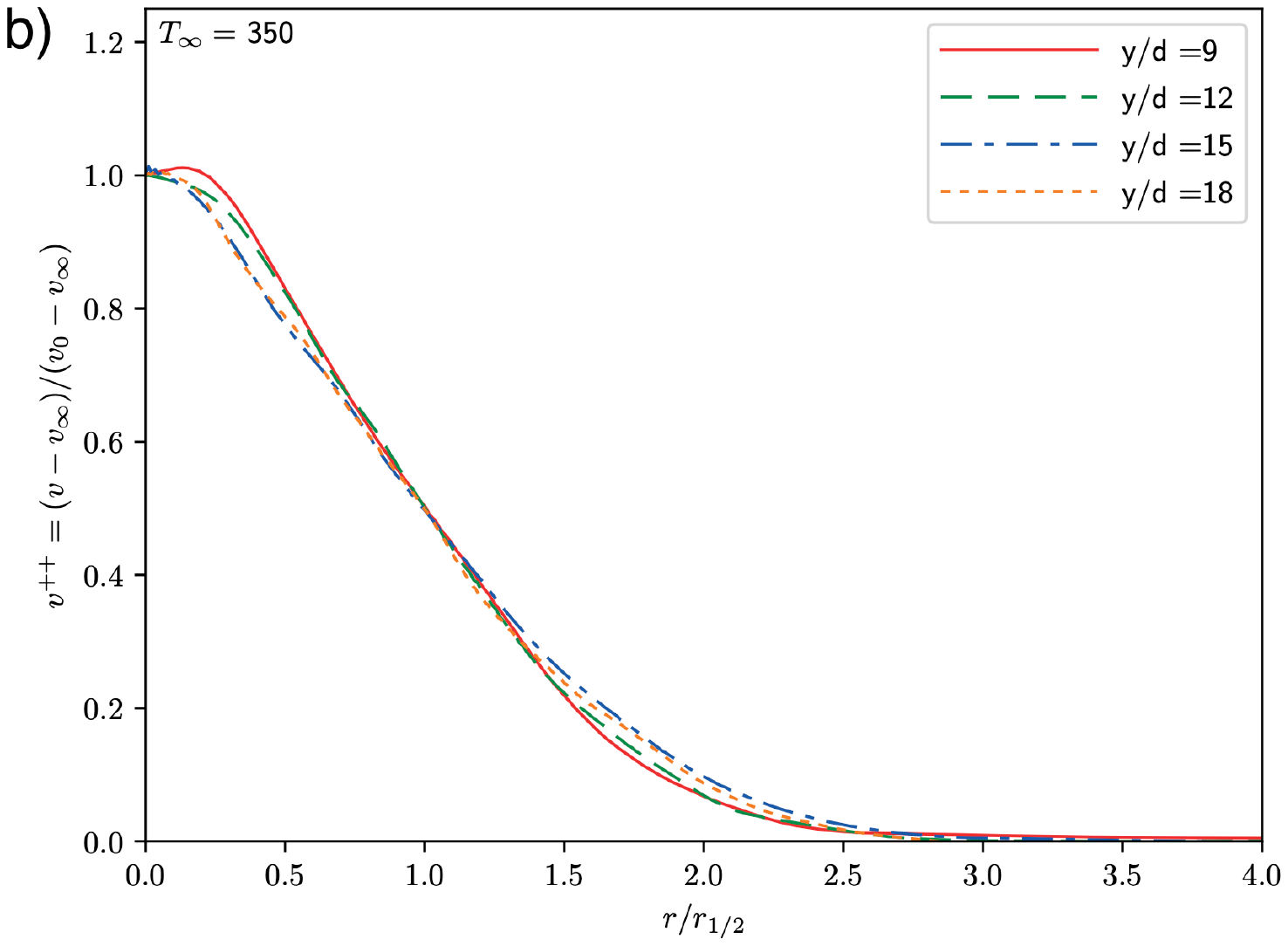}
    \label{noniso_near_r_vs_v_2}
} \\
\vspace{-28pt}
\subfloat[] { 
	\includegraphics[scale=.27]{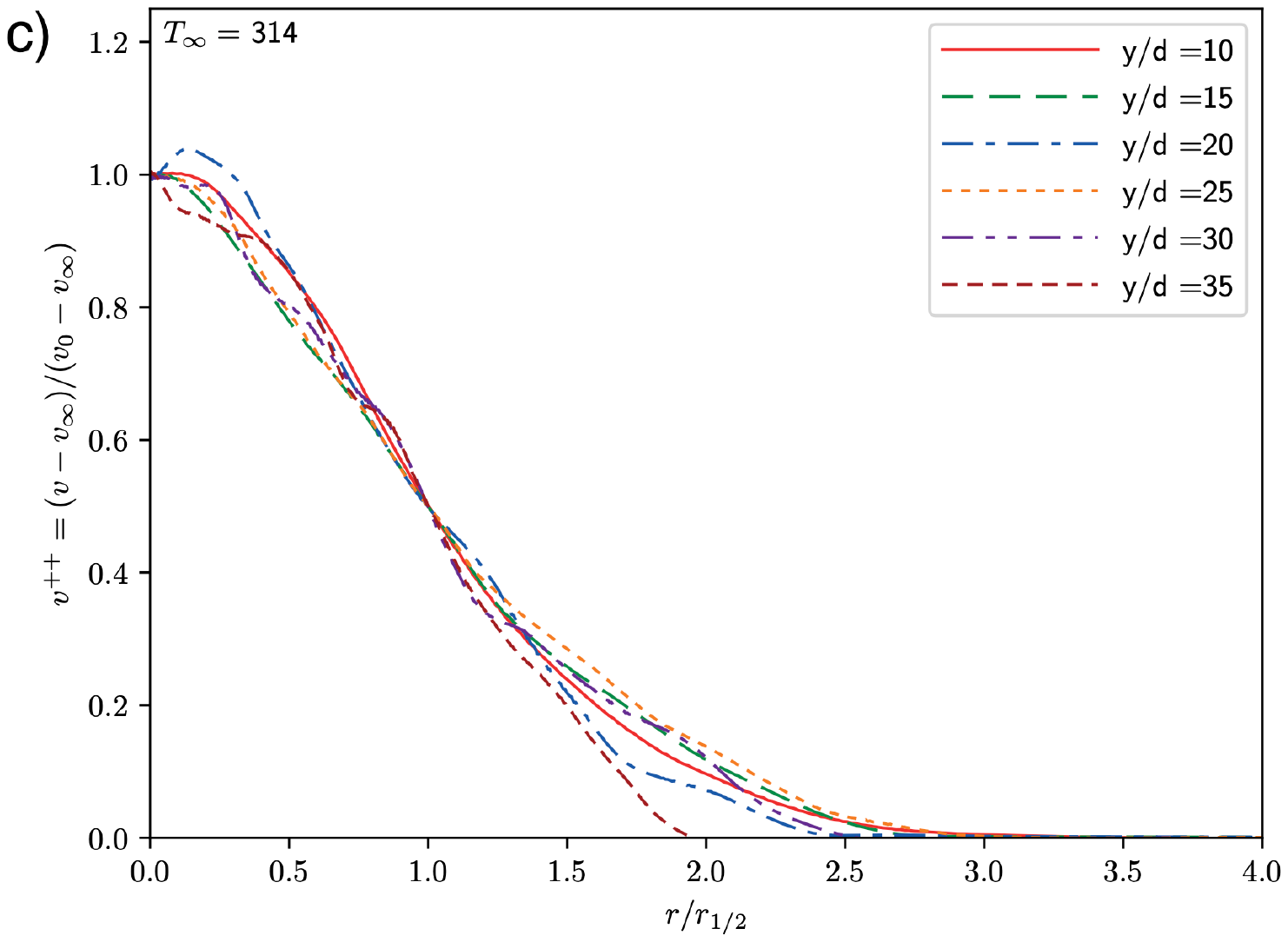}
    \label{noniso_far_r_vs_v_1}
}
\subfloat[] { 
	\includegraphics[scale=.27]{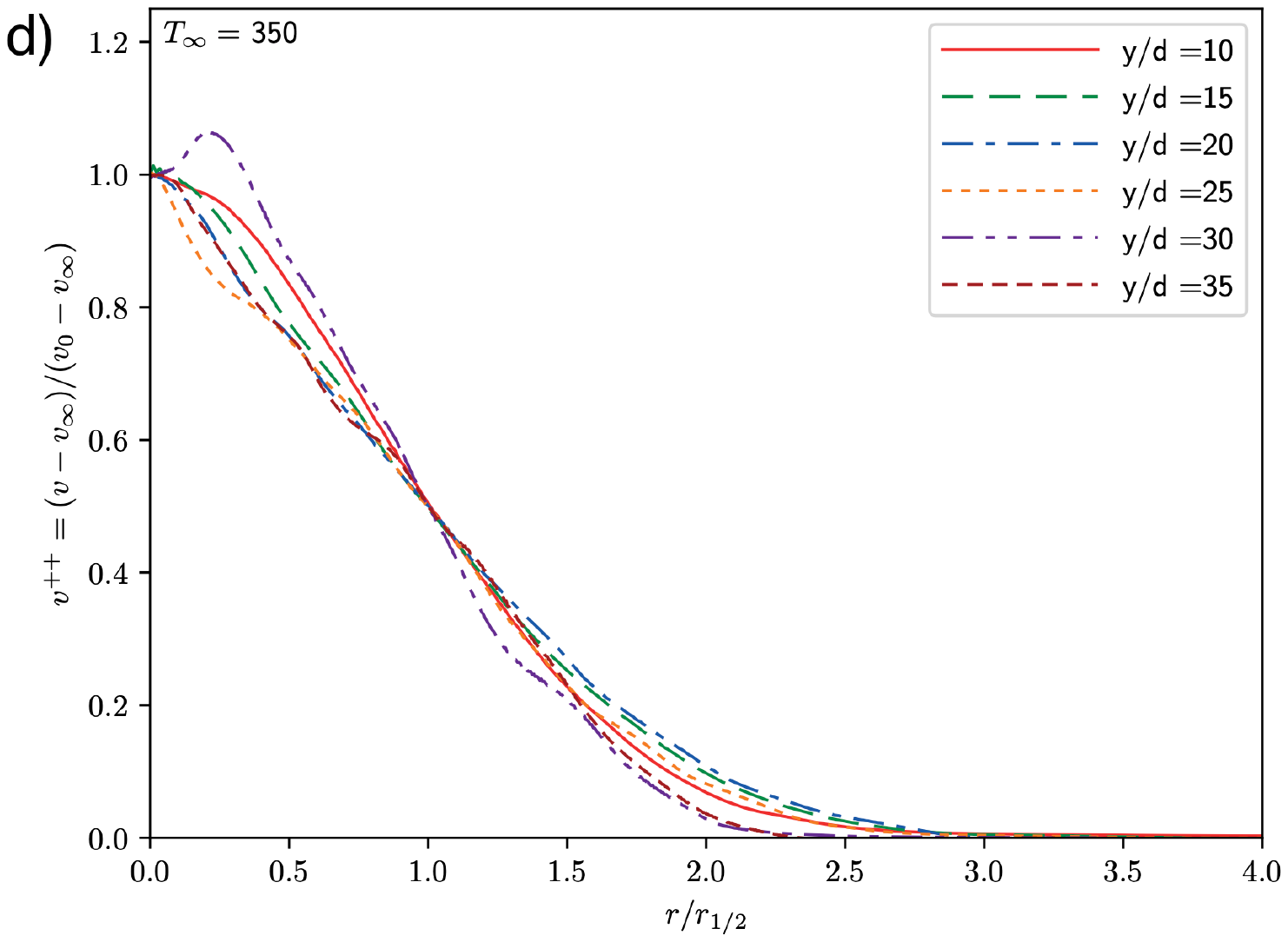}
    \label{noniso_far_r_vs_v_2}
}
\caption{Normal slices of scaled near-field velocity profiles in the (a) 314~K and (b) 350~K ambient cases. Far-field axial velocity profiles for each case, (c) and (d) respectively, are also included. Fields are averaged in both time and the radial direction. Plotted against radial direction scaled by $r_{1/2}$.}
\label{noniso_near_r_v_features}
\end{center}
\end{figure}

Figure \ref{noniso_near_r_v_features} shows a different scaling of the axial velocity profiles depicted in the previous figure. Here, the axial velocity is scaled by the centerline value at each normal slice while the radial distance is scaled by the $r_{1/2}$, as was done in Figure \ref{330_r_v_features}. Figures \ref{noniso_near_r_vs_v_1} and \ref{noniso_near_r_vs_v_2} show the near field profiles resulting from this scaling. As was the case with the 330~K ambient case, self-similarity is seen in the near field for the non-isothermal cases, where fairly agreeable profile collapse occurs at $\nicefrac{y}{d} = 15$. The 350~K ambient case profiles are slightly more in-line with each other as compared to the 314~K ambient case, but both demonstrate a general tendency toward self-similarity still. Figures \ref{noniso_far_r_vs_v_1} and \ref{noniso_far_r_vs_v_2} however do not show as agreeable a collapse as the 330~K ambient case. The 314~K ambient case has better self-similarity structure toward the centerline of the jet as compared to the 350~K ambient case, though the outer edge of the jet has a wider disparity between profiles. 

\begin{figure}[hbtp!]
\begin{center}
\subfloat[] { 
	\includegraphics[scale=.27]{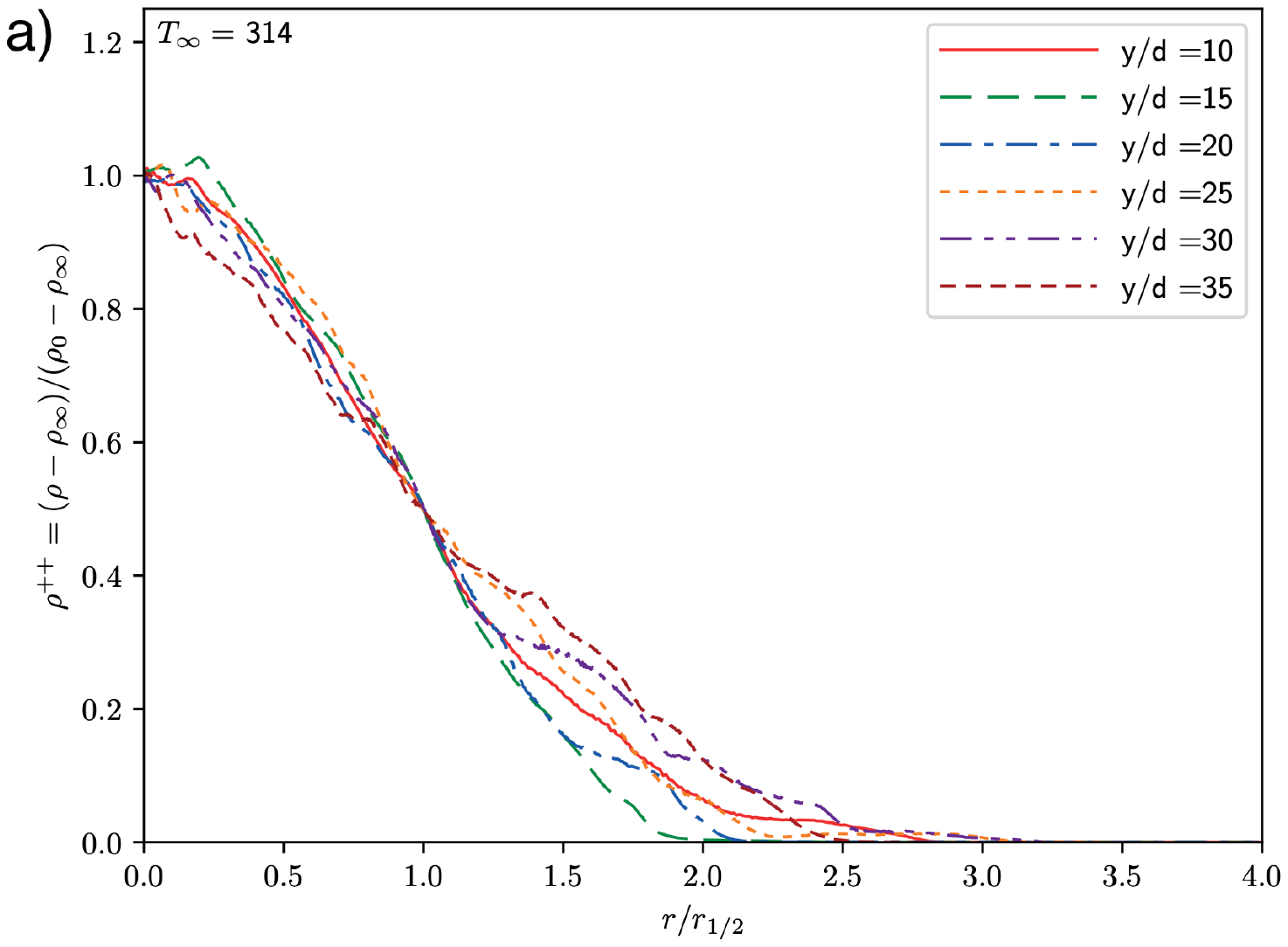}
	\label{noniso_r_vs_rho_1}
}
\subfloat[] { 
	\includegraphics[scale=.27]{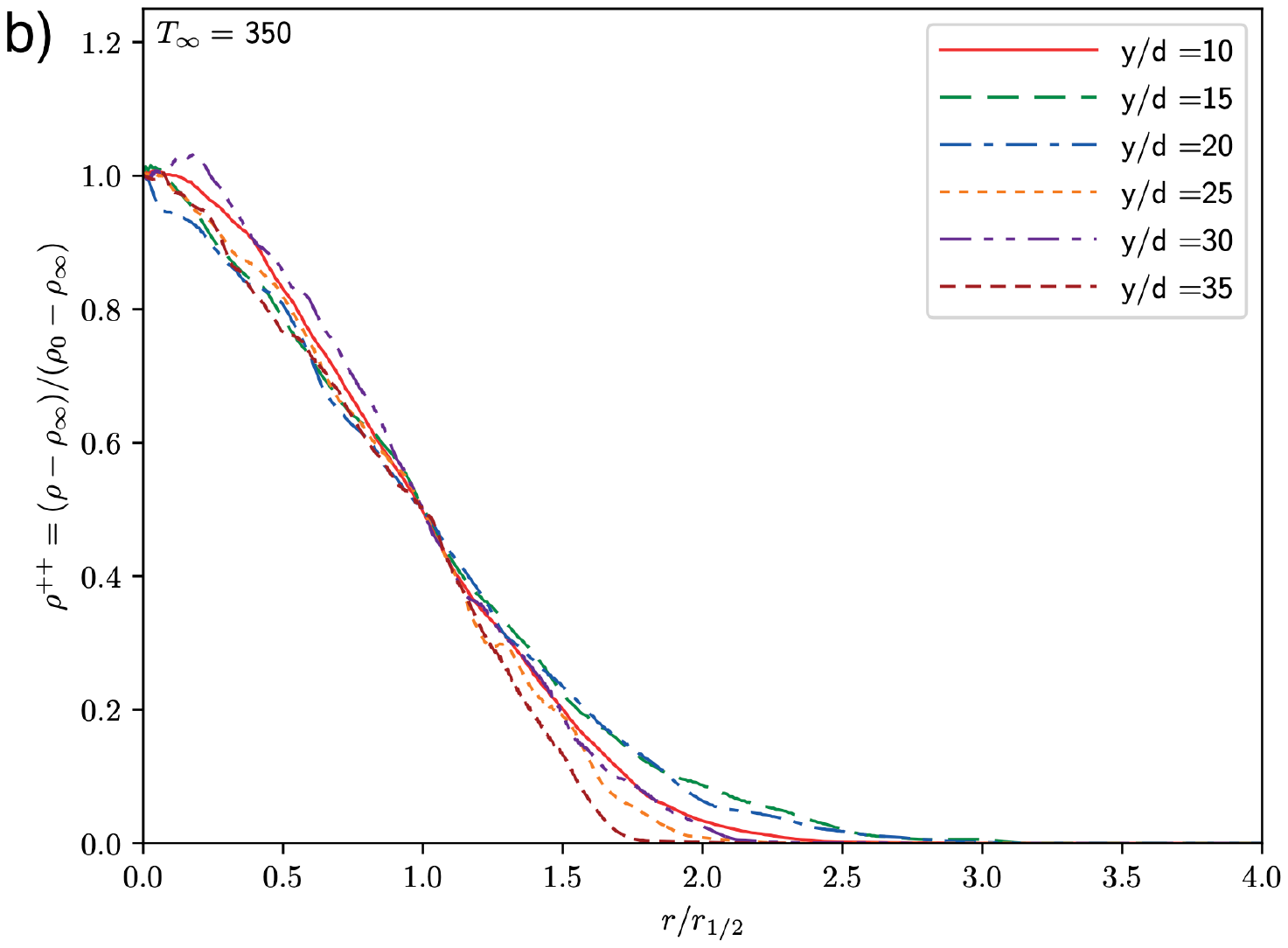}
	\label{noniso_r_vs_rho_2}
}
\caption{Normal slices of scaled far-field density, averaged in both time and the radial direction, plotted in the radial direction scaled by $r_{1/2}$ for (a) 314~K and (b) 350~K ambient case.}
\label{noniso_far_r_rho_features}
\end{center}
\end{figure}

Density self-similarity is not as evident as is in axial velocity profiles, as can be seen in Figure \ref{noniso_far_r_rho_features}. The 350~K ambient case profile spread is not as tight as it was with the axial velocity. The 314~K case has more prominent fluctuations in the curves than the previous profile depictions. Far-field density in this case has the widest spread of all previous comparisons. These density fluctuations and lack of self-similarity could be related to the increased Kelvin-Helmholtz instabilities seen in the 314~K ambient case and the subsequent density decay as seen in Figure \ref{noniso_rho_1}. One interesting feature of these plots can be seen on the outer edge of the jet, past the \gls{hmhw} mark. For the 314~K ambient case, profiles exhibit an initial decay as they dip below the $\nicefrac{y}{d}=10$ profile before increasing to fall above this initial curve at later slices downstream. This is opposite to what is seen in the 350~K ambient case, where earlier slices rise above the initial curve at $\nicefrac{y}{d} = 10$ before decaying below it farther downstream. This later decay is consistent with what is seen in the axial velocity self-similarity curves for both cases, while the density profiles for the 314~K ambient case break from this pattern. 

\begin{figure}[hbtp!]
\begin{center}
\subfloat[] { 
	\includegraphics[scale=.27]{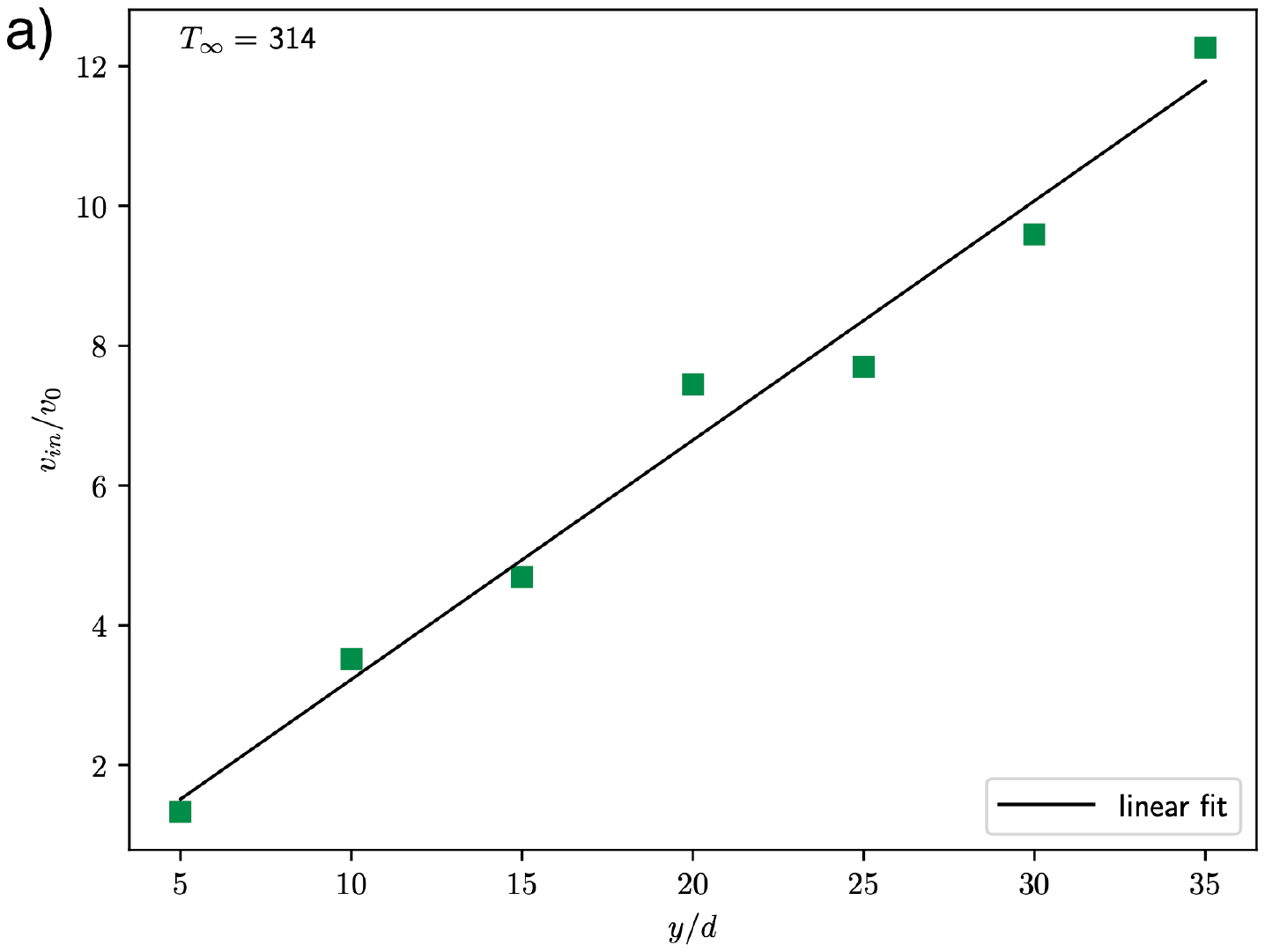}
	\label{noniso_uin_u0_x_c_1}
}
\subfloat[] { 
	\includegraphics[scale=.27]{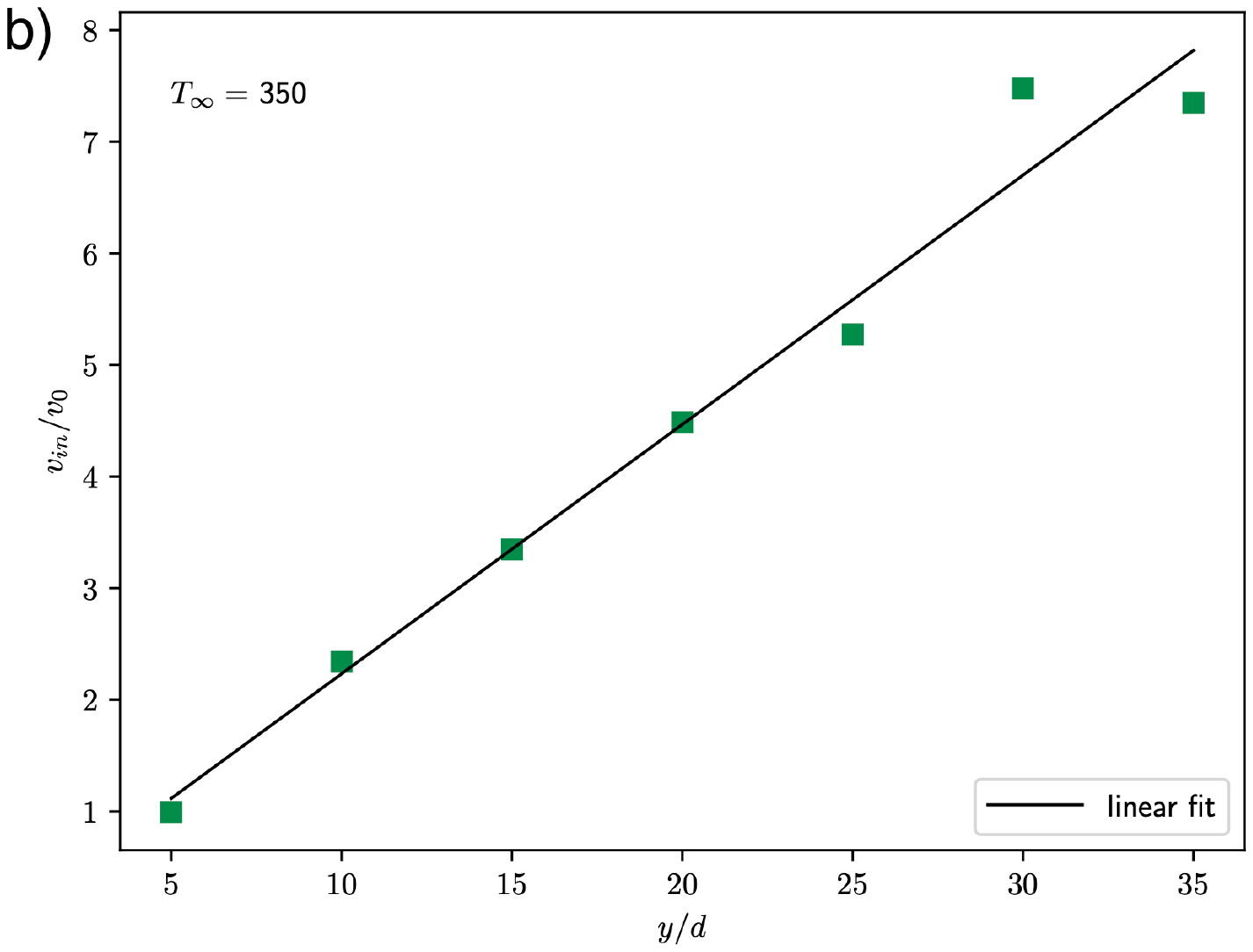}
	\label{noniso_uin_u0_x_d_2}
}
\caption{Axial inlet velocity scaled by centerline values along the axial direction for (a) 314~K and (b) 350~K ambient case. When distance downstream is scaled by jet diameter, linear decay of the centerline axial velocity is observed.}
\label{noniso_uin_u0_x_d_features}
\end{center}
\end{figure}

Figure \ref{noniso_uin_u0_x_d_features} shows the recovery of linear decay in the axial velocity along the centerline of the jet for each case. The 350~K ambient case has a less steep slope compared to the 330~K ambient case seen in Figure \ref{330_centerline_scaling}, where the decay rate according to Equation \eqref{decay_rate} is $B_v = 4.48$. This corresponds to slower decay of the axial velocity along the centerline. The 314~K ambient case has a steeper slope compared to the 330~K ambient case, with $B_v = 2.92$, corresponding to a faster decay in axial velocity along the centerline. These findings are consistent with the analysis from Figure \ref{noniso_v_vin_r_d_features}. 

\begin{figure}[hbtp!]
\begin{center}
\subfloat[] { 
	\includegraphics[scale=.27]{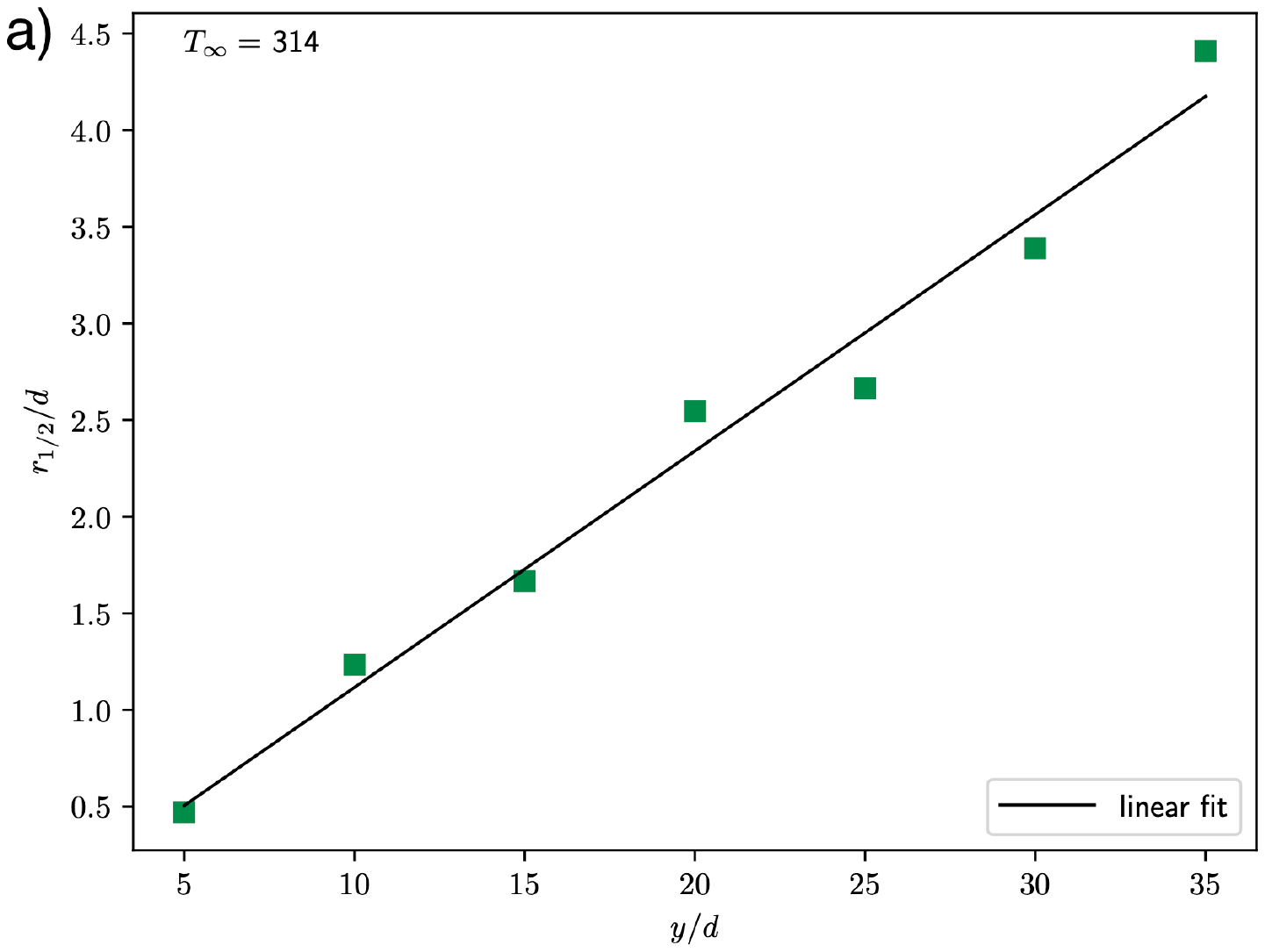}
	 \label{noniso_uin_u0_x_c_1}
}
\subfloat[] { 
	\includegraphics[scale=.27]{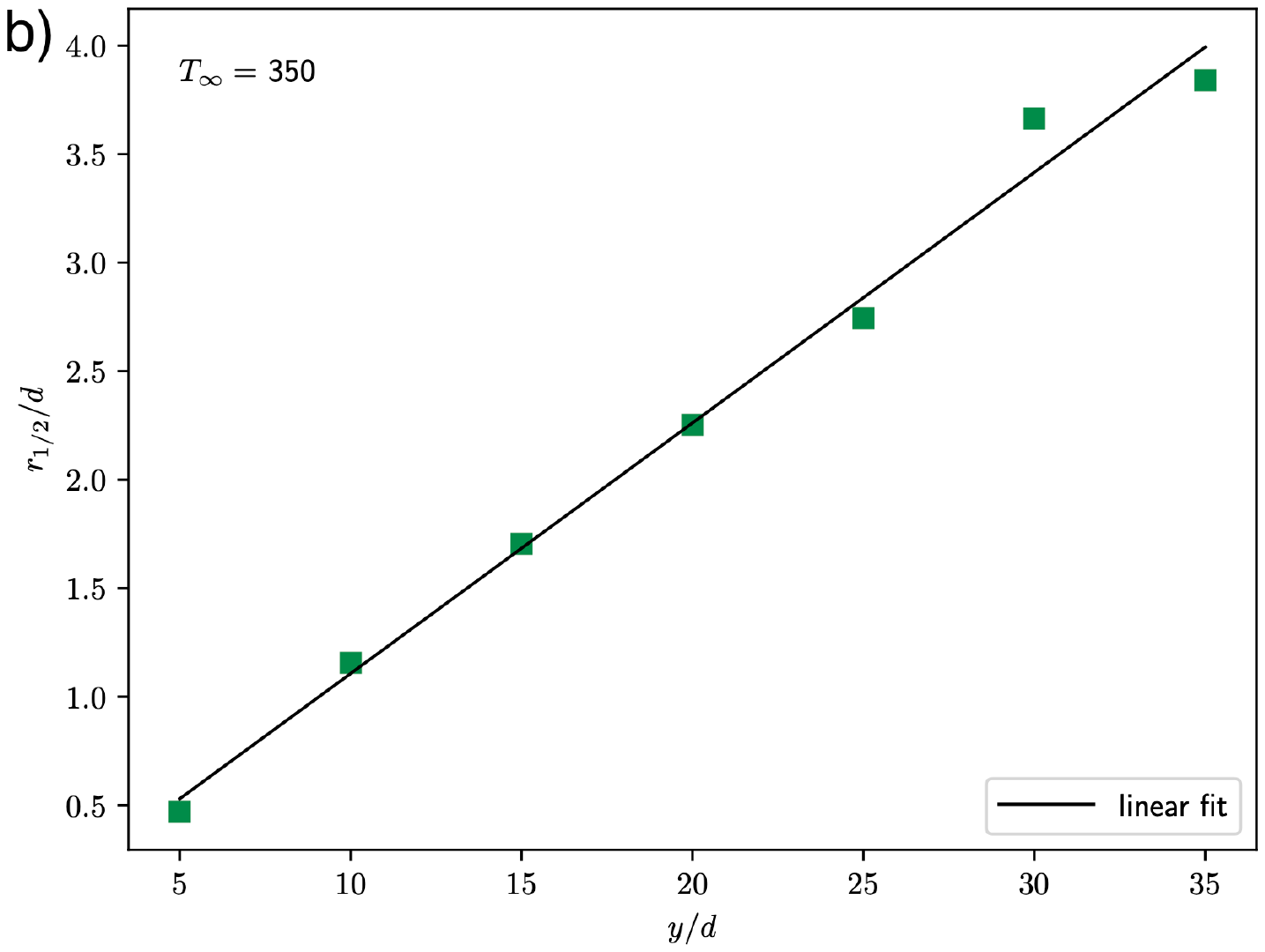}
	\label{noniso_uin_u0_x_d_2}
}
\caption{\gls{hmhw} values scaled by jet diameter along the axial direction for (a) 314~K and (b) 350~K ambient case. When distance downstream is scaled by jet diameter, linear growth of the half-widths is observed.}
\label{noniso_r12_d_x_d_features}
\end{center}
\end{figure}

Similarly, Figure \ref{noniso_r12_d_x_d_features} shows the recovery of linear spreading rate along the centerline of the jet for each case. The 350~K ambient case has a steeper slope compared to the 330~K ambient case seen in Figure \ref{330_centerline_scaling}, where the decay rate according to Equation \eqref{decay_rate} is $C_v = 0.115$. This corresponds to slower decay of the axial velocity along the centerline. The 314~K ambient case has a steeper slope compared to the 330~K ambient case, with $C_v = 0.122$, corresponding to a faster decay in axial velocity along the centerline. These findings are consistent with the analysis from Figure \ref{noniso_v_vin_r_d_features}. Overall the jet in the 314~K ambient case decays faster in the axial direction while spreading further in the radial direction compared to that of the 350~K ambient case. 

\begin{figure}[hbtp!]
\begin{center}
\subfloat[] { 
\centering
	\includegraphics[scale=.23]{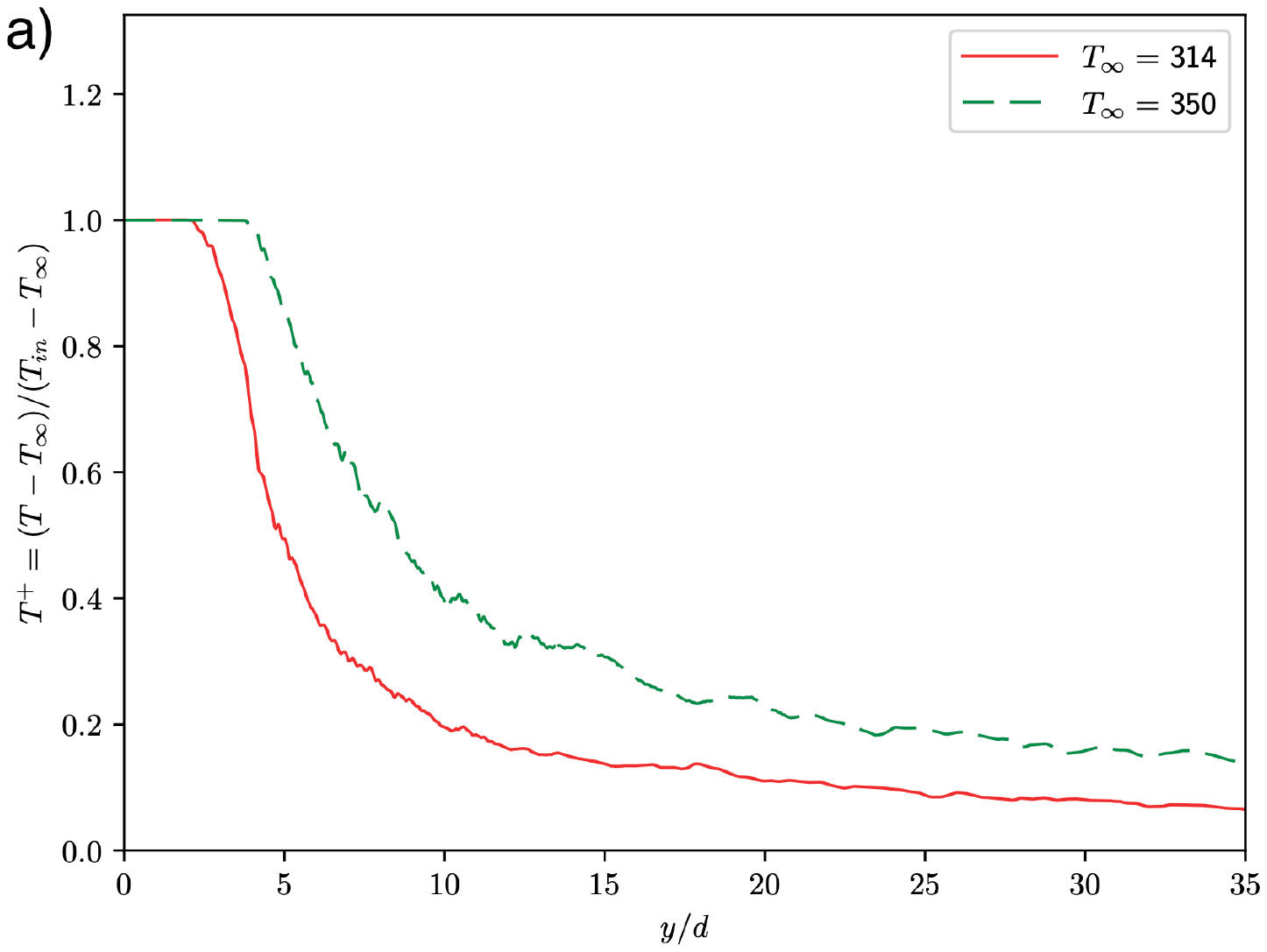}
	\label{noniso_temp_centerline_1}
}
\vspace{-16pt}
\subfloat[] { 
\centering
	\includegraphics[scale=.23]{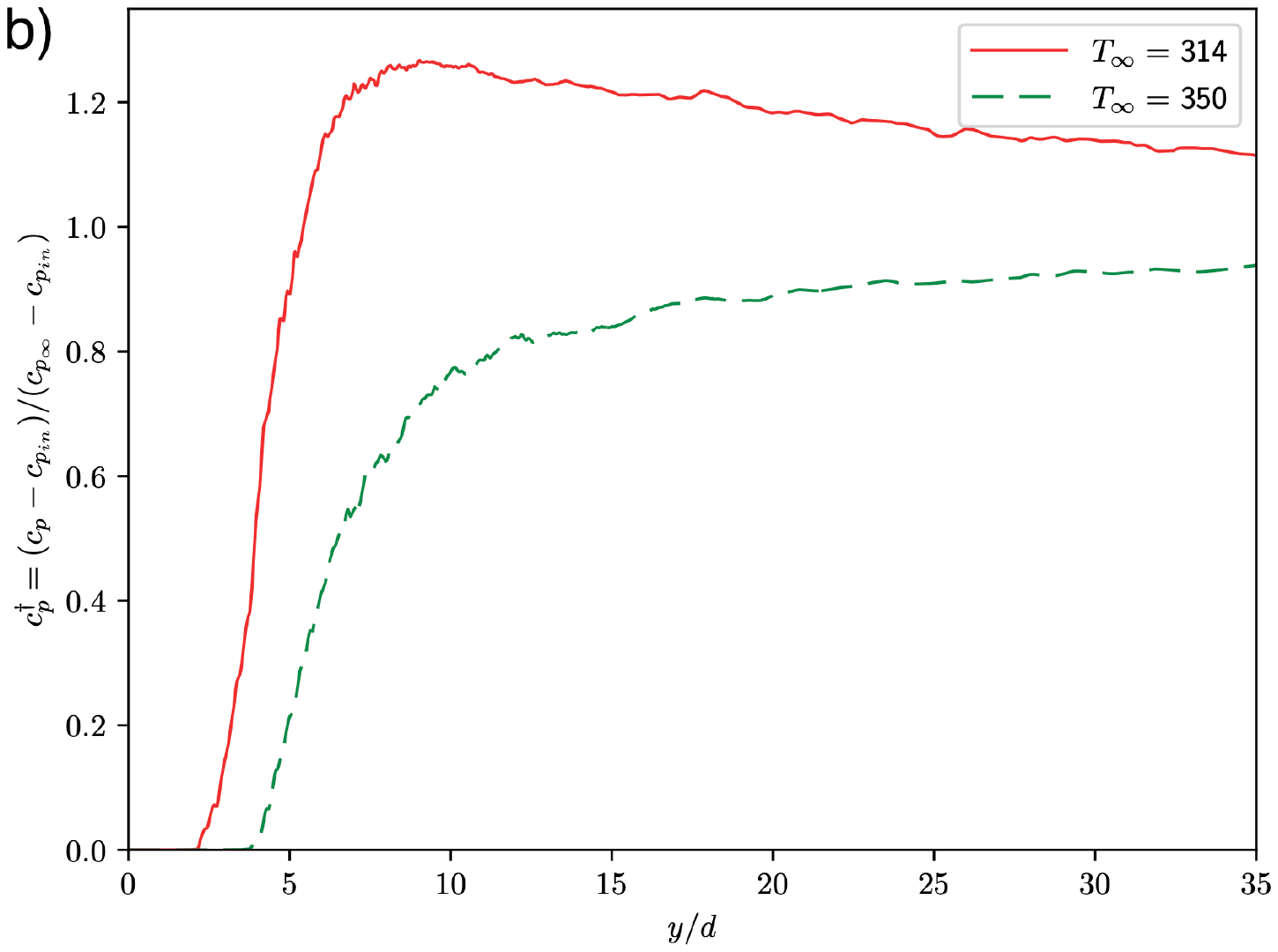}
	\label{noniso_cp_centerline_1}
} \\
\vspace{-16pt}
\subfloat[] { 
\centering
	\includegraphics[scale=.23]{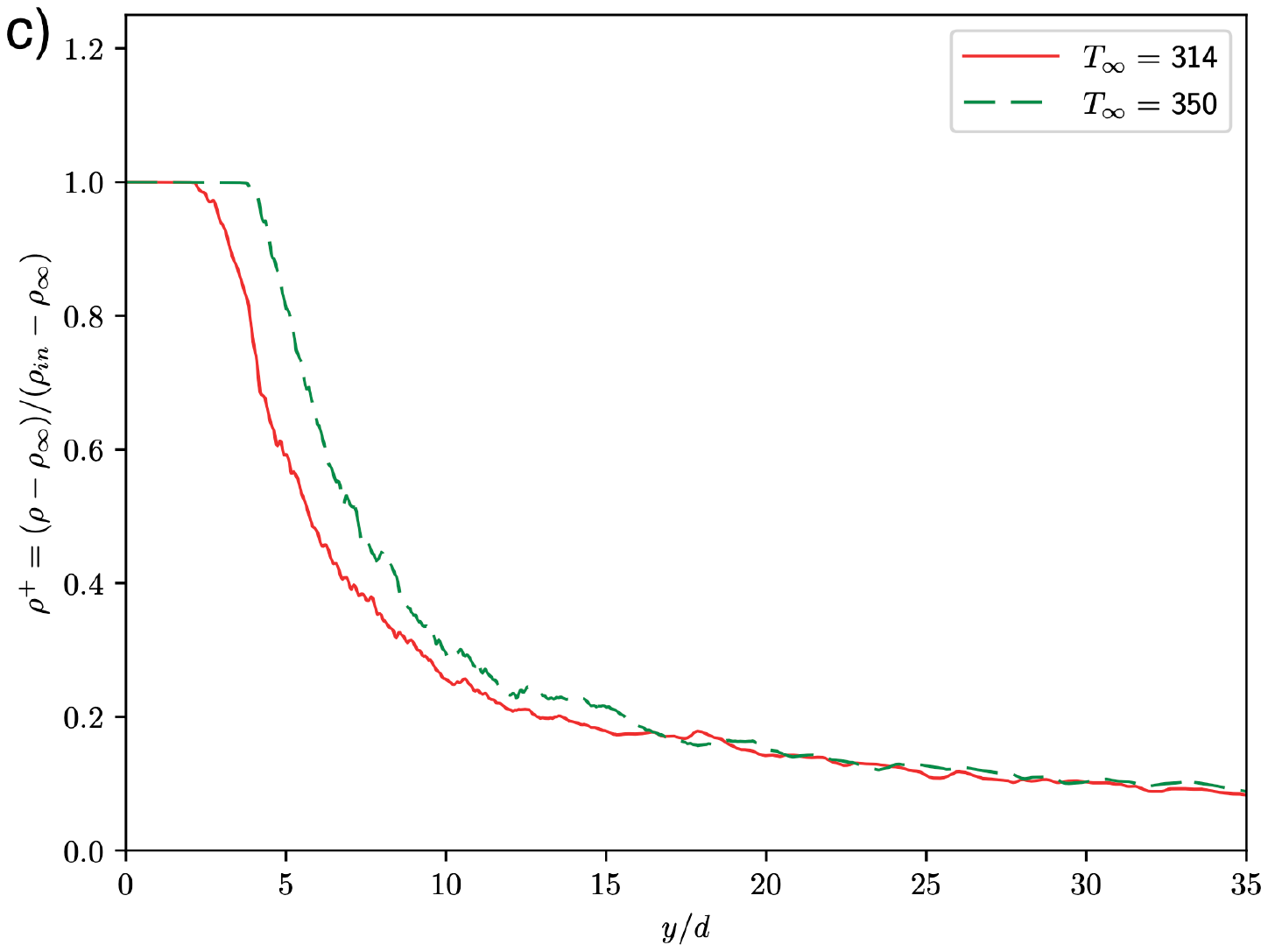}
	\label{noniso_rho_centerline_1}
} 
\vspace{-16pt}
\subfloat[] { 
\centering
	\includegraphics[scale=.23]{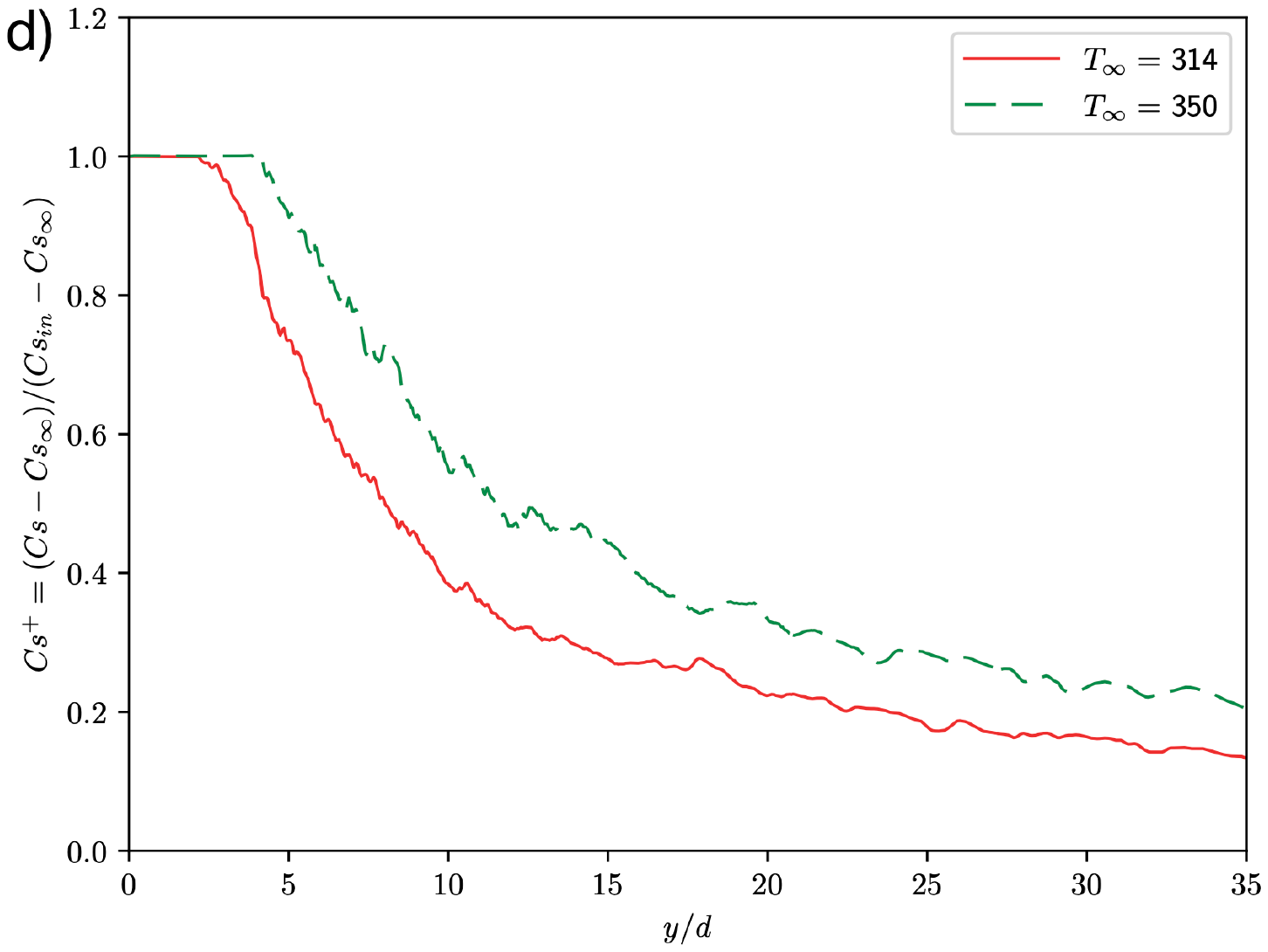}
	\label{noniso_Cs_centerline_1}
} \\
\vspace{-16pt}
\subfloat[] { 
\centering
	\includegraphics[scale=.23]{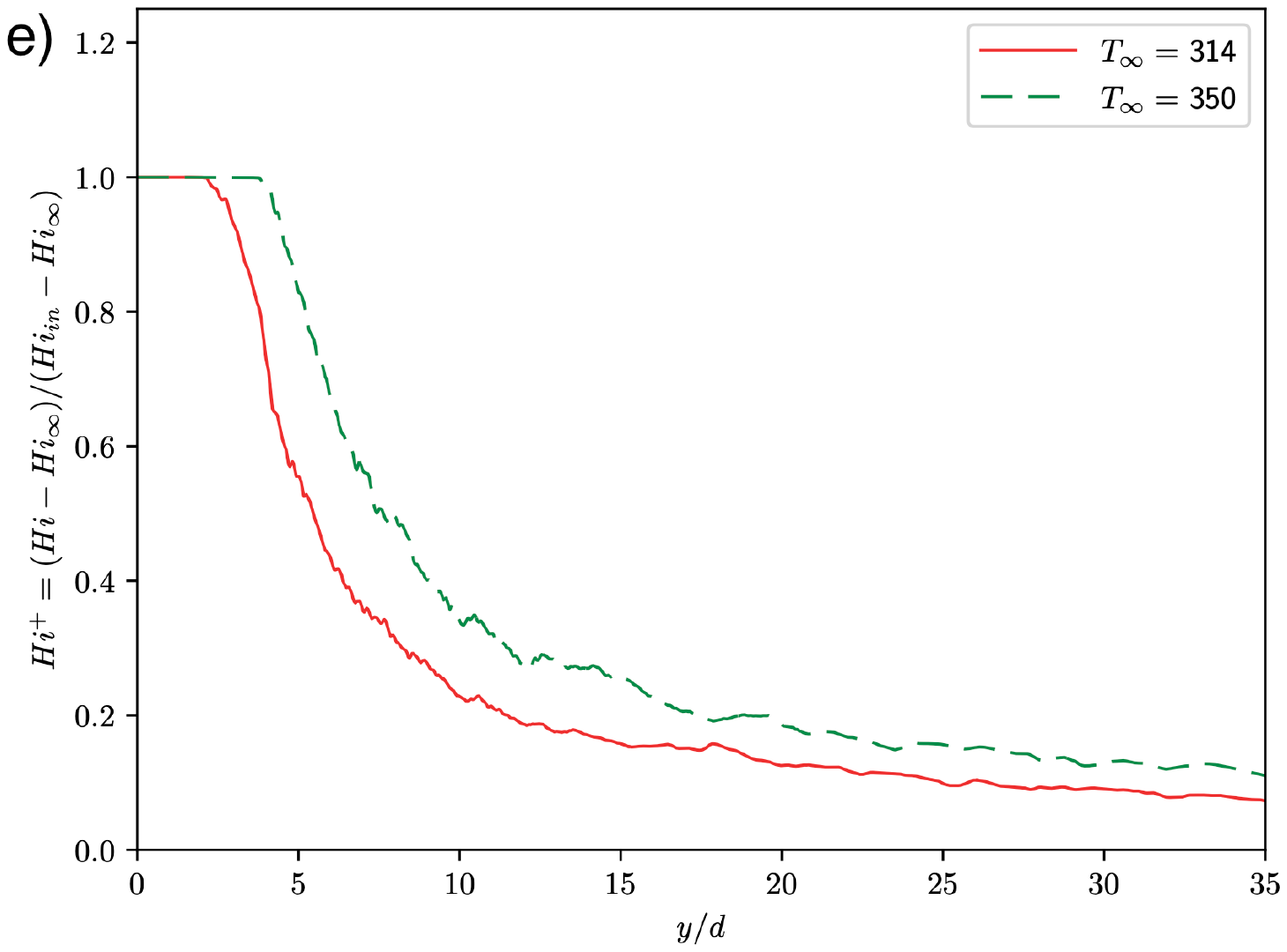}
	\label{noniso_Hi_centerline_1}
}
\vspace{-16pt}
\subfloat[] { 
\centering
	\includegraphics[scale=.23]{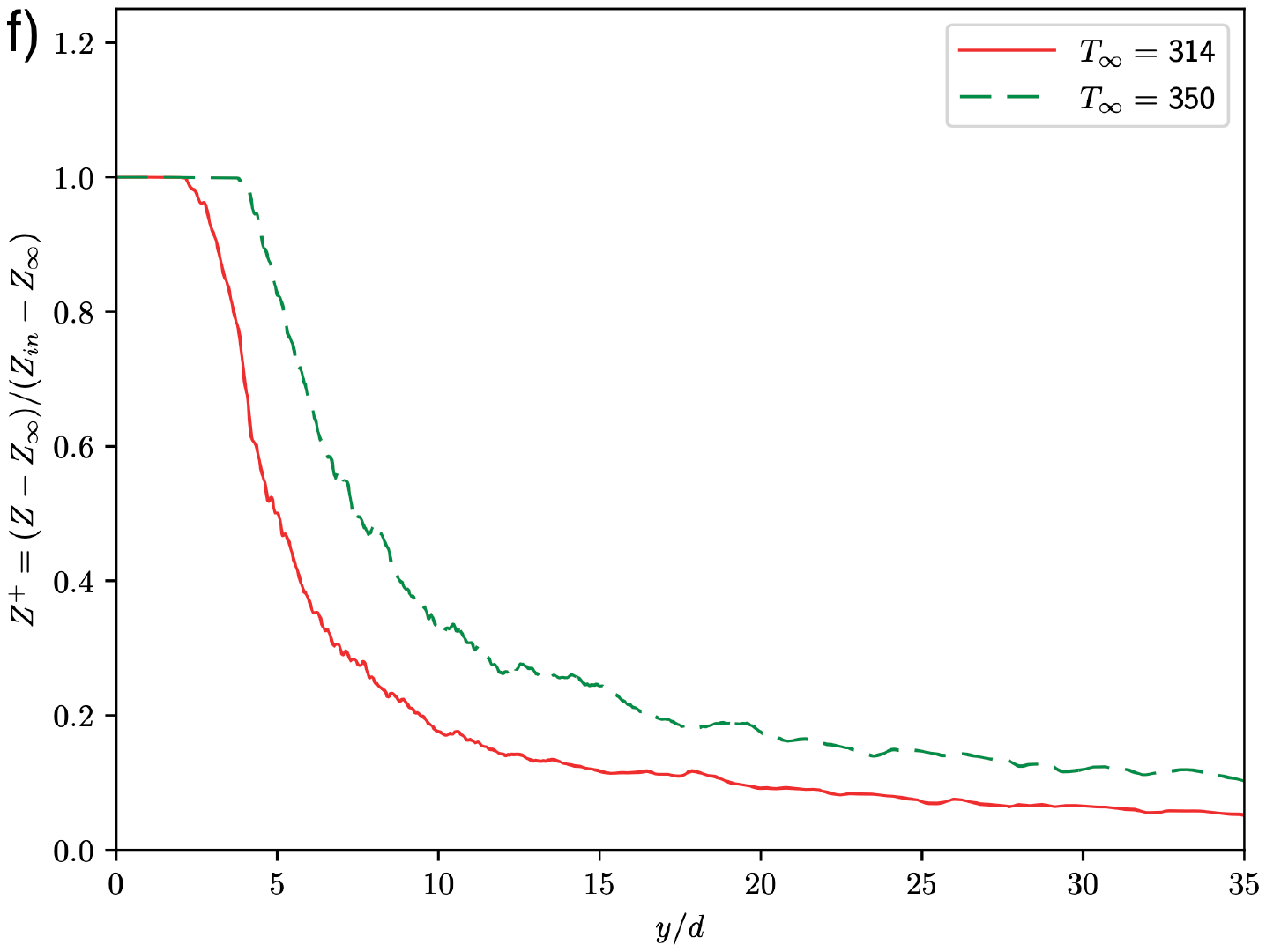}
	\label{noniso_Z_centerline_1}
} \\
\vspace{-16pt}
\subfloat[] { 
\centering
	\includegraphics[scale=.23]{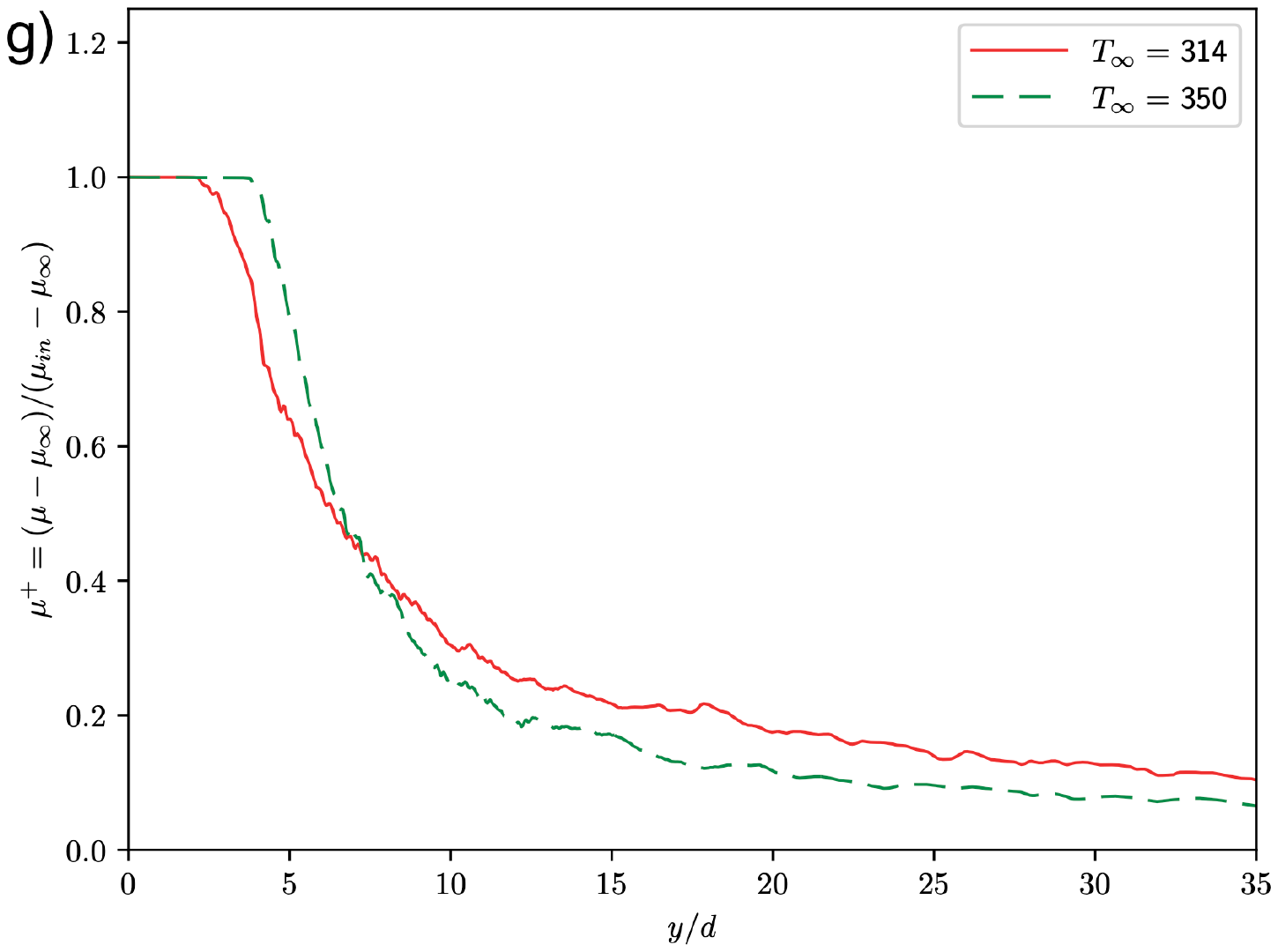}
	\label{noniso_mu_centerline_1}
}
\subfloat[] { 
\centering
	\includegraphics[scale=.23]{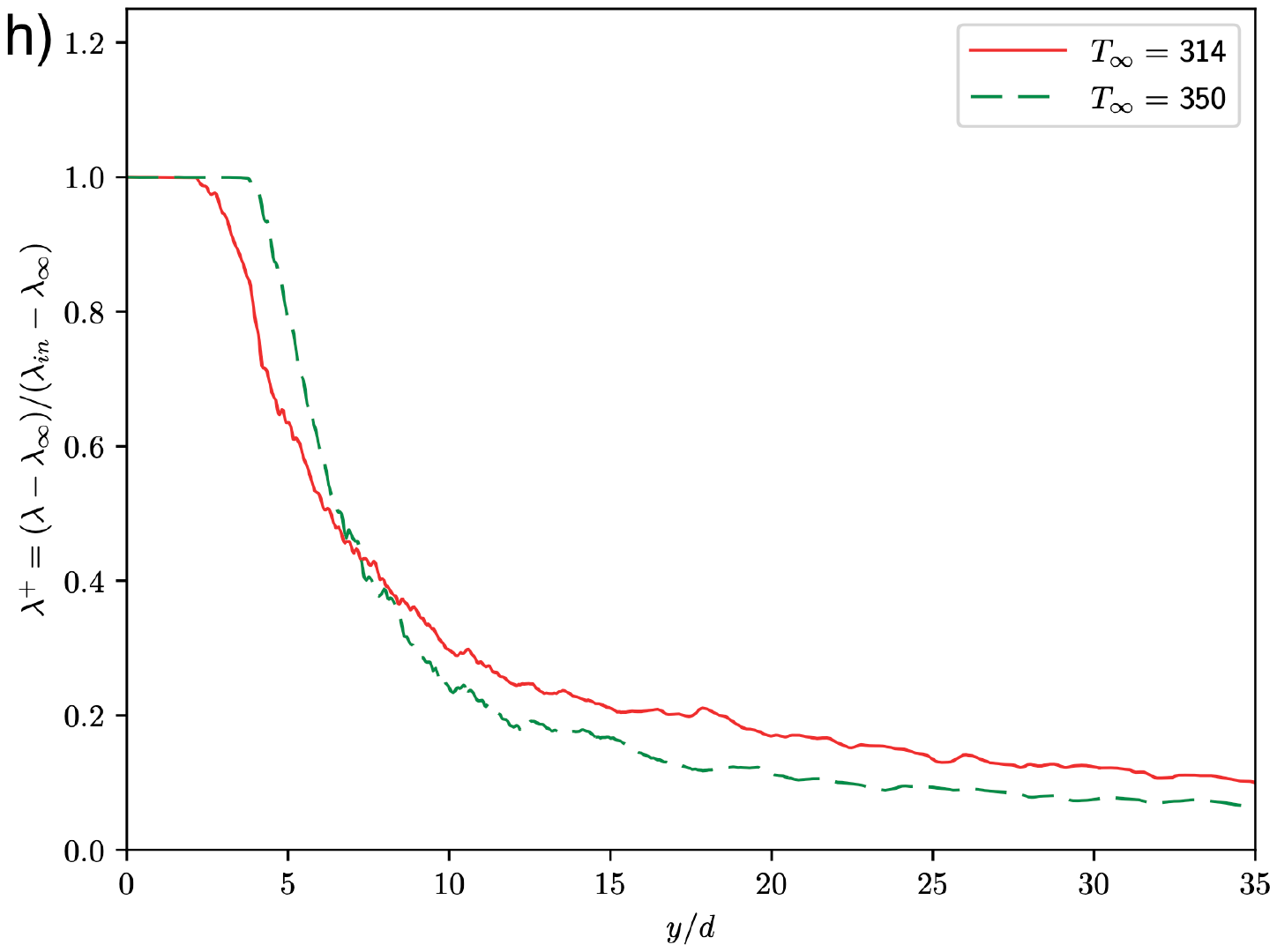}
	\label{noniso_lam_centerline_1}
}

\caption{Comparison of centerline decay for (a) Temperature, (b) Constant-Pressure Specific Heat, (c) Density, (d) Sound Speed, (e) Enthalpy, (f) Compressibility Factor, (g) Viscosity, and (h) Thermal Conductivity.}
\label{noniso_centerline_features}
\end{center}
\end{figure}

Finally, Figure \ref{noniso_centerline_features} shows various quantities along the centerline for both non-isothermal cases. For all quantities presented here, centerline plots help depict the different regions of the jet. The potential core is maintained until around $\nicefrac{y}{d} = 2.5$ for the 314~K ambient case and until around $\nicefrac{y}{d}= 5$ for the 350~K ambient case. Then there is a transition region where quantities experience rapid change. This occurs up until about $\nicefrac{y}{d}= 10$ for the 314~K ambient case and $\nicefrac{y}{d} = 12$ for the 350~K ambient case. After that, a gradual leveling out occurs as the quantity of interest continues to adjust as it mixes with the ambient fluid through the remainder of the domain. Figure \ref{noniso_temp_centerline_1} shows the temperature change for each case along the centerline of the jet. The 314~K ambient case exhibits a more rapid decay from the inflow temperature as compared to the 350~K ambient case, as can be seen in the separation between the curves between $4 \leq \nicefrac{y}{d} \leq 16$. Afterwards, they continue their respective transition toward ambient conditions at roughly the same rate. The 350~K ambient case is never as close to ambient conditions as the 314~K ambient case due to the slower initial decline, hovering instead around $T^+=0.18$ as compared to the $T^+=0.14$ that the 314~K ambient case sits at. Enthalpy and the compressibility factor in Figures \ref{noniso_Hi_centerline_1} and \ref{noniso_Z_centerline_1} respectively follow the same trend though with less separation between the two cases as compared to the temperature trajectories. Figure \ref{noniso_Cs_centerline_1} shows the sound speed transition along the centerline. Compared to the previously discussed quantities, the sound speed decay appears much more gradual, with little rate difference between the two cases. The initial transition region is less steep and the region thereafter more of a continued gradual decline as opposed to a leveling out like the other quantities. Figure \ref{noniso_cp_centerline_1} shows the constant-pressure specific heat for the two non-isothermal cases. Again, neither fully transitions to the ambient condition as the jet persists through the domain. Here the jump in specific heat over the peak present near the pseudo-boiling point can be seen for the 314~K ambient case. This case also has a steeper initial transition than the other case does, though both have similar rates of approach toward ambient conditions after this initial shift. Figure \ref{noniso_rho_centerline_1} shows the density of the two cases. This quantity follows a unique trajectory between the two cases compared to those previously discussed. Here, the initial transition rate of the two cases from the potential core to the fully developed region is nearly the same. Toward the end of this transition though,  the 314~K ambient case begins to level out at a faster rate than the 350~K ambient case, allowing the curves to collapse onto each other. Then both centerline densities continue on past $\nicefrac{y}{d} = 16$ at approximately $\rho^+ = .12$ and steadily decline toward ambient conditions thereafter. 

\subsubsection{Turbulence Dynamics}
\begin{figure}[ht!]
\begin{center}
\subfloat[] { 
	\includegraphics[scale=.27]{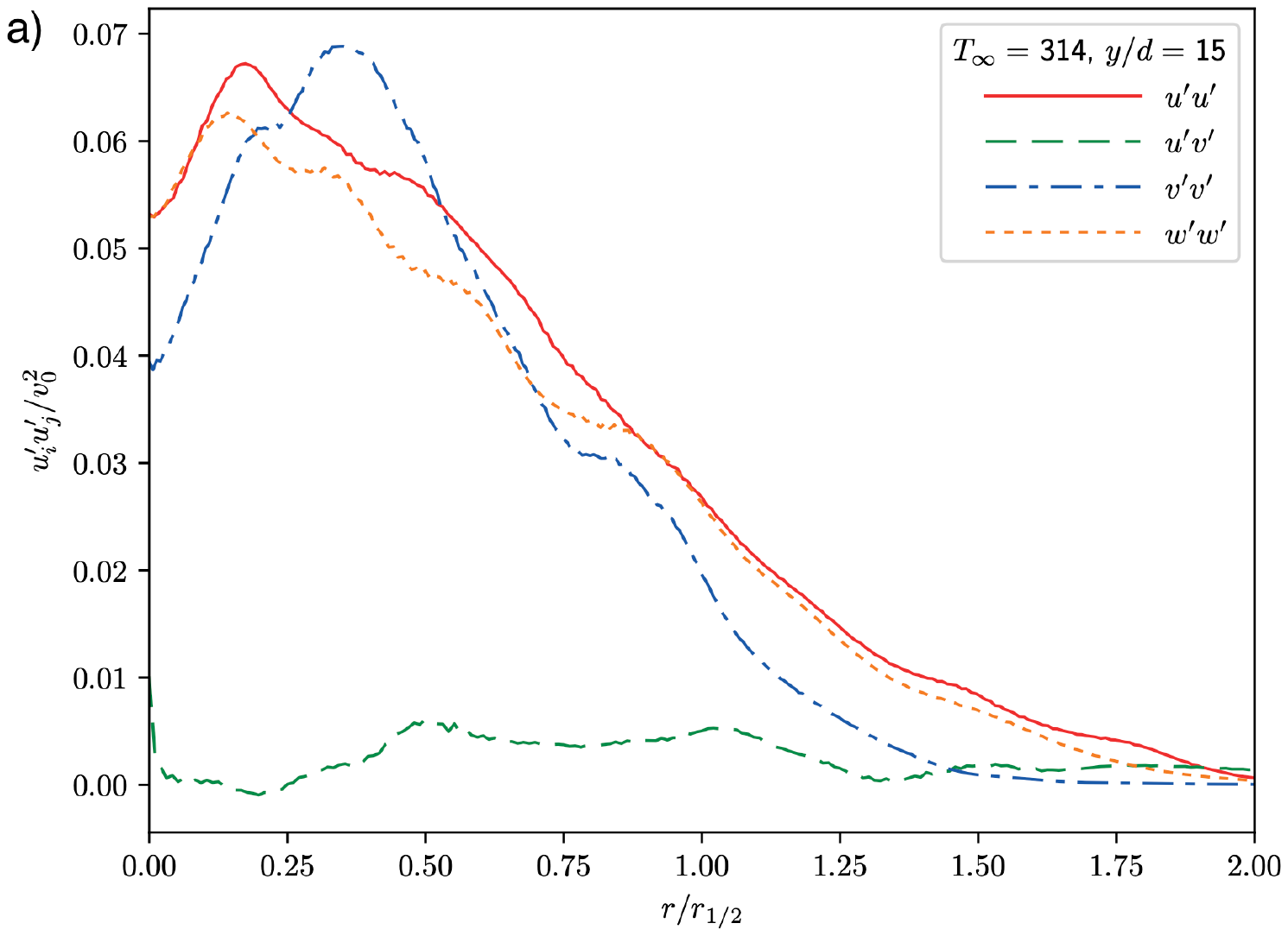}
	\label{314_rey_15}
}
\subfloat[] { 
	\includegraphics[scale=.27]{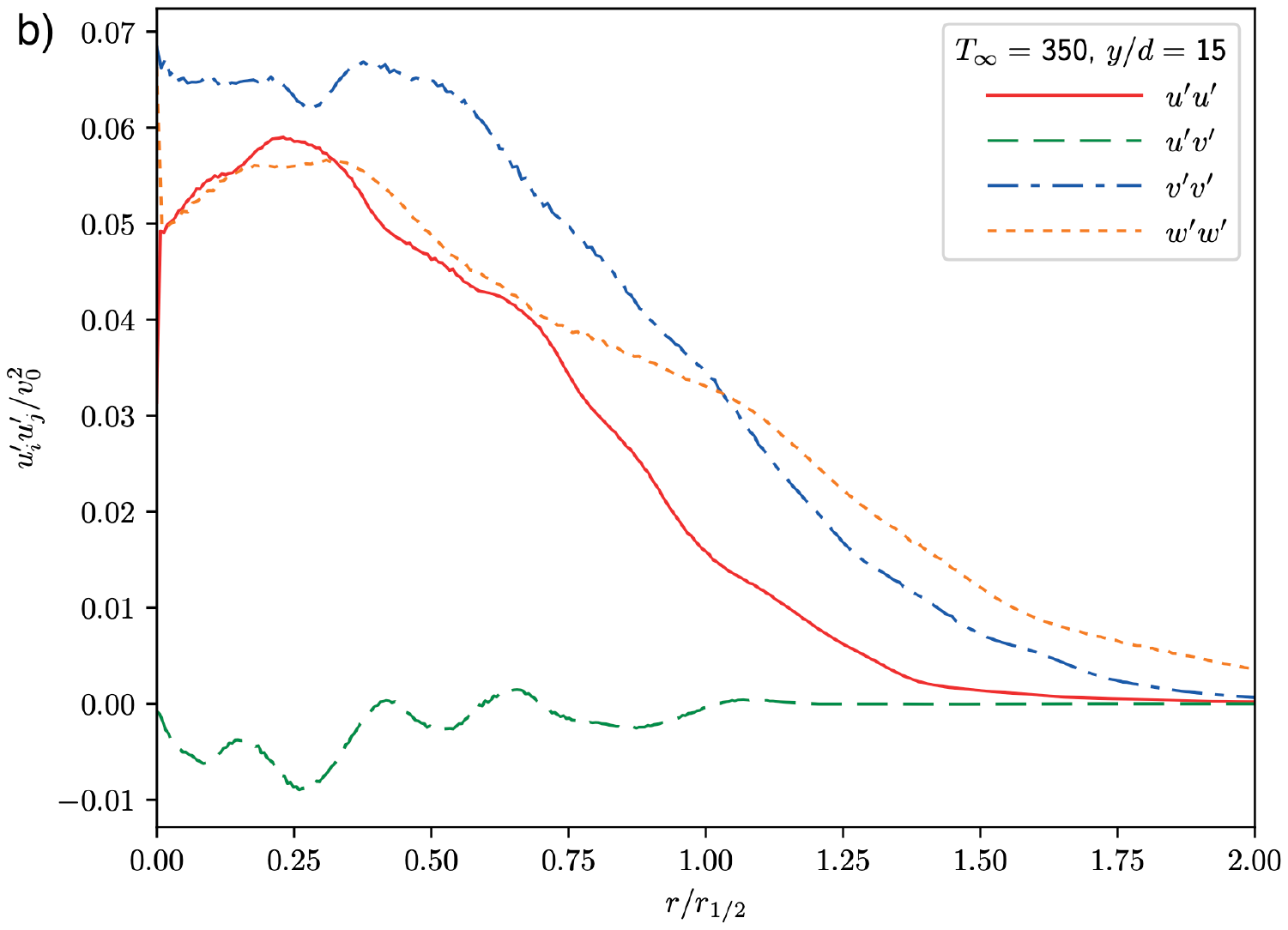}
	\label{350_rey_15}
}
\vspace{-32pt}
\subfloat[] { 
	\includegraphics[scale=.27]{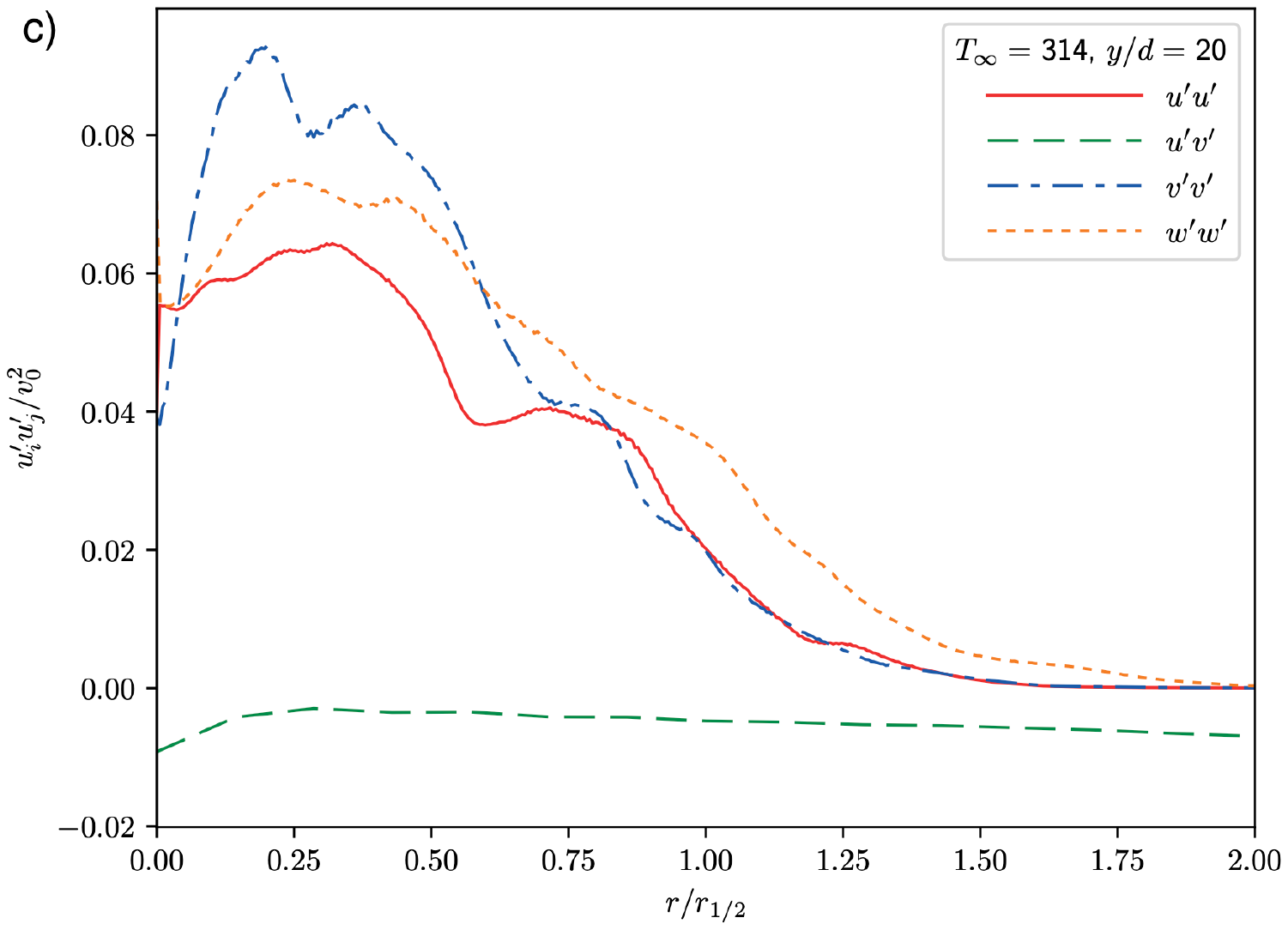}
	\label{314_rey_20}
}
\subfloat[] { 
	\includegraphics[scale=.27]{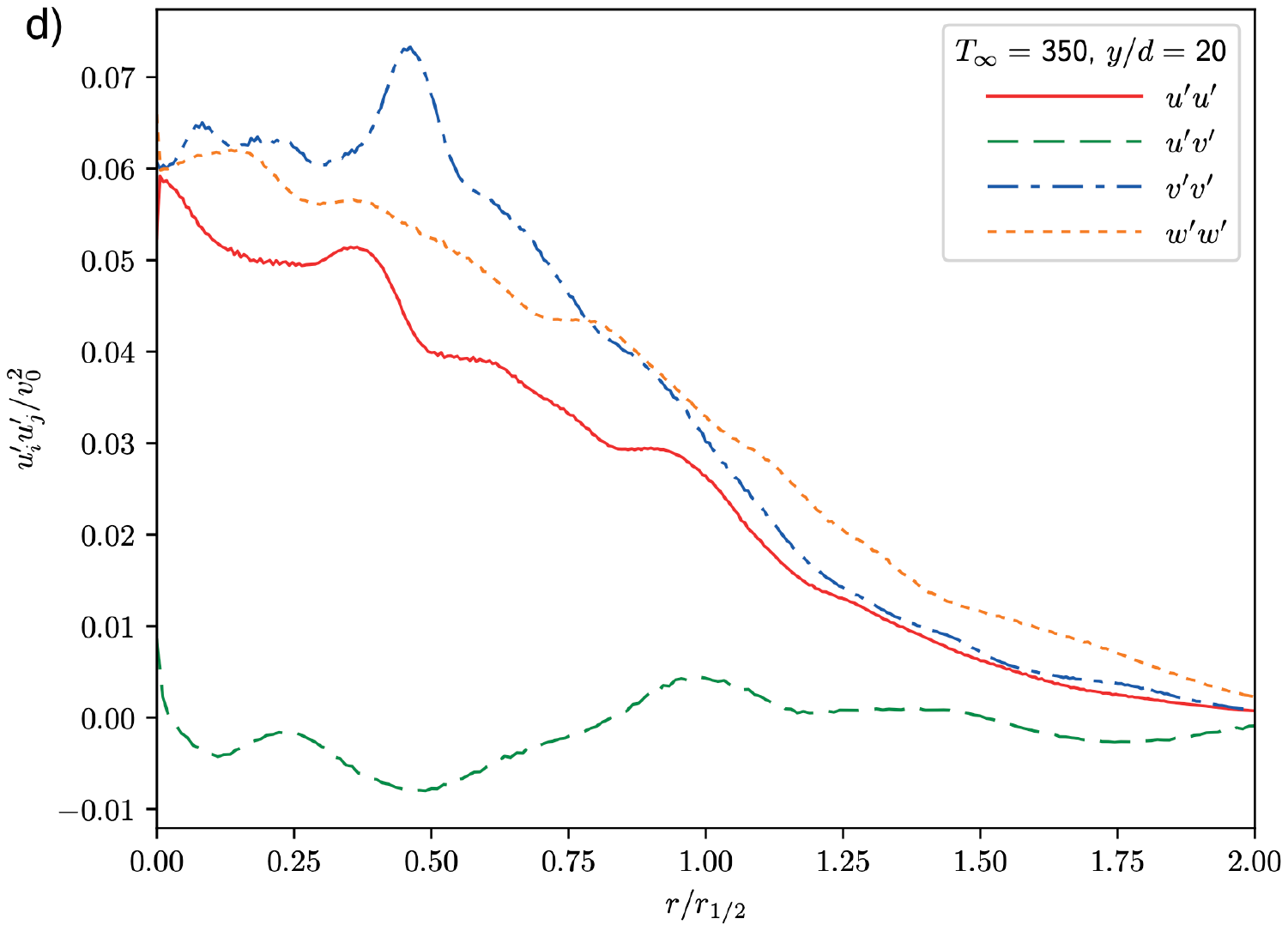}
	\label{350_rey_20}
}
\caption{Time and radially averaged Reynolds stresses at $y/d=15$ for (a) 314~K and (b) 350~K and $y/d=20$ for (c) 314~K and (d) 350~K.}
\label{noniso_reynolds_features}
\end{center}
\end{figure}

Figure \ref{noniso_reynolds_features} shows the resolved Reynolds stresses of each non-isothermal case at two different locations downstream. As was the case with the 330~K ambient resolved Reynolds stresses in Figure \ref{330_reynolds_features}, self-similarity in the axial direction is not recovered. The Reynolds stresses for both of the non-isothermal cases are generally smaller than their counterparts in the isothermal case. The stresses in Figure \ref{350_rey_15} generally follow the trends outlined in the isothermal case, but the remaining slices for these two cases do not follow suit, with little to no separation between the axial direction stresses and the other stresses. This could be due the effects of the \gls{sgs} modeling as mentioned in \cite{doi:10.1063/1.4937948}.  

\begin{figure}[ht!]
\begin{center}
\subfloat[] { 
	\includegraphics[scale=.27]{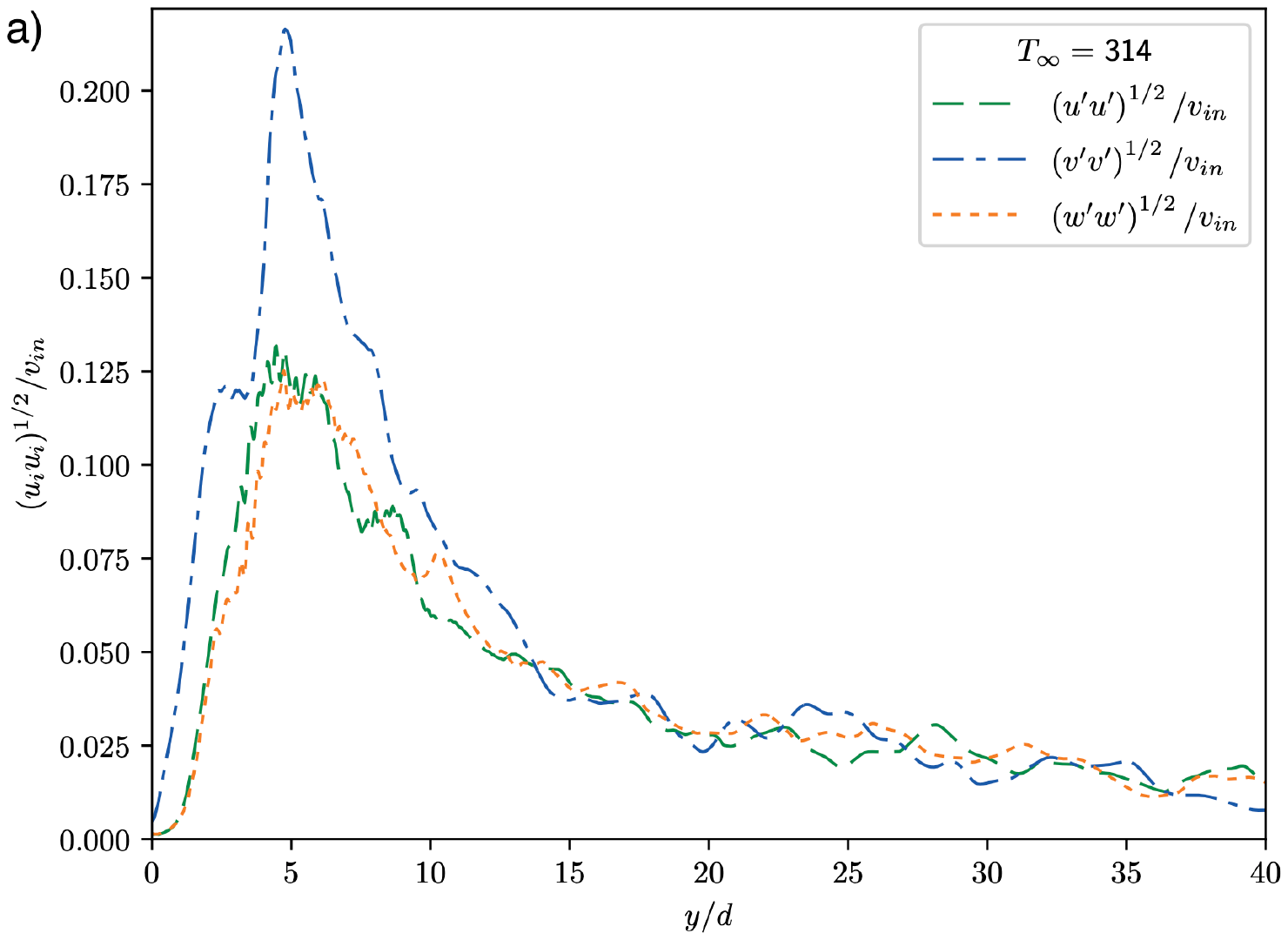}
	\label{314_TKEcomp_1}
}
\subfloat[] { 
	\includegraphics[scale=.27]{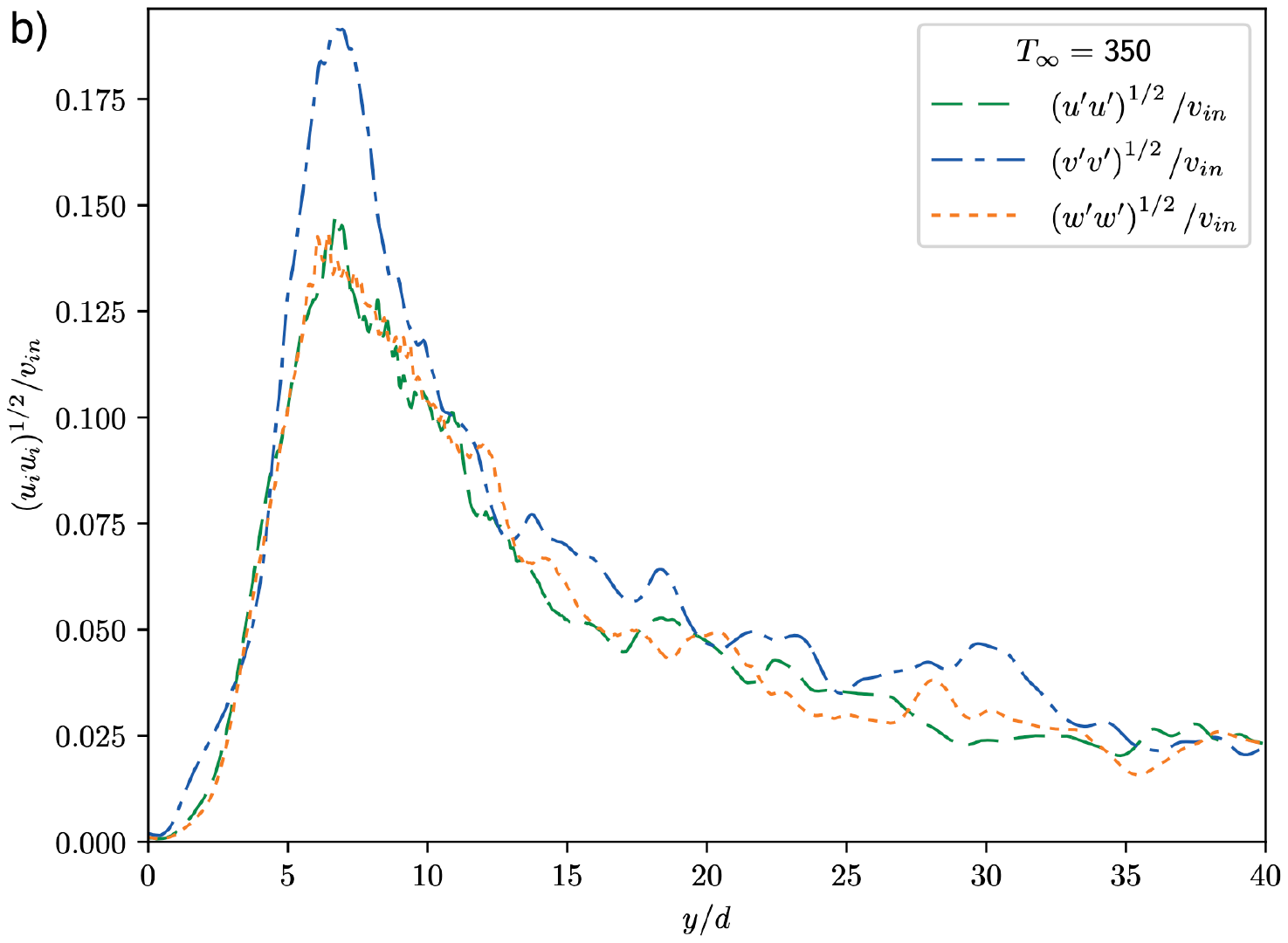}
	\label{350_TKEcomp_1}
}
\caption{Comparison of average \gls{tke} components along centerline for (a) 314~K and (b) 350~K ambient case.}
\label{noniso_TKE_component_features}
\end{center}
\end{figure}

Figure \ref{noniso_TKE_component_features} shows the components of the resolved \gls{tke} for each non-isothermal case. Both cases follow the same general trends seen in the 330~K ambient case, with the axial component being the strongest and the other two components being of the same magnitude. The 350~K ambient case is the most similar to the 330~K ambient case between the two. For this case, peak \gls{tke} occurs at approximately $\nicefrac{y}{d} = 7$, which is slightly later than the isothermal case. Additionally, this peak is aligned with the peak seen in the other components, where for the isothermal case the axial component peak came slightly before the peak of the other two. The magnitude of the peak axial component is larger in the 350~K ambient case while the magnitudes of the other directional components are similar to their counterparts in the 330~K ambient case. The decay rate is also slightly higher in the 350~K ambient case compared to the isothermal case. The \gls{tke} components of the 314~K ambient case reach their peaks slightly earlier than the 330~K ambient case at about $\nicefrac{y}{d} = 5$. The peak of the axial component is much larger than that of the other two cases. Additionally, the other \gls{tke} components of the 314~K ambient case have a smaller magnitude than their counterparts in the other cases, resulting in a larger disparity between the magnitudes of the components in this case. There is also stronger overall decay in the 314~K ambient case as compared to the 350~K ambient case.

\begin{figure}[H]
\begin{center}
\subfloat[] { 
	\includegraphics[scale=.27]{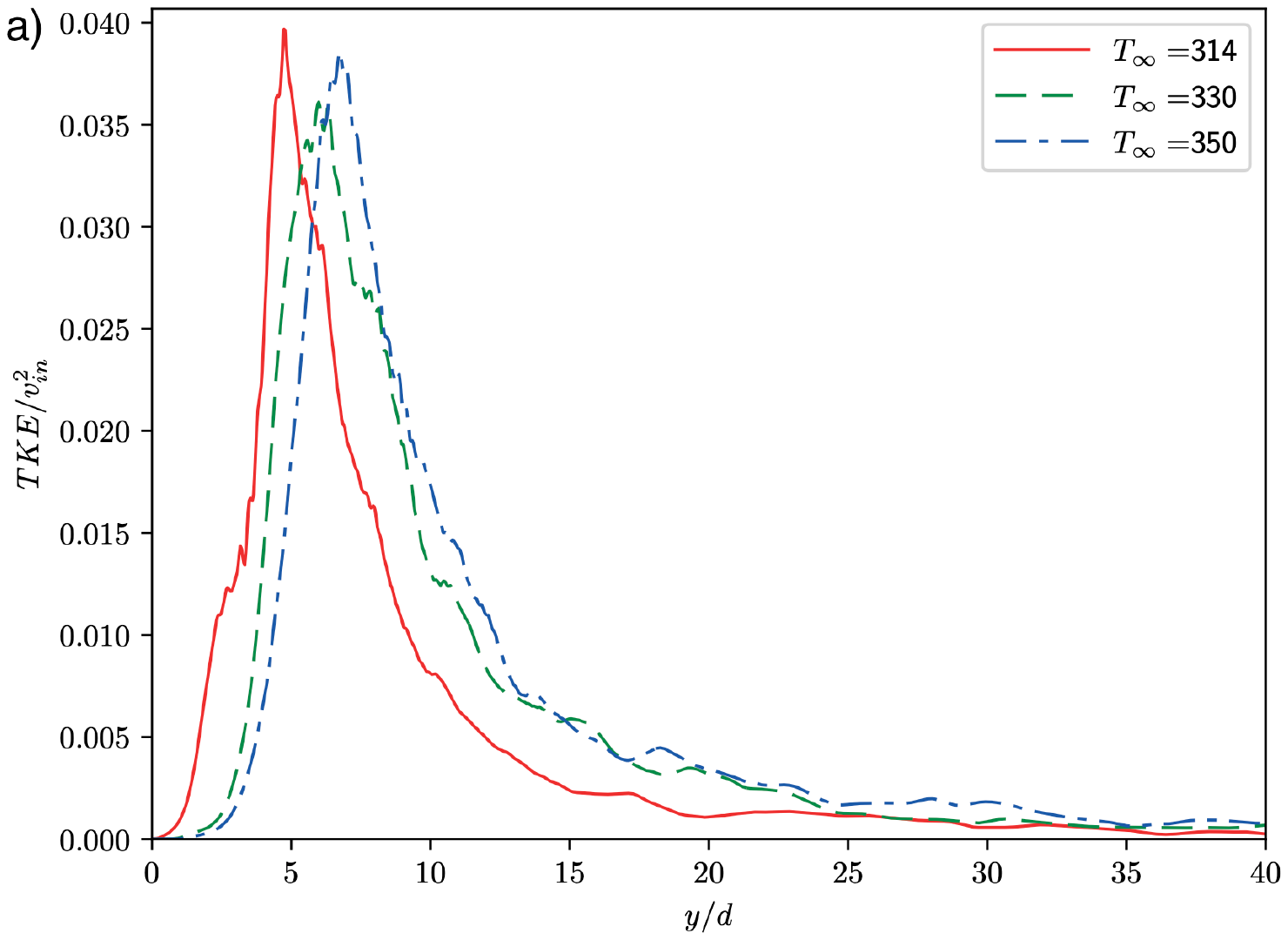}
	\label{TKE_centerline_1}
}
\subfloat[] { 
	\includegraphics[scale=.27]{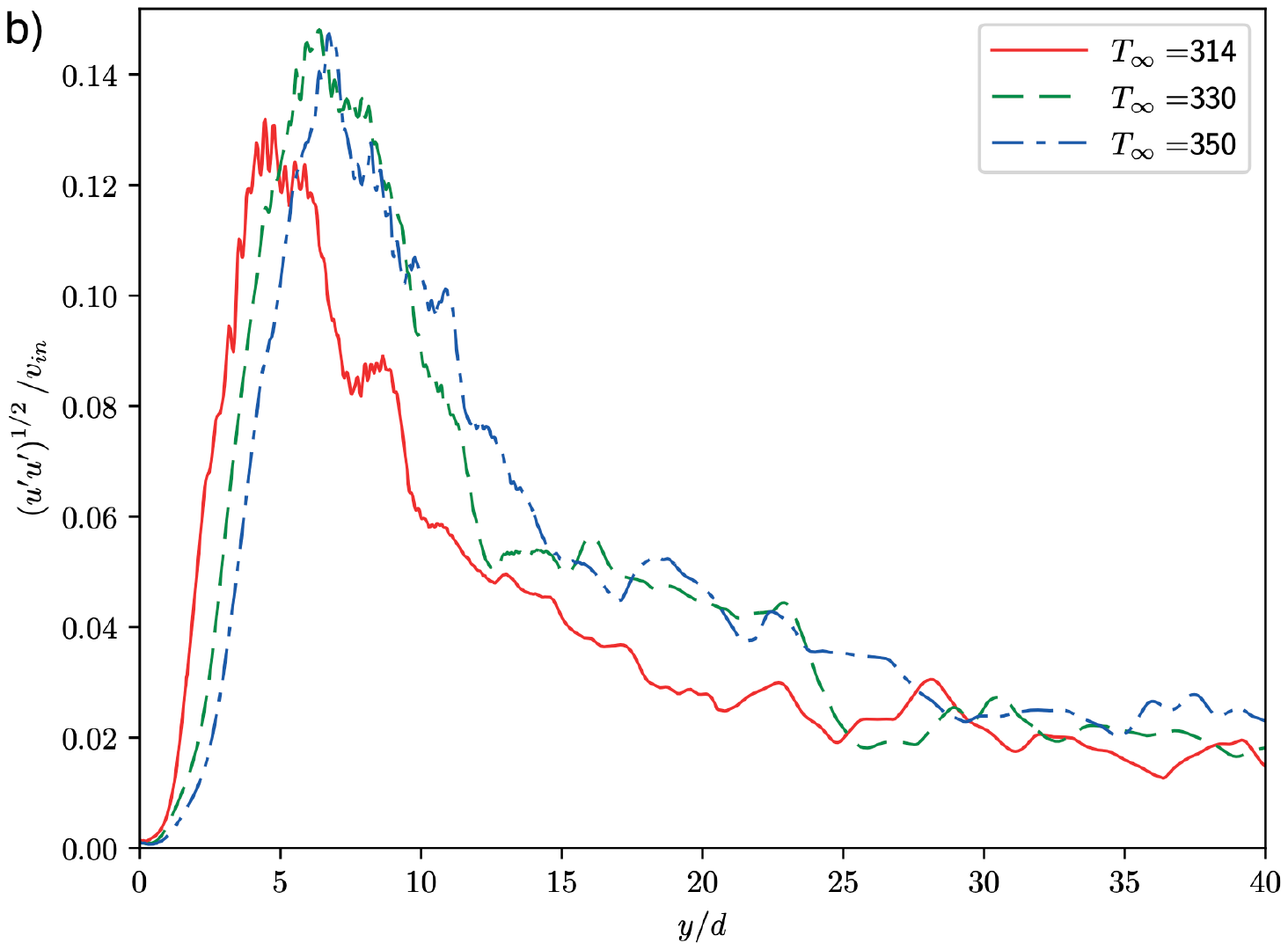}
	 \label{u_fa_1}
}
\vspace{-32pt}
\subfloat[] { 
	\includegraphics[scale=.27]{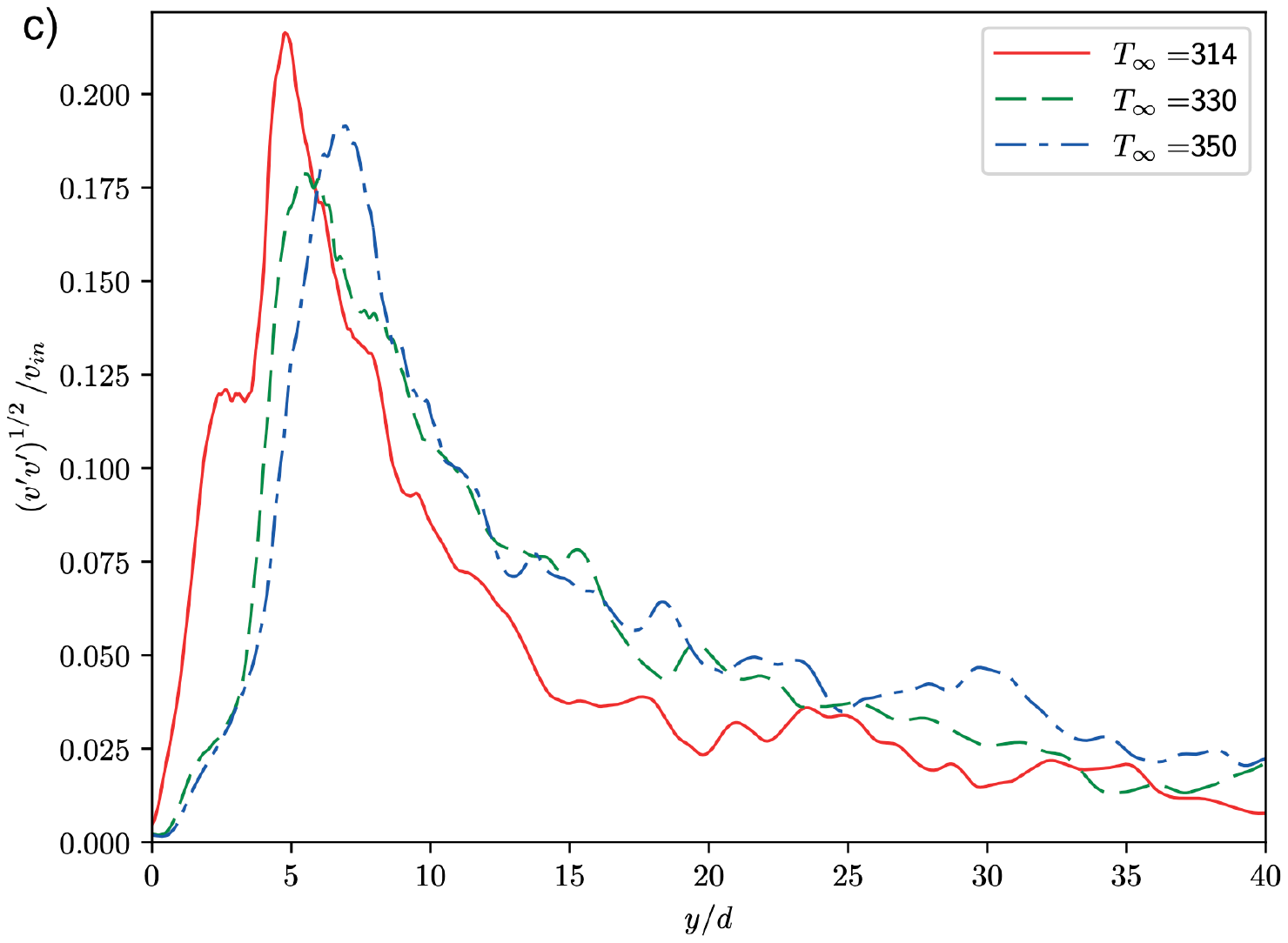}
	\label{v_fa_1}
}
\subfloat[] { 
	\includegraphics[scale=.27]{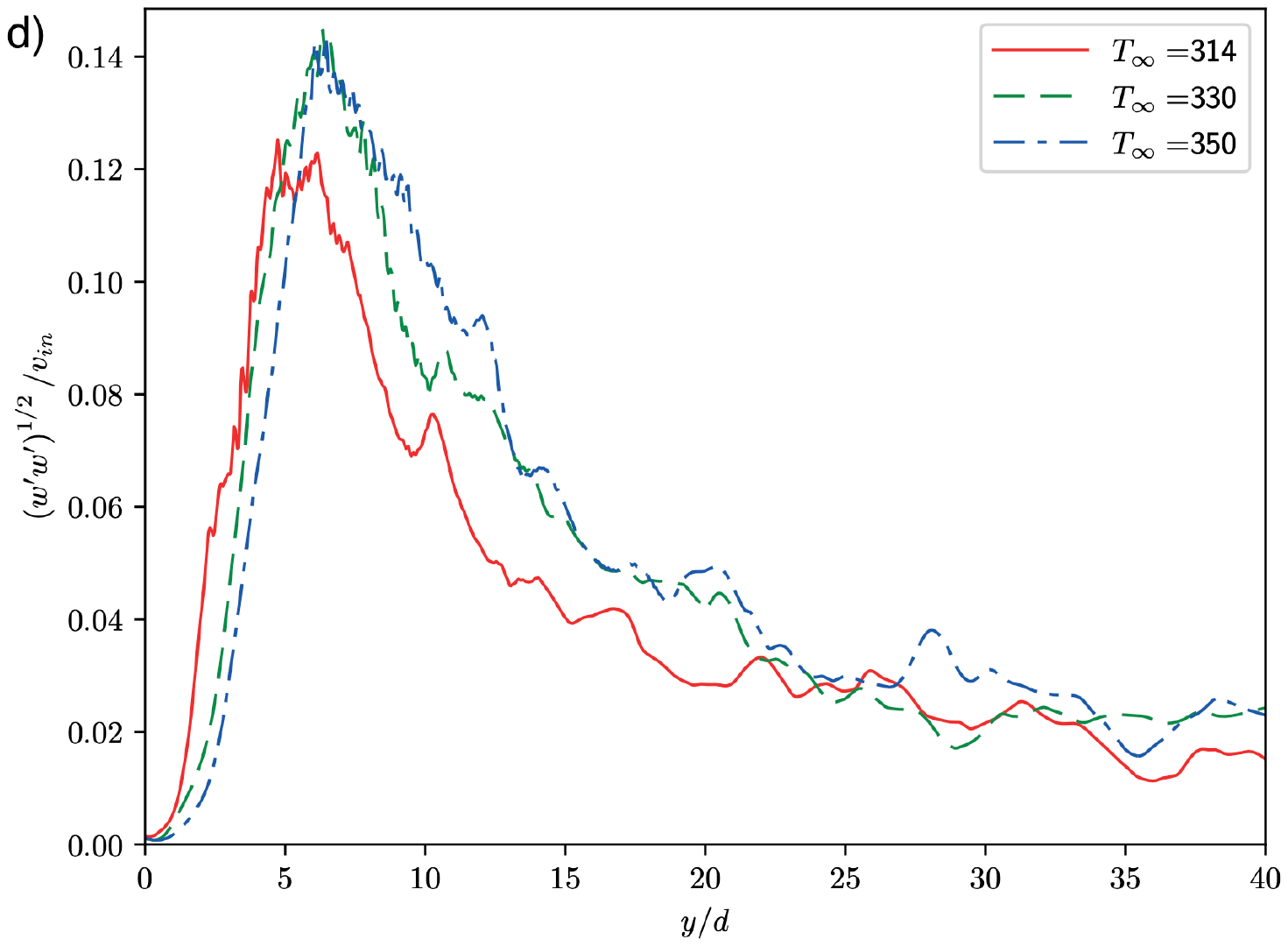}
	 \label{w_fa_1}
}
\caption{Cross-case comparisons of average \gls{tke} components along centerline including (a) Resolved \gls{tke} and (b) u-directional, (c) v-directional, and (d) w-directional \gls{tke} components.}
\label{noniso_TKE_features}
\end{center}
\end{figure}

Figure \ref{noniso_TKE_features} further emphasizes the differences between the \gls{tke} components and overall \gls{tke} by directly comparing each quantity across all cases. For all quantities, the 314~K ambient case peaks before the other two cases. For the overall resolved \gls{tke} and the axial component, the 330~K ambient case is next to the peak, followed by the 350~K ambient case. For the cross directional components, the peaks of these two cases coincide. The 314~K ambient cases experiences the strongest decay along the centerline immediately following the peak for each quantity, however, the other two cases eventually decay to the same value so that all cases exhibit a roughly equivalent leveling off as the jet progresses through the remainder of the domain. 

\subsubsection{Discussion}
Key differences in the flow structure between the 350~K and 314~K ambient cases of supercritical fluid injection are noted. Vertical slice imaging in Figures \ref{all_v_features}, \ref{all_pressure_features}, \ref{all_magvort_features}, \ref{noniso_various_features} depicts qualitative differences between the non-isothermal cases. In general, the 350~K ambient case has a very similar flow structure compared to the isothermal case. Thus, the introduction of variable density dynamics does not heavily impact jet behavior when moving away from the pseudo-boiling point. This is not the case when moving across the pseudo-boiling point. 

One of the main effects of crossing the pseudo-boiling point that can be seen in the vertical slice images and centerline plots is the formation of barrier of high constant-pressure specific heat. The effects of this ``thermal-shielding'' \cite{10.1063/1.5054797} on a denser jet can be seen in the experimental work of Mayer \emph{et al.} \cite{mayer2003raman} and Chehroudi \emph{et al.} \cite{chehroudi2002cryogenic}, and some of the numerical counterparts of these studies \cite{10.1063/1.5054797, doi:10.1063/1.1795011}. In these studies, the thermal-shield contributes to a sharp density gradient at the jet-ambient interface, which stabilizes the shear layer and allows for further jet penetration before jet breakup \cite{10.1063/1.5054797}. Density gradient stratification in these cases also contributes to finger-like structures formed downstream in the pseudo-boiling case, which are not evident in the case moving away from the pseudo-boiling point \cite{10.1063/1.5054797}. In this study, however, a lighter, warmer jet is propelled into a denser, cooler ambient fluid for the case that crosses the pseudo-boiling point, so the specific heat peak enhances a flipped density gradient. The stabilizing effects of density stratification here may be overcome by the volume dilatation \cite{Li2012}, not only enabling but amplifying the vortex formation related to a Kelvin-Helmholtz-like instability. 

Spreading rate and decay of the jet is ultimately also enhanced by the pseudo-boiling region, as compared to both the non-isothermal case here and typical values for spreading rate and decay seen in the literature for subcritical jets \cite{iso_comp_1_ref_1}. As larger vortex pairs deteriorate at the end of the potential core, high levels of mixing occur as denser ambient fluid is entrained in the lighter jet. This entrained fluid with higher thermal conductivity and constant-pressure specific heat allows for better transfer of thermal energy and faster density adjustment. As this mixing occurs, jet penetration is slowed and the lower-density liquid builds and expands spanwise in the denser-ambient surrounding. This enhanced spreading and decay is also found in similar numerical simulations \cite{Li2012} and some recent experiments \cite{10.1063/5.0072291}.

A feature not found in the literature that is explored here is the increase in anisotropy regarding \gls{tke} when crossing the pseudo-boiling point. In general, anisotropy of the resolved \gls{tke} in the isothermal and other non-isothermal cases is comparable to that seen in higher Mach or lower Reynolds flow subcritical compressible flows as is explored in \cite{iso_comp_2}. This implies that compressibility may play an important role in the enhanced mixing features seen in not only supercritical fluids in general, but especially in the case where pseudo-boiling is involved.

\section{Conclusions}

In this work, we studied three cases of a supercritical round turbulent jet injected into a supercritical environment: an isothermal case, a non-isothermal case with ambient temperature farther away from the critical point, and a non-isothermal case with ambient temperature closer to the critical point and specifically crossing a region of intense thermodynamic fluctuation known as the pseudo-boiling point. It was shown that the isothermal case had many similarities with ideal incompressible round jets. Many properties found in classical round turbulent jets were recovered in this case, such as self-similarity, linear decay along the centerline, and general trends associated with Reynolds stresses (although self-similarity in the Reynolds stresses was not recovered; general trends were still agreeable). The non-isothermal case farther away from the critical point behaved similarly to the isothermal case with slight differences being noted in potential core length and spreading rate of the jet. The non-isothermal case that transited the pseudo-boiling point exhibited noticeably different behavior.

The effects of pseudo-boiling density stratification was demonstrated for low-to-high density supercritical jet injection, in contrast to the more common high-to-low density injection scheme found in existing literature. Earlier onset of Kelvin-Helmholtz-like instabilities in the jet-ambient shear interface ultimately contributes to faster mixing and jet decay as compared to the other non-isothermal case presented here. Of note with this regard, resolved \gls{tke} in this case is redirected from the spanwise direction to the streamwise direction, displaying enhanced isotropy in the pseudo-boiling case. To the authors' knowledge, this feature has not been explored in the current literature regarding pseudo-boiling in supercritical jet configurations. Jet decay and spreading rate are both enhanced compared to the other non-isothermal case, but may be even more so due to pseudo-boiling, indicating a potential for larger heat transfer and more rapid combustion for applications of interest. A further investigation regarding lighter density supercritical injection could be useful in exploring this effect further. 

Some limitations of this work are noted. The \gls{sgs} models used in this work do not take into consideration real gas behavior, being limited to assumptions based on the ideal gas equation of state. There is room for improvement in turbulence modeling with the aim of incorporating higher order equations of state to the model formulation. As with any numerical simulation, discretization and resolution inherently lead to error. Ideally with more computational power, finer resolution can be achieved to help improve simulations. Finally, minor artifacts were noted in the non-isothermal case closest to the critical point likely due to the presence of steep gradients and issues with post-processing procedures. Though the impact of these are believed to be mild, they are still present and add to cumulative error. These artifacts are limited to the mesh interface transition from the inlet region as described in Figure~\ref{fig:domain} and are not seen in other regions where high shear or gradients might be present. Data collected for analysis begins beyond this region, with no noticeable impact seen on centerline slices. Evidence of self-similarity trends, as seen in Figure~\ref{noniso_near_r_v_features}, along with the centerline decay in Figures \ref{noniso_v_vin_r_d_features} and \ref{noniso_uin_u0_x_d_features}, as relates to general turbulent jet theory add confidence toward the validity of the results presented here. 

Future work aims to incorporate multiple species to the flow system and to investigate higher Mach flows. The impact of more specific application-oriented configurations with complex geometries on the flow field is also of interest. 

\section*{Acknowledgments}

This research was supported in part by an appointment with the National Science Foundation (NSF) Mathematical Sciences Graduate Internship (MSGI) Program sponsored by the NSF Division of Mathematical Sciences. This program is administered by the Oak Ridge Institute for Science and Education (ORISE) through an interagency agreement between the U.S. Department of Energy (DOE) and NSF. ORISE is managed for DOE by ORAU. All opinions expressed in this paper are the author's and do not necessarily reflect the policies and views of NSF, ORAU/ORISE, or DOE. 

This work was authored in part by NREL for the U.S. Department of Energy (DOE), operated under Contract No. DE-AC36-08GO28308. Funding was provided by the Exascale Computing Project (17-SC-20-SC), a collaborative effort of two U.S. Department of Energy (DOE) organizations, the Office of Science and the National Nuclear Security Administration. The views expressed in the article do not necessarily represent the views of the DOE or the U.S. Government. The U.S. Government retains and the publisher, by accepting the article for publication, acknowledges that the U.S. Government retains a nonexclusive, paid-up, irrevocable, worldwide license to publish or reproduce the published form of this work, or allow others to do so, for U.S. Government purposes. The research was performed using computational resources sponsored by the Department of Energy's Office of Energy Efficiency and Renewable Energy and located at the National Renewable Energy Laboratory.

The authors wish to thank Nicholas Wimer of NREL for his helpful comments on initial versions of this manuscript. Additionally, the authors thank Andrew Myers, Ann Almgren, and Marc Day for help in investigating and resolving early post-processing issues.

\appendix

\section{Transport Coefficient Formulation}

As stated in Eqn. \eqref{trans_decomp}, transport components can be separated into a low-pressure component incorporating ideal gas behavior and a high-pressure component that captures deviation from the ideal as formulated by Chung et al. \cite{chung:1988}. These formulations as implemented in \textit{PelePhysics} are presented here for convenience.

For viscosity, these components are calculated as follows:
\begin{equation}
\begin{split}
\mu_k = \mu_0 \left( \dfrac{1}{G_2} + A_6 Y \right), \\ 
\mu_p = \left(\dfrac{\num{36.344e-6}(MT_c)^{1/2}}{V_c^{2/3}}\right)A_7Y^2G_2\exp(A_8 + \dfrac{A_9}{T^*} + \dfrac{A_{10}}{T^{*2}}), 
\end{split}
\end{equation}
where $M$ is the molecular weight, $V_c$ is the critical molar volume, $T^* = T/\epsilon_k$ is a dimensionless temperature scaling using the Lennard-Jones potential well depth \cite{LennardJones_1931}, $Y = (\rho V_c)/6$, and $G_1 = (1-0.5Y)/(1-Y)^3$, and $G_2 = \left\{A_1\left[   1-\exp(-A_4Y)\right]/Y + A_2G_1\exp(A_5Y) + A_3G_1 \right\} \\ /(A_1A_4 + A_2 + A_3)$. The constants $A_{1-10}$ are linear functions calculated as follows: 
\begin{equation}
A_i = a_{i0} + a_{i1} \omega + a_{i2} \mu_r^4 + a_{i3} \kappa \quad i = 1,..., 10,
\end{equation} 
where $\mu_r$ is the reduced dipole moment of the species, $\kappa$ is the association factor of the species, and $a_{ij}$ are constants (see Table \ref{chung-aij}).  

Similarly, thermal conductivity components are given by: 
\begin{equation}
\begin{split}
\lambda_k = \lambda_0 \left( \dfrac{1}{H_2} + B_6 Y \right), \\ 
\lambda_p = \left(\dfrac{\num{3.039e-4}(T_c/M)^{1/2}}{V_c^{2/3}}\right)B_7Y^2H_2T_r^{1/2},
\end{split}
\end{equation}
where $H_2 = \left\{B_1\left[   1-\exp(-B_4Y)\right]/Y + B_2G_1\exp(B_5Y) + B_3G_1 \right\}/(B_1B_4 + B_2 + B_3)$ and $B_{1-7}$ are defined as: 
\begin{equation}
B_i = b_{i0} + b_{i1} \omega + b_{i2} \mu_r^4 + b_{i3} \kappa \quad i = 1,..., 7,
\end{equation}
where $b_{ij}$ are constants (see Table \ref{chung-bij}). All species-related constants mentioned in this section can also be found in Table \ref{transport-coeffs-other}. 

\begin{table}[!hbpt]
\caption{Quantities used in part for Chung's High Pressure Corrections. Values gathered from EGLib approximations \cite{ERN1995105}}
\label{transport-coeffs-other}
\begin{center}
\begin{tabular}{ r | r }
Transport Quantity, Symbol (Units)& Value  \\
\hline
Molar Mass, $M$ (g/mol) & $\num{4.400995e1}$  \\
Lennard Jones Potential Well Depth, $\epsilon_k$ (K)& $\num{2.440000e2}$  \\
Lennard Jones Collision Diameter, $\sigma$ (\AA) & $3.763$  \\
Reduced Diple Moment, $\mu_r$ (D) & $0.0$  \\
Association Factor, $\kappa$ & $0.0$
\end{tabular}
\end{center}
\end{table}

Coefficients used in the ideal gas component of each transport coefficient as formulated in Eqn. \eqref{EGLIB} are listed in the Table \ref{EGLibCoeffs}.

\begin{table}[!hbpt]
\caption{Coefficients for EGLib Polynomials \cite{ERN1995105} used to model transport quantities as described in Equation \eqref{EGLIB}}
\label{EGLibCoeffs}
\begin{center}
\begin{tabular}{ r | r r }
$i$ & $a_{\mu,i}$& $a_{\lambda,i}$  \\
\hline
0 & $\num{-2.28110345e1}$ & $-8.74831432$  \\
1 & $4.62954710$ & $4.79275291$  \\
2 & $\num{-5.00689001e-1}$ & $\num{-4.18685061e-1}$  \\
3 & $\num{2.10012969e-2}$ & $\num{1.35210242e-2}$  \\
\end{tabular}
\end{center}
\end{table}

\begin{table}[hbpt!] 
\caption{Linear Coefficients Used in Calculating High Pressure Viscosity Corrections as found by Chung et al. \cite{chung:1988}}
\label{chung-aij}
\begin{center}
\begin{tabular}{ r | r r r r }
$i$ & $a_{i0}$ & $a_{i1}$ & $a_{i2}$ & $a_{i3}$ \\
\hline
1 & $6.32402$ & $50.41190$ & $-51.68010$ & $1189.02000$ \\
2 & $\num{0.12102e-2}$ & $\num{-0.11536e-2}$ & $\num{-0.62571e-2}$ & $\num{0.37283e-1}$ \\
3 & $5.28346$ & $254.20900$ & $-168.48100$ & $3898.27000$ \\
4 & $6.62263$ & $38.09570$ & $-8.46414$ & $31.41780$ \\
5 & $19.74540$ & $7.63034$ & $-14.35440$ & $31.52670$ \\
6 & $-1.89992$ & $-12.53670$ & $4.98529$ & $-18.15070$ \\
7 & $24.27450$ & $3.44945$ & $-11.29130$ & $69.34660$ \\
8 & $0.79716$ & $1.11764$ & $\num{0.12348e-1}$ & $-4.11661$ \\
9 & $-0.23816$ & $\num{0.67695e-1}$ & $-0.81630$ & $4.02528$ \\
10 & $\num{0.68629e-1}$ & $0.34793$ & $0.59256$ & $-0.72663$ \\
\end{tabular}
\end{center}
\end{table}
\begin{table} \label{chung_conductivity_bij}
\caption{Linear Coefficients Used in Calculating High Pressure Thermal Conductivity Corrections as found by Chung et al. \cite{chung:1988}}
\label{chung-bij}
\begin{center}
\begin{tabular}{ r | r r r r }
$i$ & $b_{i0}$ & $b_{i1}$ & $b_{i2}$ & $b_{i3}$ \\
\hline
1 & $2.41657$ & $0.74824$ & $-0.91858$ & $121.72100$ \\
2 & $-0.50924$ & $-1.50936$ & $-49.99120$ & $69.98340$ \\
3 & $6.61069$ & $5.62073$ & $64.75990$ & $27.03890$ \\
4 & $14.54250$ & $-8.91387$ & $-5.63794$ & $74.34350$ \\
5 & $0.79274$ & $0.82019$ & $-0.69369$ & $6.31734$ \\
6 & $-5.86340$ & $12.80050$ & $9.58926$ & $-65.52920$ \\
7 & $81.17100$ & $114.15800$ & $-60.84100$ & $466.77500$ \\
\end{tabular}
\end{center}
\end{table}

\section{Thermodynamic and Transport Model Validation}

\begin{figure}[hbpt!]
\subfloat[] { 
	\centering
	\includegraphics[scale=.25]{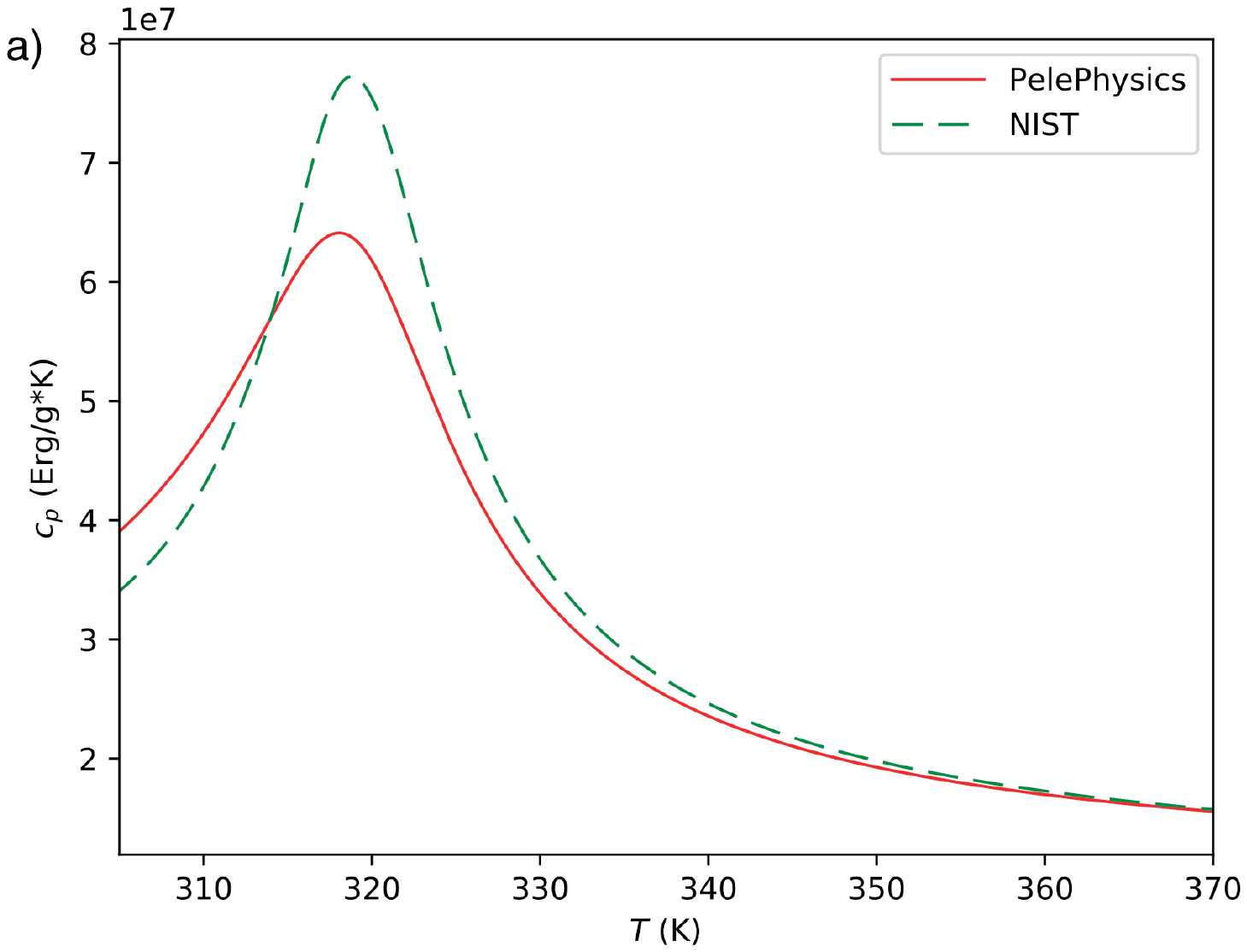}
}
\subfloat[] { 
	\centering
	\includegraphics[scale=.25]{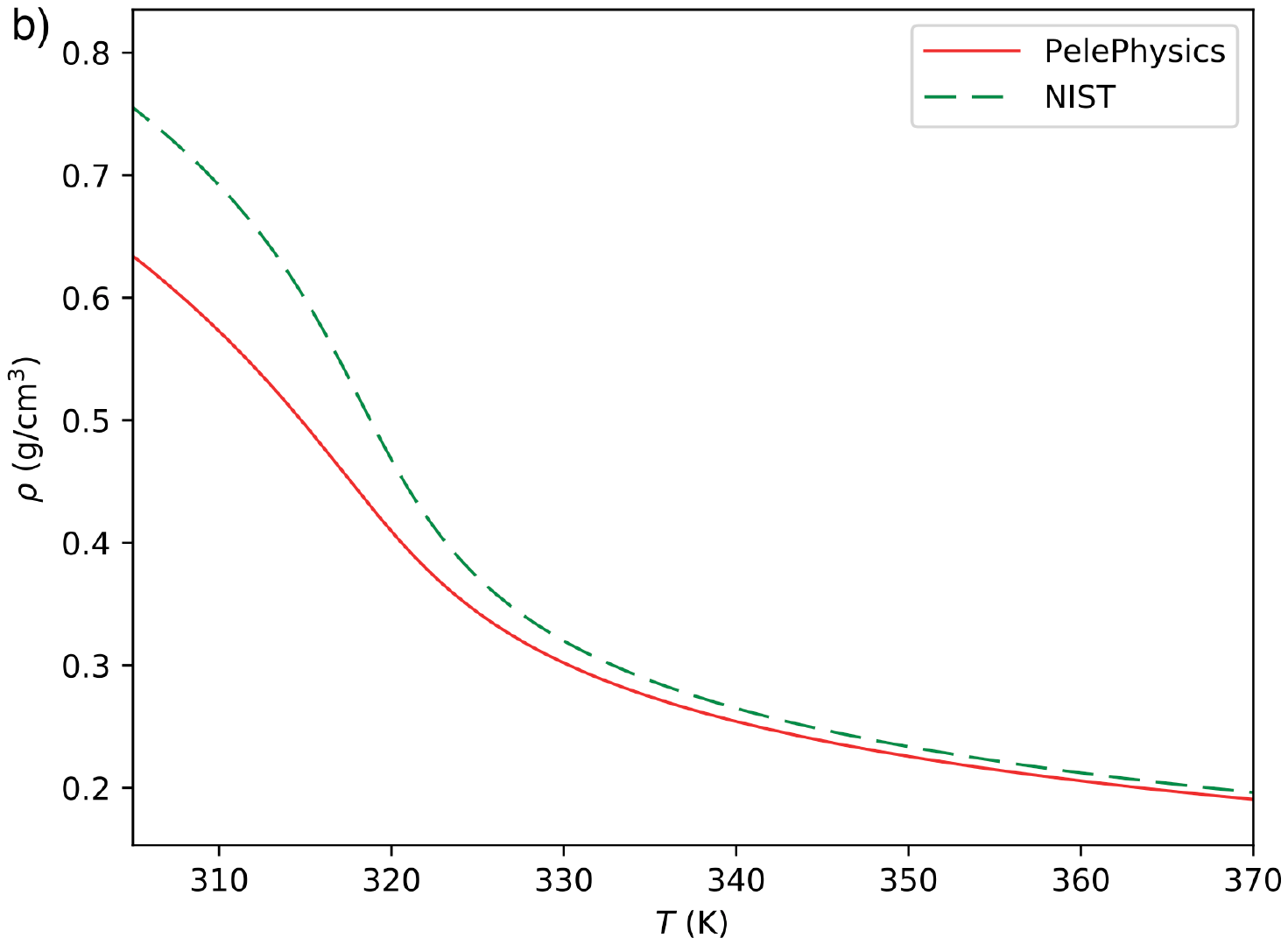}
}
\vspace{-32pt}
\subfloat[] { 
	\centering
	\includegraphics[scale=.25]{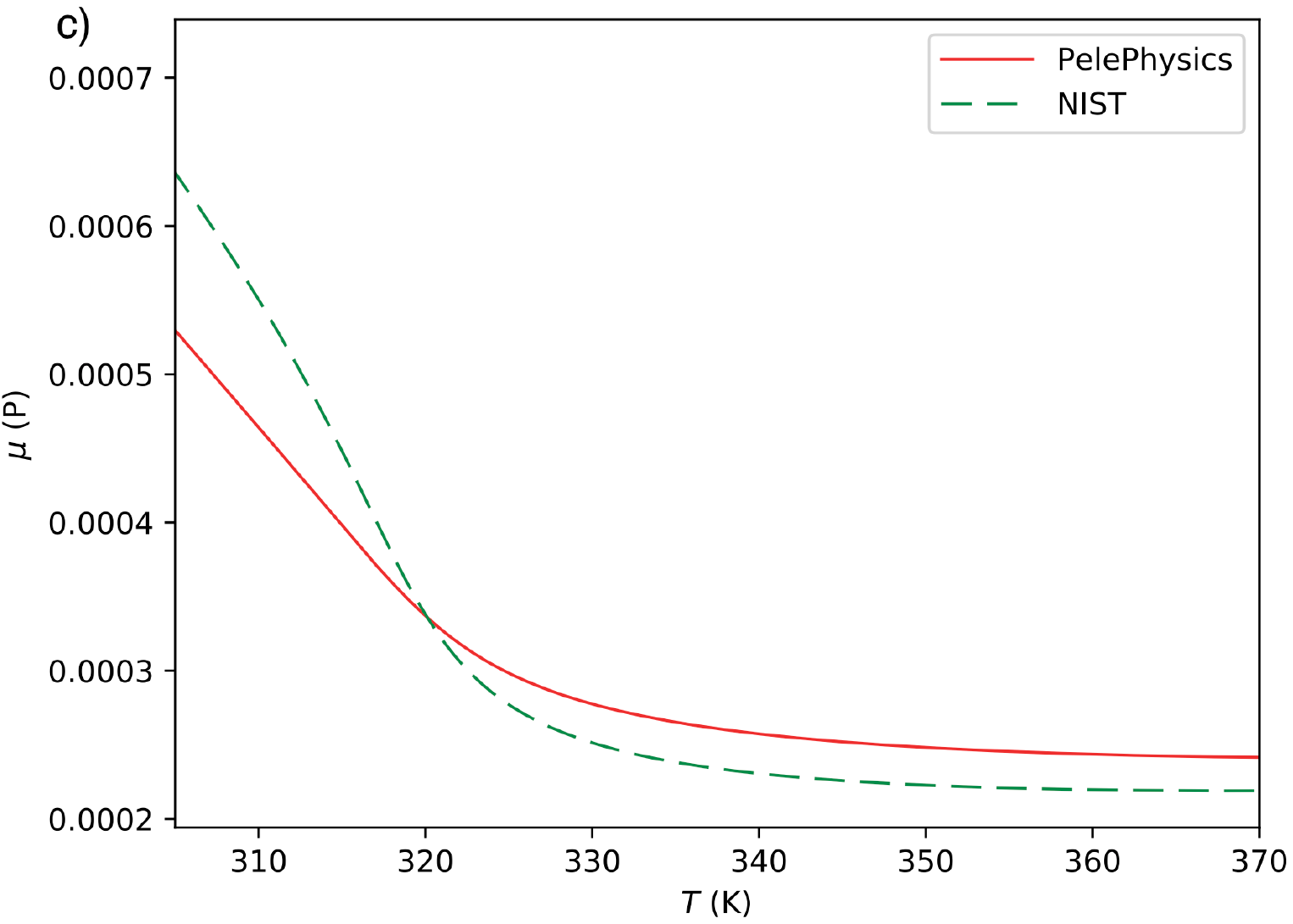}
}
\subfloat[] { 
	\centering
	\includegraphics[scale=.25]{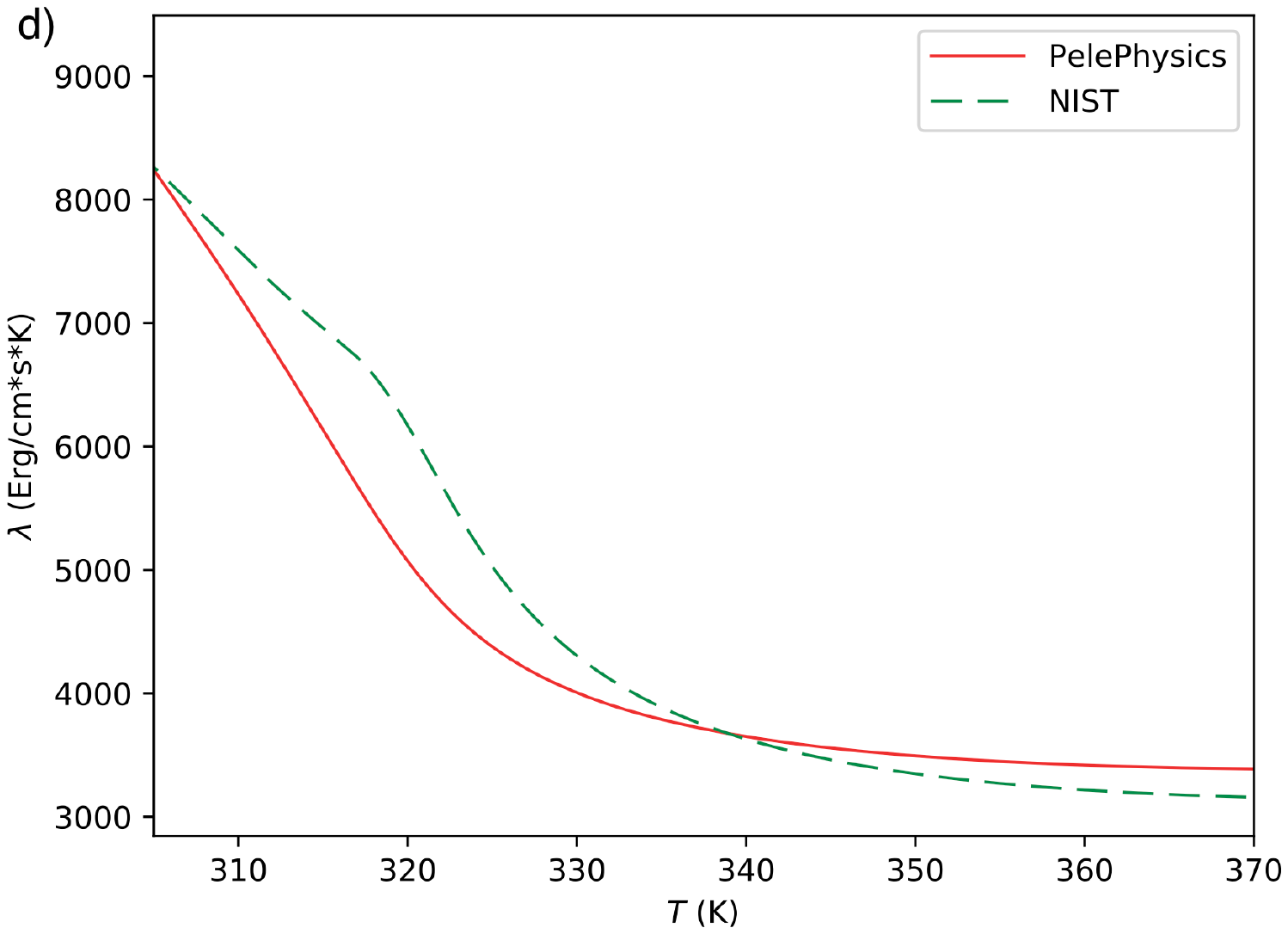}
}
\caption{Comparison of NIST WebBook values \cite{NIST} with output values from the thermodynamic and transport models described here as implemented in \textit{PelePhysics} for (a) constant-pressure specific heat, (b) density, (c) viscosity, and (d) thermal conductivity.}
\label{NIST_quantities_compare}
\end{figure}

We compare some thermodynamic and transport quantities with highly accurate data from the NIST WebBook \cite{NIST} to quantify the error present in our model within the regime of interest. Overall, quantities of interest relevant to this work follow the same general trends as the NIST data, but with larger discrepancies between the two noted near the pseudo-boiling region denoted by the peak in specific heat, as can be seen in Figure \ref{NIST_quantities_compare}. Peak error values occurring in the pseudo-critical region are also noted by \cite{rasmussen_how_2021}. 

Most of the largest error values can be found in the $314 - 330$ K range, which means modeling errors will have the largest impact on the pseudo-boiling case presented in this work. Table \ref{max_error} contains the largest percent errors over the operating range for all cases for each quantity along with the temperature value of that peak error location. Future studies will explore methods of achieving higher accuracy in thermodynamic and transport modeling in this challenging parameter regime. Since general trends are still followed by the models, results from this study still provide insight into this region.

\begin{table}[H]
\caption{Maximum percent error across operating range between NIST WebBook Data and simulation models for constant-pressure specific heat, density, viscosity, and thermal conductivity.}
\label{max_error}
\begin{center}
\begin{tabular}{ r | r r }
Parameter & Percent Error & Temperature (K)  \\
\hline
$c_p$ & 18.11 & 320.25  \\
$\rho$ & 17.41 & 314  \\
$\mu$ & 12.62 & 314  \\
$\lambda$ & 17.81 & 319.88  \\

\end{tabular}
\end{center}
\end{table}

\clearpage

\bibliography{sCO2}

\end{document}